\documentclass[prd,nofootinbib,onecolumn]{revtex4}
\usepackage{amsmath} %
\usepackage{graphicx} %

\usepackage[breaklinks=true]{hyperref} %

\IfFileExists{hypernat.sty}{\usepackage{hypernat}}

\usepackage{bm,paralist,xspace}

\newcommand{\lya}{Ly-$\alpha$\xspace}
\begin{document}

\title{A lower bound on the mass of Dark Matter particles} %
\author{Alexey~Boyarsky$^{a,b}$, Oleg~Ruchayskiy$^c$, and
  Dmytro~Iakubovskyi$^b$}

\affiliation{\it $^{a}$ETHZ, Z\"urich, CH-8093,
  Switzerland\\
  $^{b}$Bogolyubov Institute for Theoretical Physics, Kiev 03680, Ukraine\\
  $^{c}$\'Ecole Polytechnique F\'ed\'erale de Lausanne, Institute of
  Theoretical Physics, FSB/ITP/LPPC, BSP 726, CH-1015, Lausanne, Switzerland }

\newcommand{\erf}{\mathop{\rm erf}\nolimits} \newcommand{\tr}{\mathop{\rm
    tr}\nolimits} \newcommand{\sgn}{\mathop{\rm sgn}\nolimits}
\newcommand{\p}{\partial}
\newcommand{\Sp}{\mathop{\rm Sp}\nolimits}
\newcommand{\dm}{{\textsc{dm}}} 
\renewcommand{\deg}{{\textsc{deg}}} 
\newcommand{\fd}{{\textsc{fd}}} 
\newcommand{\dw}{{\textsc{nrp}}} 
\newcommand{\tg}{{\textsc{tg}}} 
\renewcommand{\sf}{{\textsc{rp}}} 
\newcommand{\mdm}{\ensuremath{m_\textsc{dm}}\xspace} 
\newcommand{\lfs}{{\Lambda_\textsc{FS}}} 
\newcommand{\qhd}{{Q}} 
\newcommand{\m}{\ensuremath{m}\xspace} 
\newcommand{\numsm}{$\nu$MSM\xspace} 
\newcommand{\ev}{\:\mathrm{eV}} 
\newcommand{\kev}{\:\mathrm{keV}} 
\newcommand{\mev}{\:\mathrm{MeV}} 
\newcommand{\gev}{\:\mathrm{GeV}} 
\newcommand{\pc} {\:\mathrm{pc}} 
\newcommand{\kpc}{\:\mathrm{kpc}} 
\newcommand{\mpc}{\:\mathrm{Mpc}} 
\newcommand{\gpc}{\:\mathrm{Gpc}} 
\newcommand{\const}{\mathrm{const}} %
\newcommand{\parfrac}[2]{\left(\frac{#1}{#2}\right)}
\newcommand{\eq}[1]{\begin{equation} #1 \end{equation}}
\newcommand{\ml}[1]{\begin{multline} #1 \end{multline}}
\newcommand{\eV} {\ensuremath{\:\mathrm{eV}}}   %
\newcommand{\keV}{\ensuremath{\:\mathrm{keV}}} %
\newcommand{\MeV}{\ensuremath{\:\mathrm{MeV}}} %
\newcommand{\GeV}{\ensuremath{\:\mathrm{GeV}}} %
\newcommand{\cm}{\:\mathrm{cm}} 
\newcommand{\km}{\:\mathrm{km}} 
\newcommand{\s}{\:\mathrm{sec}} 
\newcommand{\ks}{\:\mathrm{ksec}} 
\newcommand{\sr}{\:\mathrm{sr}} 
\newcommand{\ph}{\:\mathrm{ph}} 
\newcommand{\g}{\:\mathrm{g}} 
\newcommand{\cts}{\:\mathrm{cts}} 
\newcommand{\arcmin}{\:\mathrm{arcmin}} 
\newcommand{\xmm}{\textsl{XMM-Newton}\xspace} %


\begin{abstract}

  We discuss the bounds on the mass of Dark Matter (DM) particles, coming from
  the analysis of DM phase-space distribution in dwarf spheroidal galaxies
  (dSphs). After reviewing the existing approaches, we choose two methods to
  derive such a bound. The first one depends on the information about the
  current phase space distribution of DM particles only, while the second one
  uses both the initial and final distributions. We discuss the recent data on
  dSphs as well as astronomical uncertainties in relevant parameters. As an
  application, we present lower bounds on the mass of DM particles, coming
  from various dSphs, using both methods. The model-independent bound holds
  for any type of fermionic DM. Stronger, model-dependent bounds are quoted
  for several DM models (thermal relics, non-resonantly and resonantly
  produced sterile neutrinos, etc.). The latter bounds rely on the assumption
  that baryonic feedback cannot significantly increase the maximum of a
  distribution function of DM particles. For the scenario in which
    all the DM is made of sterile neutrinos produced via non-resonant mixing
    with the active neutrinos (NRP) this gives $m_{\dw}>1.7$ keV.  Combining
    these results in their most conservative form with the X-ray bounds of DM
    decay lines, we conclude that the NRP scenario remains allowed in a very
    narrow parameter window only.
  This conclusion is independent of the results of the Lyman-alpha analysis.
  The DM model in which sterile neutrinos are resonantly produced in the presence of
  lepton asymmetry remains viable. Within the minimal neutrino extension of the
  Standard Model (the $\nu$MSM), both mass and the mixing angle of the DM
  sterile neutrino are bounded from above and below, which suggests the
  possibility for its experimental search.

\end{abstract}


\maketitle

\section{Introduction}
\label{sec:introduction}

The nature of Dark Matter is one of the most intriguing questions of particle
astrophysics. Its resolution would have a profound impact on the development
of particle physics beyond the Standard Model.

Although the possibility of having massive compact halo objects (MACHOs) as a
dominant form of DM is still under debate (see recent discussion
in~\cite{Calchi:07} and references therein), it is widely believed that Dark
Matter is composed of non-baryonic particles.  However, the Standard Model of
elementary particles does not contain a viable Dark Matter particle candidate
-- a massive, neutral and long-lived particle. Active neutrinos, which are both
neutral and stable, form structures in a top-down
fashion~\cite{Zeldovich:70,Bisnovatyi:80,Bond:80,Doroshkevich:81,Bond:83}, and
thus cannot produce the observed quantity of early-type galaxies~\cite[see
e.g.][]{White:83,Peebles:84a}.  Therefore, the DM particle hypothesis implies
the extension of the Standard Model (SM).

The DM particle candidates may have very different masses (for reviews of DM
candidates see e.g.~\cite{Bergstrom:00,Bertone:05,Carr:06,Taoso:07}): massive
gravitons with the mass $\sim 10^{-19}\ev$~\cite{Dubovsky:04}, axions with the
mass $\sim 10^{-6}\ev$~\cite{Holman:83}, sterile neutrinos having mass in the
keV range~\cite{Dodelson:93}, sypersymmetric (SUSY) particles
(gravitinos~\cite{Pagels:82}, neutralinos~\cite{Haber:85},
axinos~\cite{Covi:99} with their masses ranging from eV to hundreds GeV,
supersymmetric Q-balls~\cite{Kusenko:97b}, WIMPZILLAs with the mass $\sim
10^{13}\gev$~\cite{Kuzmin:98,Chung:99}, and many others). Thus, the mass of DM
particles becomes an important characteristic which may help to distinguish
between various DM candidates and, more importantly, may help to differentiate
among different models beyond the SM.

It was suggested in~\cite{Tremaine:79} that quite a robust and
model-independent \emph{lower bound} on the mass of DM particles can be
obtained by considering phase space density evolution of compact astrophysical
objects, most notably dwarf spheroidal satellites (dSphs) of the Milky Ways.
The idea was developed further in a number of works (see
e.g.~\cite{Madsen:84,Madsen:91,Madsen:90,Dalcanton:00,Hogan:00,Madsen:00}).

Another way to distinguish between various DM models, and particularly to put a
bound on the DM mass, is the analysis of the Lyman-$\alpha$ (Ly-$\alpha$)
forest data~\cite{Hui:97,Gnedin:01,Weinberg:03}.\footnote{Absorption feature
  by neutral hydrogen at $\lambda = 1216$~\AA~at different redshifts in
  the spectra of distant quasars.} %
This method essentially constrains the possible shape of the power spectrum of
density fluctuations at comoving scales $\sim$Mpc.
Assuming \emph{a DM model}, (i.e. a particular primordial velocity distribution
of DM particles), one can obtain a relationship between the DM particle mass
\emph{in this model} and the shape of the power spectrum, probed by
Ly-$\alpha$.

Although very promising, the Ly-$\alpha$ method is very complicated and
indirect. First of all, under the assumption that the distribution of the
neutral hydrogen traces that of the DM, one can reconstruct the power spectrum
of 
density fluctuations at redshifts $z \sim 2-5$ from the statistics of
Lyman-$\alpha$ absorption lines.  One can then perform a fit of the
Lyman-$\alpha$ data (often together with the measurements of anisotropy of
temperature of cosmic microwave background and the data of large-scale
structure surveys), to extract the information about cosmological models. This
is usually done by using the Monte-Carlo Markov chain
technique~\cite{Lewis:02}.  At redshifts probed by Ly-$\alpha$, the evolution
of structure has already entered the (mildly) non-linear stage. Therefore, to properly
relate the measured power spectrum with the parameters of a given cosmological
model one would have to perform a prohibitively large number of hydrodynamic
numerical simulations. Therefore, various simplifying approximations have to
be
realized~\cite{Theuns:98,Gnedin:01,McDonald:05,Viel:04,Viel:05,Viel:05b,Viel:05c,Regan:06a}.

Apart from these computational difficulties, the physics entering the \lya
analysis is complicated, and not yet fully understood~(see
e.g.~\cite{Kim:07a,Bolton:07a,Viel:2002ui,Viel:2001hd,Viel:2003fx}). Moreover,
the DM particles can significantly influence the background physics, further
complicating the Ly-$\alpha$ analysis~\cite{Biermann:06,Gao:07,Stasielak:07}.
For a recent overview of the \lya method see
  e.g.~\cite{Boyarsky:08c}.

The systematic uncertainties associated with both computational difficulties
and complicated physics of \lya systems are not fully explored.  Therefore,
\emph{it is very important to have a lower mass bound on DM particles from
  more direct and simple considerations}.  In this paper we discuss the
Tremaine-Gunn and related DM mass bounds based on the phase-space density
considerations as well as possible ways to strengthen them for several DM
models.  The obtained phase-space density bounds are weaker yet comparable
with Ly-$\alpha$ bounds and therefore provide an interesting alternative.
We consider a class of the so-called ``generic'' DM models, where DM particles
are produced thermally and decouple while being relativistic, thus having the
(relativistic) Fermi-Dirac momentum spectrum.  We also consider models of
non-thermal DM production. In this case the primordial velocity spectrum of DM
particles depends on the details of the production mechanism.  We analyze the
case when the velocity spectrum can be approximated by the \emph{rescaled} Fermi-Dirac
spectrum, or has two such components (a colder and a warmer one).

A very important example of such a DM particle is the \emph{sterile}
(right-handed) neutrino.  Although known as a DM candidate for some 15
years~\cite{Dodelson:93}, recently sterile neutrinos have attracted a lot of
attention. It was shown~\cite{Asaka:05a} that if one adds three right-handed
(sterile) neutrinos to the Standard Model, it is possible to explain
simultaneously the data on neutrino oscillations (see
e.g.~\cite{Fogli:05,Strumia:06,Giunti:06} for a review) and the Dark Matter in
the Universe, without introducing any new physics \emph{above electro-weak
  scale} $M_W \sim 100\gev$.  Moreover, if the masses of two of these
particles are between $\sim 100 \MeV$ and electro-weak scale and are almost
degenerate, it is also possible~\cite{Asaka:2005pn} to generate the correct baryon
asymmetry of the Universe (see e.g.~\cite{Dolgov:97,Riotto:98}). The third
(lightest) sterile neutrino can have mass in keV-MeV range\footnote{There are
  several interesting astrophysical applications of $\keV$ sterile neutrinos
  (see e.g.
  \cite{Sommer:99,Kusenko:06a,Biermann:06,Hidaka:06,Hidaka:07,Stasielak:06}
  and references therein).} %
and be coupled to the rest of the matter weakly enough to provide a viable
(\emph{cold} or \emph{warm)} DM candidate.

This theory, explaining the three observed phenomena ``beyond the SM'' within
one consistent framework, is called the
\emph{$\nu$MSM}~\cite{Asaka:2005pn,Asaka:05a} (see
also~\cite{Shaposhnikov:07b}).

Although weakly coupled, the DM sterile neutrino in the $\nu$MSM can be
produced in the correct quanities to account for all of the DM. There are several
mechanisms of production: non-resonant active-sterile neutrino oscillations
(\textbf{\emph{non-resonant production}} mechanism, \textbf{NRP})
~\cite{Dodelson:93,Dolgov:00,Abazajian:01a,Asaka:06b,Asaka:06c},
resonant active-sterile neutrino oscillations in the presence of lepton
asymmetry (\textbf{\emph{resonant production}} mechanism,
\textbf{RP})~\cite{Shi:98,Abazajian:01a,Shaposhnikov:08a,Laine:08a}, decay of
the gauge-singlet scalar field~\cite{Shaposhnikov:06} (see also
\cite{Kusenko:06a,Petraki:07,Petraki:08}).  The \lya analysis of the
  sterile neutrino DM, produced via NRP scenario
%
  was performed in a number of works~\cite{Viel:06,Seljak:06,Viel:08}. These
  bounds were recently revisited in~\cite{Boyarsky:08c}, using the SDSS \lya
  dataset together with WMAP5~\cite{Dunkley:2008ie}.  The lower bound on the
  DM mass was found to be in this case $8\kev$ (at $99.7\%$~CL).
  Ref.~\cite{Boyarsky:08c} also analyzed a more general case CWDM case : a
  mixture of NRP sterile neutrino with cold DM (see also~\cite{Palazzo:07}).
  These results were applied to the RP produced sterile neutrino
  in~\cite{Boyarsky:08d}. It was shown that the mass as low as $2\kev$ is
  compatible with \lya data.

In this paper we will analyze in detail restrictions on the sterile
neutrinos, produced via first two production mechanisms and briefly comment on
the third one in the Discussion.  In the case of non-resonant production the
primordial velocity spectrum is approximately \emph{proportional} to the
Fermi-Dirac distribution~\cite{Dolgov:00,Hansen:01} (the exact spectrum was
calculated in ~\cite{Asaka:06b,Asaka:06c}).

The paper is organized as follows. In Section~\ref{sec:overview} we review DM
mass bounds, based on the phase-space density arguments. In
Section~\ref{sec:maxim-coarse-grain} we introduce the concept of maximal
coarse-graining and propose a conservative modification of the original
Tremaine-Gunn bound. In Section~\ref{sec:analys-meas-valu} we analyze new
observational data on recently discovered dSphs
(see~\cite{Gilmore:07,Simon:07} and references therein), and use it to
determine the phase-space density of these objects.  Special attention is paid
to determine various systematic uncertainties of measured values. Our
results are summarized in Section~\ref{sec:bounds}. We conclude with the
discussion of the results, analysis of possible uncertainties and outlook for
the further improvement of the mass bounds in Section~\ref{sec:discussion}.

\section{DM mass limits}
\label{sec:overview}


If the DM particles are fermions, there is a very robust bound on their mass.
Namely, due to the Pauli exclusion principle, there exists the densest
``packing'' of the fermions in a given region of the phase space.  Decreasing
the mass of DM particles, one increases the number of them in a given
gravitationally bound object, containing DM. The requirement that the
phase-space density of the DM does not exceed that of the degenerate Fermi gas
leads to the \emph{lower mass bound}.  For example, for a spherically
symmetric DM-dominated object with the mass $M$ within the region $R$, one
obtains the lower bound $m_\deg$ on the DM mass by demanding that the maximal
(Fermi) velocity of the degenerate fermionic gravitating gas of mass $M$ in
the volume $\frac43 \pi R^3$ does not exceed the escape velocity $v_\infty =
\left(\frac{2 G_N M}{R}\right)^{1/2}$:
\begin{equation}
  \label{eq:23}
  \hbar \left(\frac{9\pi
      M}{2gm_{\deg}^{4}R^{3}}\right)^{1/3} \le \sqrt{\frac{2G_{N}M}{R}}
  \Rightarrow m_{\deg}^{4} \ge \frac{9\pi\hbar^3}{4\sqrt{2}gM^{1/2}R^{3/2}G_{N}^{3/2}}.
\end{equation}
Here and below $g$ denotes the number of internal degrees of freedom of DM
particles, and $G_N$ is the Newton's constant.  Such a consideration, applied
to various DM dominated objects, leads to the mass bound, which we will call
$m_\deg$ in what follows (see Table~\ref{tab:gilmore} below).\footnote{The
  spatially homogeneous DM distribution is only an approximation. In reality
  one should consider self-gravitating degenerate fermionic gas. It is
  possible to show that, under some external conditions, the system of weakly
  interating fermions undergoes a first-orger phase transition to a nearly
  degenerate ``fermion star''~\cite{Bilic:97}. The existence of such objects
  may also have insteresting astrophysical applications~\cite{Viollier:94}.}

The above considerations assume that the dSphs are \emph{purely
    spherical} systems. Analysis of~\cite{Martin:08} shows that ellipticity of
  stars in dSphs vary from $0.22^{+0.18}_{-0.22}$ for Leo IV to $0.80\pm 0.04$
  for Ursa Major I. 
  Simulated DM halos on the other hand tend to have
    rather moderate ellipticity, $\epsilon_{DM}\lesssim 0.32$
    \cite{Kuhlen:07}.\footnote{Therefore it is hard to explain the ellipticity of
    stars in the most elongated dSphs, see the discussion in
    \cite{Martin:08}.}
  According to Appendix~\ref{App:aspher}, the ellipticity of DM halos can
  lower the resulting limit on $m_\deg$ by $\lesssim$ 10\%.

The limit, obtained in such a way, is very robust, as it is independent of the
details of the formation history of the system. The only uncertainties
associated with it are those of astronomical nature: systematic errors in the
determination of velocity and density distribution. All these issues will be
discussed below
(Section~\ref{sec:analys-meas-valu},~\ref{sec:maxim-coarse-grain}).

\bigskip

For particular DM models (with the known primordial velocity dispersion) and
under certain assumptions about the evolution of the system which led to the
observed final state, this limit can be strengthened~%
\cite{Tremaine:79,Tremaine:86,Cowsik:87,Madsen:84,Madsen:90,Madsen:91,Madsen:01,Boyanovsky:08}.
The argument is based on the \emph{Liouville's theorem}~(see e.g.
\cite{Tremaine:86,Binney-Tremaine}) and assumes that the collapse of the
system is disipationless and collisionless.  The Liouville theorem states that
the phase-space distribution function $f(t,x,v)$ does not change in the course
of disipationless collisionless dynamics. The consequence of the Liouville
theorem is that the function $f(t,x,v)$ ``moves'' in the phase-space,
according to the Hamiltonian flow, and therefore its maximum (over the phase
space) remains unchanged. Therefore, if one could determine the
characteristics of a phase-space distribution function from astronomically
observed quantities (in the first place average density $\bar\rho$ and
velocity dispersion $\sigma$)\footnote{%
  The quantity directly measured in astrophysical observations is the
  projection of stars' velocities $\vec v(R)$ along the line of sight. We will
  denote such a projection by $V(R)$ to distinguish it from the absolute value
  of the 3D velocity $v(r)$. The 1D velocity dispersion is defined as
  $\sigma(R) = \langle V^2(R)\rangle^{1/2}$ and is in principle the function
  of the projected radius $R$. However in the DM dominated objects, for $R$
  greater than certain characteristic scale rotation curve flattens,
  $\sigma\approx \const$. It is this constant which is usually referred to as
  ``velocity dispersion''. } in dSphs (or any other DM dominated objects), the
Liouville theorem would allow to connect the measured values with the
primordial properties of DM particles.

One such characteristics of the phase-space distribution is its
\emph{maximum}.  Any physical measurement can probe only the phase-space
distribution, averaged over some phase-space region -- a \emph{coarse-grained}
phase-space density (PSD) (as opposed to exact or \emph{fine-grained} PSD).
Such a coarse-grained PSD, averaged over phase-space cells $\Delta \Pi(x,v)$
centered around points $(x,v)$ in the phase space, is defined via
\begin{equation}
  \label{eq:32}
  \bar f(t,x,v) = \frac1{\mathrm{vol}(\Delta\Pi)}\int\limits_{\Delta\Pi(x,v)} d\Pi' \,f(t,x',v')
\end{equation}
(here $\mathrm{vol}(\Delta\Pi)$ is the volume of the phase-space cell).
From the definition (\ref{eq:32}) it is clear that the maximal (over the whole
phase space) value of the coarse-grained PSD $\bar f_{max}(t)$ cannot exceed
the maximal value of the corresponding fine-grained PSD.
On the other hand, as a consequence of the Liouville theorem, the maximum of
the fine-grained PSD $f_{max}$ does not change in time. Thus, one arrives to
the following inequality
\begin{equation}
  \bar f_{max}(t) \le f_{max}\;.
\label{ff}
\end{equation}
The inequality~(\ref{ff}) allows to relate the properties of DM at present
time $t$ with its primordial properties, encoded in $f_{max}$.
For example, if one assumes that initially DM particles possess relativistic
Fermi-Dirac distribution function with some temperature $T_\fd$
(relativistically decoupled thermal relics):
\begin{equation}
  \label{eq:3}
  f_\fd(p) = \frac{g}{(2\pi\hbar)^3}\frac 1{e^{p/T_{\fd}}+1}
\end{equation}
and recovers from astronomical measurements that in the final state the
coarse-grained PSD of the system is described by the isothermal sphere (see
e.g.~\cite{Binney-Tremaine})
%
with a core radius $r_{c}$ and a 1D velocity dispersion $\sigma$, whose
maximum is given by
\begin{equation}
  \label{eq:4}
  \bar f_{iso,max} = \frac{9\sigma^2}{4\pi G_N (2\pi \sigma^2)^{3/2}r_c^2}
\end{equation}
the comparison of the maximum of the coarse-grained PSD~(\ref{eq:4}) with its
primordial (fine-grained) value leads to the so-called Tremaine-Gunn mass
bound~\cite{Tremaine:79}:
\begin{equation}
  m_\fd \ge m_\tg, \quad \text{where}\quad m_\tg^4
  \equiv \frac{9(2\pi\hbar)^{3}}{(2\pi)^{5/2}gG_N\sigma \, r_{c}^{2}}.
  \label{eq:1}
\end{equation}
For the case of initial distribution~(\ref{eq:3}) this bound is stronger than
the one, based on the Pauli exclusion principle, by a factor
$2^{1/4}$~\cite{Tremaine:79}. For different primordial DM distributions this
difference can be significant (as we will demonstrate later). We would like to
stress, though, that these stronger bounds make assumptions about the
evolution of phase-space density, while the one, based on the Pauli exclusion
principle does not assume anything about either primordial velocity
distribution of the particles, or the formation history of the observed object
and simply compares measured phase-space density with the maximally allowed
for fermions.

\bigskip

Another characteristics of the phase-space distribution function is the
``average phase-space density''
\begin{equation}
  \qhd \equiv
  \frac{\bar\rho }{\langle v^{2} \rangle^{3/2}} \;,
  \label{eq:2}
\end{equation}
introduced in \cite{Hogan:00,Dalcanton:00}.
The value of $\qhd_f$ (average PSD today) is simply defined in terms of the
observed quantities $\bar\rho$ and $\langle v^2\rangle = 3 \sigma^2$ and
therefore serves as a convenient estimator of the PSD for any DM dominated
object.  One can calculate primordial $Q_i$ for an arbitrary homogeneous
distribution function $f(p)$
\begin{equation}
  \label{eq:46}
  Q_i = \frac{g\, m^4}{(2\pi\hbar)^3}\frac{\Bigl(\int f(p)d^3p\Bigr)^{5/2}}{\Bigl(\int f({p}){p}^{2}d^3{p}\Bigr)^{3/2}}
\end{equation}
and compare it with its value today $Q_f$. It was claimed
in~\cite{Hogan:00,Dalcanton:00} that $\qhd$ cannot increase during the
evolution of DM:
\begin{equation}
  Q_i \ge Q_f.
  \label{eq:59}
\end{equation}
Applying this inequality to the dSphs, one
obtains several times stronger mass bound, than that of~\cite{Tremaine:79}.

To illustrate the origin of the inequality~(\ref{eq:59}), authors
of~\cite{Hogan:00,Dalcanton:00} noticed that in the case of the
\textit{uniform monoatomic ideal gas}, $\qhd$ is related to the usual
thermodynamic entropy per particle (see
Appendix~\ref{sec:ideal-boltzmann-gas}) and the inequality for $\qhd$
becomes a consequence of the second law of thermodynamics.  Indeed, in this
case one can see that
\begin{equation}\frac{S[f]}{N} =
  -\log\left(\frac{\qhd(\bar\rho,\sigma)\hbar^3}{m^{4}}\right) + \log
  C[f]\;,
  \label{CQ_corr}
\end{equation}
where in the right hand side of~(\ref{CQ_corr}) functional $C[f]$ does not depend on
the average density and velocity of the DM particles.

However, because of the long-range interaction of DM particles, the notion of
Boltzmann entropy is well-defined only for the primordial DM distribution and
not for the final state of DM evolution (see e.g. the discussion in
\cite{Hansen:04}).  Moreover, we will show below that in general \emph{the
  increase of entropy does not imply the decrease of} $Q$.  Indeed, the values
of $C[f]$ are different for different types of phase-space distributions $f$
and therefore they can change with time if the shape of the (coarse-grained)
distribution changes.  Namely, even if initial ($i$) and final ($f$) states
both satisfy relation~(\ref{CQ_corr}) between the entropy and $\qhd$ ($S_{i,f}
= \log C_{i,f} - \log\frac{Q_{i,f}\hbar^3}{\mdm^4}$) from the second law of
thermodynamics
\begin{equation}
  S_f\ge\
S_i \label{eq:12}
\end{equation}
it only follows that
\begin{equation}
  \label{eq:9}
  Q_i \ge Q_f\frac{C_i}{C_f}\;.
\end{equation}
Therefore, in general, the inequality~(\ref{eq:59}) does not follow from
entropic considerations.


Moreover, the simple relation~(\ref{CQ_corr}) between the entropy and $\qhd$
does not hold for the distributions we are interested in. For example, for the
Fermi-Dirac distribution~(\ref{eq:3}) one has:
\begin{equation}
  \label{eq:25}
  \frac SN = \const\frac{m_\fd^4}{\qhd \hbar^3}
\end{equation}
(see Appendix~\ref{sec:dw} for details).  The relation becomes even more
complicated, if one considers DM candidates (e.g., sterile neutrinos,
gravitinos), which are produced out of thermal equilibrium. In general, when
the primordial distribution function depends on several parameters, both
$\qhd$ and entropy are expressed through these parameters in a non-trivial way
and the simple relation~(\ref{CQ_corr}) does not hold. For example, this is
the case when DM is produces in two stages and the DM distribution shape has
two components: colder and warmer one. Physically interesting examples
include: production of sterile neutrino in the presence of lepton
asymmetry~\cite{Shi:98,Shaposhnikov:08a,Laine:08a}; production of gravitino
thermally at high temperatures (see e.g.~\cite{Bolz:00,Rychkov:07})
accompanied by non-thermal production via late decays of next-to-lightest
supersymmetric particles~(see e.g.~\cite{Borgani:96}).

Keeping in mind the above considerations, one might be tempted to use the
entropy of the system as an estimator of PSD and utilize the entropy
increase~(\ref{eq:12}) instead of the inequality on $\qhd$ to put a lower
bound on the DM mass.  However, unlike $\qhd$, which by definition is
expressed solely in terms of measured quantities $\bar\rho$ and $\sigma$, the
inequality~(\ref{eq:12}) requires the knowledge of the phase-space
distribution function in the final state (e.g. to determine the $C_f$ in the
right-hand side of Eq.~(\ref{CQ_corr}) or, more generally to express the
entropy of the final state in terms of the observed quantities).  This
information cannot be simply deduced from astronomical observations. One
possible way to formulate a conservative, robust inequality would be to find
the \emph{maximal} possible entropy for a given system with measured
macroscopic parameters.  However, it was shown
in~\cite{Antonov:62,Lynden-Bell:68,Tremaine:86,Binney-Tremaine} that such a
maximum does not exist. Namely, for a gravitating system which usually
consists of a compact core and a widely dispersed halo of finite mass, the
total Boltzmann entropy of the system goes to infinity when the halo becomes
infinite.  Physically, the measured density and velocity dispersion
characterize the inner part of the object.
The astronomical observations do not usually probe the outskirts of
gravitating systems (such as dSphs) and phase-space distributions (such as
(\ref{eq:4})) do not describe them properly. On the other hand, to compare
with the homogenous initial system having a primordial velocity spectrum, we
need to know an entropy of the \emph{whole} system. The large (and unknown!)
fraction of this entropy can be related to the outskirts. The entropy of the
gravitating system depends on the precise state of the halo.

As a result, it is not possible to construct a simple and robust limit, using
entropy considerations.


\section{Maximal coarse-graining}
\label{sec:maxim-coarse-grain}

In view of the above arguments, to derive a conservative mass bound, in this
work we will follow the original approach of Tremaine and
Gunn~\cite{Tremaine:79} with some modification.

An important advantage of this approach is that the maximum of the phase space
density is likely to be located in the inner, dense part of an object.
Therefore, under this reasonable assumption, the results \emph{do not depend}
on the DM distribution in the outskirts (see the discussion above).

As discussed already, the coarse-grained phase-space distribution in the final
state cannot be measured directly, and one has to make assumptions to deduce
its maximum. A conservative way to minimize this uncertainty is to use the
``maximally coarse-grained distribution''. It is based on a simple fact that
the mean value of a function, averaged over an arbitrary region cannot exceed
its maximal value. Therefore, the average value of coarse-grained phase space
density in a large phase-space volume can be taken as a conservative estimate
of the $\bar F_{max}$, independent on assumptions about the actual form of
phase-space distribution.

To this end we consider an (approximately spherically symmetric) gravitating
system (having in mind a dwarf spheroidal galaxy), that has the mass $M(R)$
confined within the radius $R$.  The phase-space volume, occupied by the DM
particles, forming such a system can be approximated by
\begin{equation}
  \label{eq:37}
  \Pi_\infty = \left(\frac43\pi\right)^2 R^3 v^3_\infty\;,
\end{equation}
where we have introduced \emph{escape velocity} $v_{\infty}^2$. The
``coarsest'' PSD is such that the averaging~(\ref{eq:32}) goes over the whole
phase-space volume: $\Delta \Pi = \Pi_\infty$:
\begin{equation}
\label{eq:62}
\bar F = \frac{M}{\Pi_\infty} = \frac9{16\pi^2}\frac{M}{R^3
  v^3_\infty}=\frac{3\bar\rho}{4\pi v_\infty^3}
\end{equation}
As an estimate for $R$ we take \emph{half-light radius} $r_h$ (i.e. the radius
where surface brightness profile falls to 1/2 of its maximal value).
Neglecting possible influence of ellipticity of stellar orbits (c.f.
Appendix~\ref{App:aspher}), assuming constant DM density within $r_h$ and
isothermal distribution of stars~\cite{Pryor:90}, we obtain the following
estimate on the average DM density within $r_h$:
\begin{equation}
  \label{eq:60}
  \bar\rho = \frac{3\log 2}{2\pi} \frac{\sigma^2}{G_N r_h^2},
\end{equation}
%
%
Assuming isotropic velocity distributions,\footnote{This
    assumption seems to be correct for the DM particles, since numerical
    simulations of DM structures of different scales show that the velocity
    anisotropy $\beta(r) \equiv 1 -
    \frac{\sigma_{\theta}^2+\sigma_\phi^2}{2\sigma_r^2}$ tends to be zero
    towards the central
    region~\cite{Cole:95,Carlberg:97,Hansen:06,Zait:07,Van_Hese:08}.  It is
    not clear whether $\beta$ equals to zero for stars in dSphs.
    The assumption of isotropy of stellar velocities leads to the \emph{cored}
    density profiles~\cite{An:05,Evans:08}, therefore our estimate for
    $\bar\rho$ tends to be robust. This is confirmed by comparison of the
    estimate~(\ref{eq:60}) with those, based
    on~\cite{Strigari:07a,Strigari:07,Wu:07}, where DM density profiles were
    obtained under the assumptions of different anisotropic distributions of
    stars in dSphs.} %
the escape velocity $v_\infty$ of the DM particles is related to the
  velocity dispersion $\sigma$ via $v_\infty \simeq {\sqrt{6}}{\sigma}$.  In
such a way we obtain the averaged PSD $\bar F$:
\begin{equation}
  \label{eq:38}
  \bar F = \frac{M}{\Pi_\infty} =\frac{\bar\rho}{8\pi\sqrt{6} \sigma^3}
  \approx \frac{3\log2}{16 \sqrt{6}\pi^{2} G_N
    \sigma r_h^2} \approx 1.25 \,\frac{M_\odot}{\!\pc^{3}}
  \left(\frac{\km}{\s}\right)^{-3} \left(\frac{\km/\!\s}{\sigma}\right)
  \left(\frac{1\pc}{r_h}\right)^{2}\;,
\end{equation}
which coincides with its maximal value (being flat).  

As a consequence of Eq.~(\ref{ff}), this ``coarse-grained'' PSD $\bar F$ is
smaller than the $f_{max}$ -- the maximum value of fine-grained PSD, equal to
its primordial value:
\begin{equation}
  \bar F \le f_{max}\;.
\label{ineq1}
\end{equation}
Eq.~(\ref{ineq1}) relates the observed properties of the DM-dominated systems
(l.h.s.) with the microscopic quantity on the r.h.s. of inequality, which
depends on the production mechanism of the DM.

In this paper we are mostly interested in two types of primordial momentum
distribution. One is the relativistic Fermi-Dirac~(\ref{eq:3}) with its
$f_{max}$ being equal to
\begin{equation}
  \label{eq:15}
  f_{max,\fd} = \frac{g\,m_\fd^4}{2(2\pi\hbar)^3}
\end{equation}
(we fix the overall normalization of the phase-space distribution function by
the relation $ M = \int d^3 x\, d^3 v\, f(t,x,v)$, where $M$ is the total mass
of the system).  Another one is an (approximate) form of the momentum
distribution for sterile neutrinos, produced via non-resonant oscillations
with the active ones~\cite{Dodelson:93,Dolgov:00}. For the latter case we
consider the velocity dispersion to be\footnote{In reality the momentum
  distribution in the case of non-resonant production does not have thermal
  shape. The exact shape, taking into account contributions from primeval
  plasma at temperatures around QCD transition, can be computed only
  numerically~\cite{Asaka:06b,Asaka:06c}. The difference between the exact
  distribution and~(\ref{eq:10}) does not exceed 20\%, which does not affect
  the mass bounds.}
\begin{equation}
  \label{eq:10}
  f_\dw(p) = \frac{g\chi}{e^{p/T_{\nu}} + 1}\;.
\end{equation}
The normalization constant $\chi$ is proportional to the mixing strength
between active and sterile neutrinos and $T_\nu$ is the temperature of
neutrino background $T_\nu(z) = (1+z)T_{\nu_0}$, related to the temperature of
the CMB background today via $T_{\nu_0} = (4/11)^{1/3} T_\textsc{cmb,0}$.  For
the maximal value of distribution~(\ref{eq:10}) we find
\begin{equation}
  \label{eq:16}
  f_{max,\dw} = \frac{g \chi m_\dw^4}{2(2\pi\hbar)^3}\;.
\end{equation}
From the definition~(\ref{eq:10}) one can relate the normalization factor
$g\,\chi$ to the DM abundance (see e.g.~\cite{Hansen:01})
\begin{equation}
  \omega_\dm \equiv \Omega_{\dm} h^2 =  g\chi \frac{m_\dw [\!\ev]}{94\ev}.
  \label{chi}
\end{equation}
Therefore we can rewrite maximal value of the primordial phase-space
density~(\ref{chi}) as
\begin{equation}
  \label{eq:17}
  f_{max,\dw} = \frac{94\omega_\dm }{2(2\pi\hbar)^3}\frac{m_\dw^3}{\!\ev^3}\;.
\end{equation}
Notice, that unlike the Fermi-Dirac case, for the NRP scenario $f_{max}$
behaves as the \emph{third} power of particle's mass.

In the presence of lepton asymmetry in primeval plasma the resonant production
of sterile neutrinos becomes possible~\cite{Shi:98}. A possible lepton
asymmetry, generated in the framework of the \numsm and spectra of sterile
neutrino DM were recently computed in~\cite{Laine:08a,Shaposhnikov:08a}.
Qualitatively, these spectra contain a ``cold'' (resonant) component and a
``warm'' one, produced through non-resonant oscillations, analogously to the
NRP scenario of~\cite{Dodelson:93}. The spectra as a whole become colder than
in the NRP case (see e.g.~Fig.~6 in \cite{Laine:08a}). The maxima of
primordial phase-space distributions for these spectra are higher (sometimes
significantly) than for spectra, produced in the NRP scenario (c.f.  Fig.~5 in
\cite{Laine:08a}).  Therefore, in general mass bound for such a DM is expected
to be weaker than that of the NRP scenario.  The exact form of these spectra
can be computed only numerically.  We used a number of spectra\footnote{We are
  grateful to M.~Laine and M.~Shaposhnikov for providing these spectra to us.}
to check those which satisfy the bound~(\ref{ineq1}) or TG bound (see
Section~\ref{sec:bounds}).

\begin{figure}[t]
  \centering
  \includegraphics[width=\linewidth]{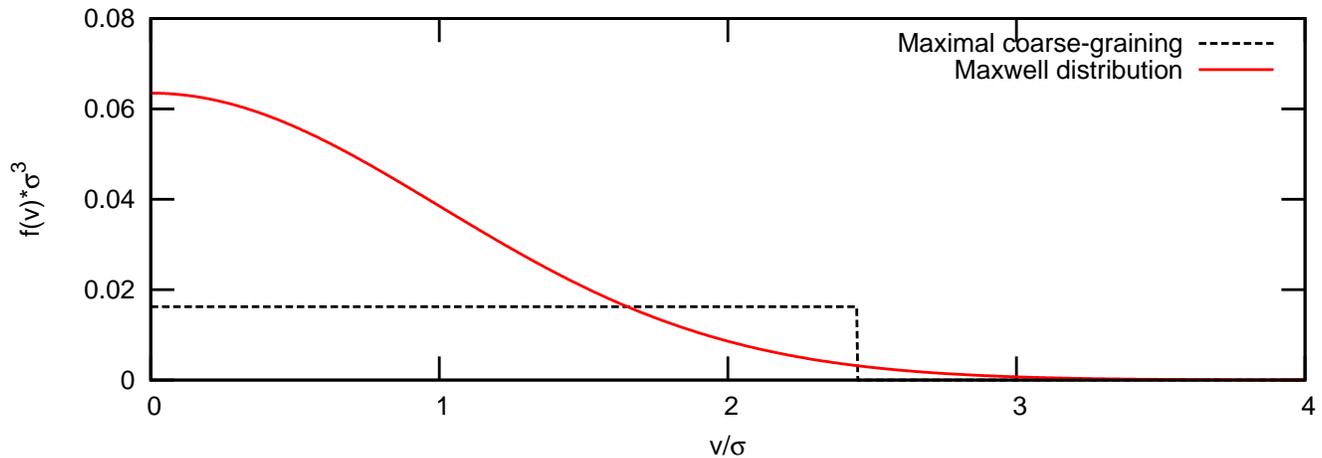}
  \caption{Comparison of velocity profiles assumed in \cite{Tremaine:79} (red
    solid line) and in this work (black dashed line).}
  \label{fig:velocities}
\end{figure}

Let us compare expression~(\ref{eq:38}) with the original Tremaine-Gunn bound
(maximum of the right hand side of Eq.~(\ref{eq:1})):
\begin{equation}
  \label{eq:8}
  F_\tg = \frac{9}{8\pi^2 \sqrt{2\pi} G_N  \sigma r_c^2}\;.
\end{equation}
The values of $\bar F$ is smaller than $F_\tg$ by
\begin{equation}
  \label{eq:21}
  \frac{\bar F}{F_\tg} = \frac{\log2\sqrt\pi}{6\sqrt{3}} \left(\frac{r_c}{r_h}\right)^2  \approx 0.118 \left(\frac{r_c}{r_h}\right)^2\;,
\end{equation}
where $r_c$ and $r_h$ are the core radius of isothermal profile and the
half-light radius, correspondingly. When comparing $\bar F$ and $F_\tg$ below,
we take $r_h \simeq r_c$.
%
Essentially, the difference between $\bar F$ and $F_\tg$ is due to the
different assumed velocity distributions. While the Maxwell distribution was
assumed in~\cite{Tremaine:79} (c.f. Eq.~(\ref{eq:4})), we assume constant
velocity profile from escape velocity $v_\infty$ down to $v=0$ (as shown on
the Fig.~\ref{fig:velocities}). The numerical factor in~(\ref{eq:21}) is the
ratio of areas under two velocity curves of Fig.~\ref{fig:velocities}.
Translated into the mass bound, relation~(\ref{eq:21}) means that for DM
particles with distribution~(\ref{eq:3}) one would obtain roughly 40\%
stronger mass bound by using the original Tremaine-Gunn bound, rather than
$\bar F$ (and $\approx 60\%$ stronger mass bound for the case of the
distribution~(\ref{eq:10})).

Let us compare our new bound with the one, based of
\cite{Dalcanton:00,Hogan:00}.  Following the definition~(\ref{eq:2}), we
express the measured value $Q_f$ for a dSph through the observed quantities
\begin{equation}
  \label{eq:5}
  \qhd = \frac{\bar\rho}{\eta^3(3\sigma^2)^{3/2}} \approx 14.83 \frac{M_\odot}{\!\pc^{3}}
  \left(\frac{\km}{\s}\right)^{-3} \left(\frac{\km\s^{-1}}{\sigma}\right)
  \left(\frac{1\pc}{r_h}\right)^{2} \frac1{\eta^3}\;,
\end{equation}
where $\eta$ is the scaling factor which accounts for the fact that the dark
matter particles do not necessarily have the same velocity dispersion as the
stars, $r_{h}$ is the half-light radius, $\sigma$ is the measured
one-dimensional velocity dispersion of the stars and $\bar\rho$ is defined
in~(\ref{eq:60}).  It was estimated in~\cite{Dalcanton:00} that $\eta \approx
1$.
 In Eq.~(\ref{eq:5}) we used the same value of $\bar\rho$ as
in Eq.~(\ref{eq:38}).  For 
for the same dSph, $Q_f$ is bigger
than $\bar F$ (given by expression~(\ref{eq:38})) by a factor
$8\pi\sqrt{2}/3\approx 11.85\dots$.

On the other hand, for any initial momentum distribution $f(p)$ we should
compare $Q_i$, given by Eq.~(\ref{eq:46}), with the $f_{max}^{(i)}$. For both
types of distribution~(\ref{eq:3}) and~(\ref{eq:10}) the ratio of initial
$Q_i/f^i_{max}$ is given by
\begin{equation}
  \label{eq:18}
  \frac{{Q_i}}{f_{max}^{(i)}} =\frac{4\pi\zeta^{5/2}(3)}{5\sqrt{15}\zeta^{3/2}(5)}
  \approx 0.973\dots
\end{equation}
As a result, a bound, based on the decrease of the average PSD $Q$ is stronger
than $\bar F$ bound from the same object by a factor:
\begin{equation}
  \label{eq:19}
  \frac{f^{(i)}_{max}}{Q_i}\frac{Q_f}{\strut\bar F} \approx 12.176\dots
\end{equation}
(where again we put $\eta = 1$). This leads to $\approx 1.87$ times stronger
bound on the $m_\fd$ and $\approx 2.3 $ times stronger bound for $m_\dw$.

\section{Analysis of measured values}
\label{sec:analys-meas-valu}

Recently, a number of very faint, very dense dSphs were
detected~\cite{Belokurov:07,Koposov:07,Irwin:07,Zucker:06,Belokurov:06,Gilmore:07,Simon:07}.
To calculate the mass limits, we used the data from two recent papers:
\cite{Gilmore:07,Simon:07}. First of all, we should notice that although both
of these papers provide the estimate of $\qhd$ for each object, they use
different prescriptions for computing this value.

In \cite{Gilmore:07} the quantity $\qhd$ is estimated inside the half-light
radius $r_{h}$, using one-dimensional velocity dispersion $\sigma$ of stars:
\begin{equation}
  \label{eq:20}
  Q_{Gil} = \frac{\bar \rho}{\sigma^3} = \frac{3}{8\pi G_N r_{h}^{2}\sigma}\;.
\end{equation}
Compared to our definition~(\ref{eq:38}) $\bar F =
\frac{\log2}{2\sqrt{6}\pi}Q_{Gil}\approx 0.045 Q_{Gil}$.
Following~\cite{Mateo:91} the authors of~\cite{Simon:07} define \emph{central
  density}
\begin{equation}
  \label{eq:22}
  \rho_0 = 166 \sigma^2\eta^2/r_c^2
\end{equation}
where $\eta \sim 1$ is a numerical parameter, characterizing plausible density
profiles (for details see~\cite{Mateo:91,Simon:07}). They used $\rho_0$ to
define the quantity:
\begin{equation}
  Q_{SG}\equiv \frac{\rho_{0}}{\sigma^{3}}\;,
\end{equation}
As a result for the same object $Q_{SG}$ is by a factor of $14.60$ greater
than $Q_{Gil}$.

Using the available information about dSph galaxies
(refs.~\cite{Gilmore:07,Simon:07} and refs. therein), we calculate $\bar F$,
trying also to estimate the errors. Several factors contribute to the errors
of $\sigma$ and $r_h$.

First of all, as $\sigma$ is the dispersion of measured velocities, it has the
statistical error (which can be quite large for the ultra-faint dSphs where
the number of stars can be rather small ($\sim 10 - 100$,
c.f.~\cite[Table~3]{Simon:07}). However, the systematic error is much larger.
The authors of~\cite{Simon:07} found the systematic error on their
determination of velocity dispersion to be $2.2\km/\s$. We add this error in
quadratures to the statistical errors, found in~\cite[Table~3]{Simon:07}. The
results are shown in the column number 4 in the Table~\ref{tab:gilmore}.

The half-light radius $r_h$ is a derived quantity and there are several
contributions to its errors. First of all, the surface brightness profile is
measured in angular units and their conversion to parsecs requires the
knowledge of the distance towards the object. These distances are generally
known with uncertainties of about 10\%
(see~\cite{Mateo:98,Belokurov:07,Bonanos:03,Dall'Ora:06,Coleman:07a,Jong:08,Okamoto:08,Zucker:06,Lee:03,Rizzi:07,Bellazzini:04,Mateo:98,Bellazzini:05,Pietrzynski:08}).
Another uncertainty comes from the method of determination of $r_h$. The
surface brightness profile gets fit to various models to determine this
quantity. For several dSphs: Coma Berenices, Canes Venatici II, Hercules and
Leo IV authors used two different profiles (Plummer and exponential) for
evaluating the annular half-light radius~\cite{Belokurov:07}. Their results
are present in the Table~\ref{tab:rh}.
\begin{table}
\begin{tabular}{l|c|c}
\hline
Galaxy & $r_h$, Plummer & $r_h$, exponential \\
\hline
Coma Berenices & 5.0' & 5.9' \\
\hline
Canes Venatici II & 3.0' & 3.3' \\
\hline
Leo IV & 3.3' & 3.4' \\
\hline
Hercules & 8.0' & 8.4' \\
\hline
\end{tabular}
\caption{Uncertainties of determination of half-light radius $r_h$ for several dSphs.}
\label{tab:rh}
\end{table}
We use these results to estimate the systematic error on $r_h$ to be 20\% and
use it for all the dSph, where $r_h$ is quoted without errors. The results of
determination $r_h$ are shown in the 3rd column of the
Table~\ref{tab:gilmore}.  The obtained values of $\bar F$ with corresponding
errors are presented in the Table~\ref{tab:gilmore}, column 5. We determined
the errors on $\bar F$ by pushing the uncertainties in both $\sigma$ and $r_h$
so that the values of $\bar F$ is minimized (maximized).

\begin{table*}[t]
  \begin{tabular}{l|c|c|c|c|c|c|c|c}
    \hline
    &  & $r_{h}$ & $\sigma$ & $\bar F$ & $m_{\deg}$ & $m_{\textsc{fd}}$ & $m_{\dw}$ & $m_{\dw,\textsc{tg}}$
    \\
    \multicolumn{1}{c|}{dSph}& References & $\pc$ & $\mathrm{km/s}$
    & \footnotesize $M_\odot\pc^{-3} \:(\km/\s)^{-3}$ & \keV & \keV &
    \keV & \keV\\
    \multicolumn{1}{c|}{(1)} & (2) & (3) & (4) & (5) & (6) & (7) & (8) & (9)\\
    \hline
    \multicolumn{8}{c}{\small dSphs from ~\cite{Gilmore:07}}\\
    \hline
    Sextans   & \cite{Mateo:98,Gilmore:07} & 630$\pm$170 & 6.6$\pm$2.3 & 4.78$^{+8.97}_{-2.58}$  $\cdot 10^{-7}$ & 0.147$^{+0.044}_{-0.026}$ & 0.174$^{+0.053}_{-0.031}$ & 0.454$^{+0.192}_{-0.104}$ & 0.715$^{+0.302}_{-0.163}$ \\
Fornax    & \cite{Mateo:98,Gilmore:07} & 400$\pm$103 & 10.5$\pm$2.7& 7.45$^{+10.74}_{-3.70}$  $\cdot 10^{-7}$ & 0.164$^{+0.041}_{-0.026}$ & 0.195$^{+0.049}_{-0.031}$ & 0.527$^{+0.183}_{-0.108}$ & 0.830$^{+0.288}_{-0.170}$ \\
Leo~I     & \cite{Mateo:98,Gilmore:07} & 330$\pm$106 & 8.8$\pm$2.4 & 1.31$^{+2.59}_{-0.72}$  $\cdot 10^{-6}$ & 0.189$^{+0.059}_{-0.034}$ & 0.224$^{+0.070}_{-0.041}$ & 0.635$^{+0.279}_{-0.148}$ & 1.00$^{+0.44}_{-0.23}$ \\
UrsaMinor & \cite{Mateo:98,Gilmore:07} & 300$\pm$74 & 9.3$\pm$2.8 & 1.49$^{+2.27}_{-0.76}$  $\cdot 10^{-6}$ & 0.195$^{+0.051}_{-0.031}$ & 0.232$^{+0.060}_{-0.037}$ & 0.665$^{+0.240}_{-0.139}$ & 1.05$^{+0.38}_{-0.22}$ \\
Carina    & \cite{Mateo:98,Gilmore:07} & 290$\pm$72 & 6.8$\pm$1.6 & 2.19$^{+2.87}_{-1.05}$ $\cdot 10^{-6}$ & 0.215$^{+0.050}_{-0.032}$ & 0.255$^{+0.060}_{-0.039}$ & 0.755$^{+0.243}_{-0.148}$ & 1.19$^{+0.38}_{-0.23}$ \\
Draco     & \cite{Gilmore:07,Martin:08} & 221$\pm$16 & 9.5$\pm$1.6 & 2.70$^{+1.07}_{-0.69}$ $\cdot 10^{-6}$ & 0.226$^{+0.020}_{-0.016}$ & 0.269$^{+0.023}_{-0.019}$ & 0.809$^{+0.095}_{-0.076}$ & 1.27$^{+0.15}_{-0.12}$ \\
Bootes    & \cite{Mateo:98,Belokurov:06,Munoz:06} & 246$\pm$28& 6.5$^{+2.1}_{-1.3}$ & 3.18$^{+1.88}_{-1.24}$  $\cdot 10^{-6}$ & 0.236$^{+0.029}_{-0.027}$ & 0.280$^{+0.035}_{-0.033}$ & 0.855$^{+0.143}_{-0.130}$ & 1.35$^{+0.23}_{-0.20}$ \\
Sculptor  & \cite{Mateo:98,Gilmore:07} & 160$\pm$40 & 10.1$\pm$0.3 & 4.99$^{+3.62}_{-1.98}$ $\cdot 10^{-6}$ & 0.264$^{+0.038}_{-0.031}$ & 0.314$^{+0.046}_{-0.037}$ & 0.993$^{+0.198}_{-0.154}$ & 1.56$^{+0.312}_{-0.243}$ \\
Leo~II    & \cite{Mateo:98,Gilmore:07} & 185$\pm$48 & 6.8$\pm$0.7 & 5.38$^{+5.55}_{-2.30}$ $\cdot 10^{-6}$ & 0.269$^{+0.052}_{-0.035}$ & 0.319$^{+0.062}_{-0.042}$ & 1.02$^{+0.27}_{-0.17}$ & 1.60$^{+0.43}_{-0.27}$ \\

    \hline
    \multicolumn{8}{c}{\small dSphs from ~\cite{Simon:07}}\\
    \hline
    Canes~Venatici~I~~ & \cite{Simon:07,Martin:08} & 564$\pm$36 & 7.6$\pm$2.2 & 5.17$^{+3.46}_{-1.63}$ $\cdot 10^{-7}$ & 0.150$^{+0.020}_{-0.013}$ & 0.178$^{+0.024}_{-0.016}$ & 0.467$^{+0.087}_{-0.055}$ & 0.735$^{+0.137}_{-0.087}$ \\
Ursa~Major~I~~~~~~ & \cite{Simon:07,Martin:08} & 318$^{+50}_{-39}$ & 7.6$\pm$2.4 & 1.63$^{+1.46}_{-0.70}$ $\cdot 10^{-6}$ & 0.199$^{+0.035}_{-0.026}$ & 0.237$^{+0.041}_{-0.031}$ & 0.684$^{+0.163}_{-0.118}$ & 1.08$^{+0.26}_{-0.19}$ \\
Hercules~~~~~~~~~~ & \cite{Simon:07,Martin:08} & 330$^{+75}_{-52}$ & 5.1$\pm$2.4 & 2.25$^{+3.74}_{-1.28}$ $\cdot 10^{-6}$ & 0.216$^{+0.060}_{-0.041}$ & 0.257$^{+0.071}_{-0.049}$ & 0.762$^{+0.294}_{-0.187}$ & 1.20$^{+0.46}_{-0.29}$ \\
Leo~T~~~~~~~~~~~~~ & \cite{Simon:07,Martin:08} & 178$\pm$39 & 7.5$\pm$2.7 & 5.26$^{+8.22}_{-2.66}$  $\cdot 10^{-6}$ & 0.267$^{+0.071}_{-0.043}$ & 0.318$^{+0.084}_{-0.051}$ & 1.01$^{+0.37}_{-0.21}$ & 1.59$^{+0.59}_{-0.33}$ \\
Ursa~Major~II\footnotemark{}~~~~~ & \cite{Simon:07,Martin:08} & 140$\pm$25 & 6.7$\pm$2.6 & 9.53$^{+13.55}_{-4.59}$ $\cdot 10^{-6}$ & 0.310$^{+0.077}_{-0.047}$ & 0.369$^{+0.091}_{-0.056}$ & 1.23$^{+0.42}_{-0.24}$ & 1.94$^{+0.67}_{-0.38}$ \\
Leo~IV~~~~~~~~~~~~ & \cite{Simon:07,Martin:08} & 116 $^{+26}_{-34}$ & 3.3$\pm$2.8 & 2.82$^{+34.39}_{-1.91}$ $\cdot 10^{-5}$ & 0.406$^{+0.368}_{-0.100}$ & 0.483$^{+0.438}_{-0.119}$ & 1.77$^{+2.41}_{-0.55}$ & 2.79$^{+3.80}_{-0.87}$ \\
Coma~Berenices~~~~ & \cite{Simon:07,Martin:08} & 77 $\pm$ 10 & 4.6$\pm$2.3 & 4.59$^{+7.53}_{-2.19}$ $\cdot 10^{-5}$ & 0.459$^{+0.126}_{-0.069}$ & 0.546$^{+0.150}_{-0.082}$ & 2.08$^{+0.80}_{-0.41}$ & 3.28$^{+1.25}_{-0.64}$ \\
Canes~Venatici~II~ & \cite{Simon:07,Martin:08} & 74$^{+14}_{-10}$ & 4.6$\pm$2.4 & 4.97$^{+8.92}_{-2.66}$ $\cdot 10^{-5}$ & 0.468$^{+0.137}_{-0.082}$ & 0.557$^{+0.163}_{-0.097}$ & 2.14$^{+0.87}_{-0.48}$ & 3.36$^{+1.38}_{-0.76}$ \\

    \hline
\end{tabular}
\caption{Parameters for dSphs from~\cite{Gilmore:07,Simon:07} (columns 1--5) and derived lower mass limits for
  various types of DM (columns 6--9).
  $m_\deg$ refers to the limit from
  Pauli exclusion principle (\ref{eq:23}), $m_\fd$ is the limit for
  particles with the momentum distribution~(\ref{eq:3}), $m_\dw$ and $m_{\dw,\tg}$ -- for
  distribution~(\ref{eq:10}). All results are quoted for $g=2$ internal
  degrees of freedom. Results for NRP scenario are for $\omega_\dm=0.105$~\cite{Spergel:07}.}
\label{tab:gilmore} \footnotetext{There is an extensive evidence
  that Ursa Major~II is a tidally
  disrupted dSph~\cite[see e.g.][]{Simon:07}. Therefore, the results for it
  are provided for illustrative purposes only.}
\end{table*}

\section{Results}
\label{sec:bounds}

Our main results are compiled into the Table~\ref{tab:gilmore} (columns 6--9).
The \textbf{column 6} of Table~\ref{tab:gilmore} contains the bound on
$m_\deg$ (given by Eq.~(\ref{eq:23})) based on the Pauli exclusion principle.
It is independent of the details of the evolution of the system, is not
affected by the presence of baryons (see below) and holds for any fermionic
DM.  The \textbf{column 7} contains the mass bounds for the relativistically
decoupled DM particles (primordial distribution~(\ref{eq:3})), obtained by
combining Eqs.(\ref{eq:38})--(\ref{eq:15}).  Combining
Eqs.~(\ref{eq:38}),~(\ref{ineq1}) and (\ref{eq:17}) one obtains the result for
the case of DM with primordial velocity distribution~(\ref{eq:10}), quoted in
the \textbf{column 8}. Both bounds in columns 7 and 8 conservatively assume
maximally coarse-grained distribution function (see
Section~\ref{sec:maxim-coarse-grain}).  In instead of the maximal
coarse-graining, one assumes the isothermal distribution in the final state
(c.f. Fig.~\ref{fig:velocities}), one arrives to the original Tremaine-Gunn
bound, shown in the \textbf{9th column}. It is obtained by comparing the
expressions~(\ref{eq:16}) with~(\ref{eq:8}).\footnote{The value of $r_c$ is
  not currently known for several new, faint dSphs, from which we obtain the
  best limits on DM mass.  Therefore, to calculate the Tremaine-Gunn limit in
  Table~\ref{tab:gilmore}, we use the conservative estimate $r_c \approx r_h$
  (see comment after Eq.(\ref{eq:21})).}  We denote the corresponding mass
bound by $m_{\dw,\tg}$.

We quote all the mass bounds with the corresponding uncertainties, coming from
those of in determination of $\sigma$ and $r_h$ (see
Section~\ref{sec:analys-meas-valu}). However, for any given object there can
be unique reasons, violating the standard assumptions and therefore increasing
the uncertainties.  Therefore, although the strongest bounds in
Table~\ref{tab:gilmore} come from the Canes Venatici~II (CVnII) dSph, we
decided to take a value which independently follows from several objects as a
single number, characterizing our results (for a given type of DM). To this
end we choose the value, obtained for Leo~IV.\footnote{Notice, that the
  numbers for Leo~IV essentially coincide with the mass limits from CVnII and
  Com if all uncertainties in these dSphs are pushed to \emph{minimize} the
  mass bound.} %
Thus, the mass bounds, quoted below are excluded from three dSphs: Leo~IV,
CVnII and Coma Berenices (Com)\footnote{It is possible that Coma Berenices is
  undergoing tidal disruption (like another ultra-faint dSph, Ursa~Major~II
  (UMaII), closely resembling Com)~\cite{Simon:07}.  However, unlike UMaII (or
  the best known example of tidally disrupted dSph, Sagittarius), there are no
  known tidal streams near the position of Coma Berenices and the evidence in
  favor of tidal disruption are quite moderate \cite[c.f.  discussion
  in][\S3.6]{Simon:07}.} %
To summarize, we obtain the following lower bounds
\begin{equation}
  m_\deg >
  0.41\kev\;,
\label{eq:28}
\end{equation}
%
\begin{equation}
  \label{eq:27}
  m_\fd > 
  0.48\kev\;,
\end{equation}
\begin{equation}
  \label{eq:29}
  m_\dw > 
  1.77\kev\;,
\end{equation}
and
\begin{equation}
  \label{eq:30}
  m_{\dw,\textsc{tg}} > 
  2.79\kev\;.
\end{equation}

We can compare lower bounds~(\ref{eq:29})--(\ref{eq:30}) with the upper ones,
coming from astrophysical (X-ray) constraints on the possible flux from
sterile neutrino DM
decay~\cite{Boyarsky:05,Boyarsky:06b,Boyarsky:06c,Riemer:06,Watson:06,Boyarsky:06e,Abazajian:06b,Boyarsky:06d,Boyarsky:06f,Boyarsky:07a,Boyarsky:07b}.
\begin{figure}[t]
  \centering
  \includegraphics[width=.8\linewidth]{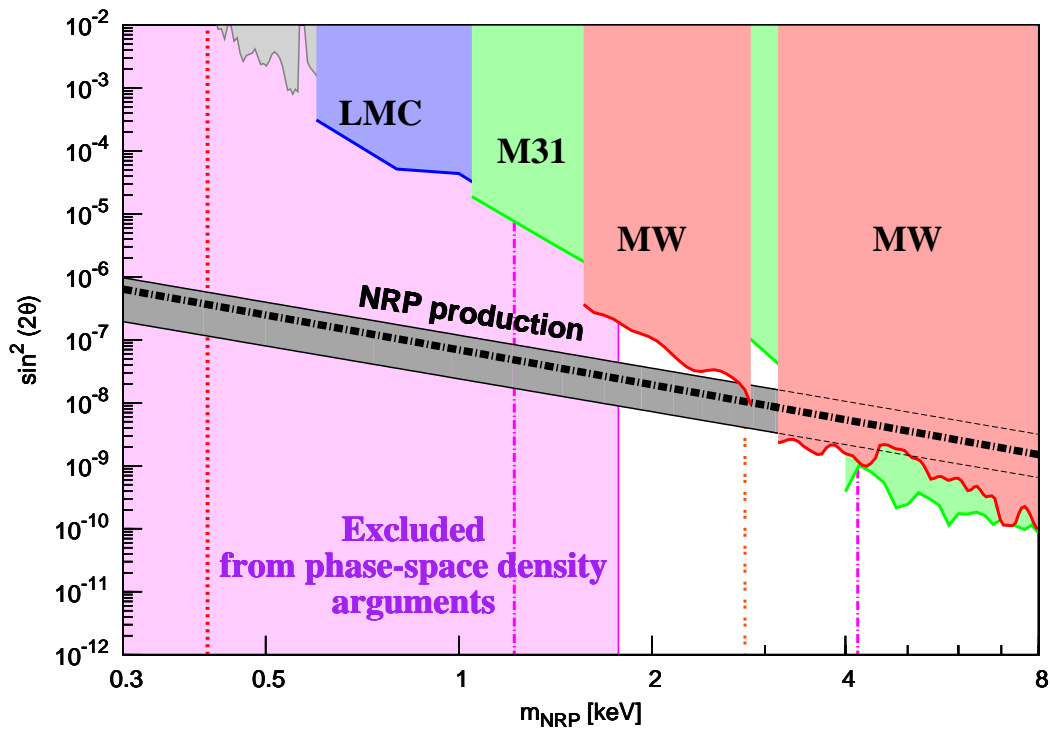}
  \caption{Restrictions on parameters of sterile neutrino (mass and mixing
    $\sin^2(2\theta)$ between sterile and active neutrinos) from X-rays
    (\cite{Boyarsky:06c,Boyarsky:06d,Boyarsky:06f,Boyarsky:07a,Boyarsky:07b})
    and phase-space density considerations (this work).  Our analysis excludes
    the region to the left of the vertical line~(\ref{eq:29}) (purple shaded
    region). Two dashed-dotted vertical lines mark the systematic
    uncertainties of this bound. The dotted line on the left marks the
    bound~(\ref{eq:28}) based on the Pauli exclusion principle. The
    double-dotted dark orange line marks the bound~(\ref{eq:30}). {The
      black dashed-dotted line is the \emph{NRP production curve} (i.e. pairs
      of $m_\dw$ and $\theta$ that lead to the correct DM
      abundance)~\cite{Asaka:06c}. The gray region marked ``NRP production''
      accounts for possible uncertainties in the abundance computations
      (see~\cite{Asaka:06b,Asaka:06c} for details).}}
  \label{fig:dw}
\end{figure}
{Taking central value~(\ref{eq:29}) and comparing it with the X-ray
  constraints, one sees that there exists a narrow window of parameters for
  which 100\% of DM can be made from the NRP sterile neutrino (c.f.
  Fig.\ref{fig:dw}). Less conservative bound~(\ref{eq:30}), based
  on~\cite{Tremaine:79} (marked by the dark orange double-dotted vertical line
  on the Fig.~\ref{fig:dw}) almost completely closes this window.}  Notice,
that these bounds are comparable with the lower mass limit $m_\dw > 5.6 \keV$,
coming from the Ly-$\alpha$ forest analysis of~\cite{Viel:08}.

\begin{figure}
  \centering
  \includegraphics[width=.7\linewidth]{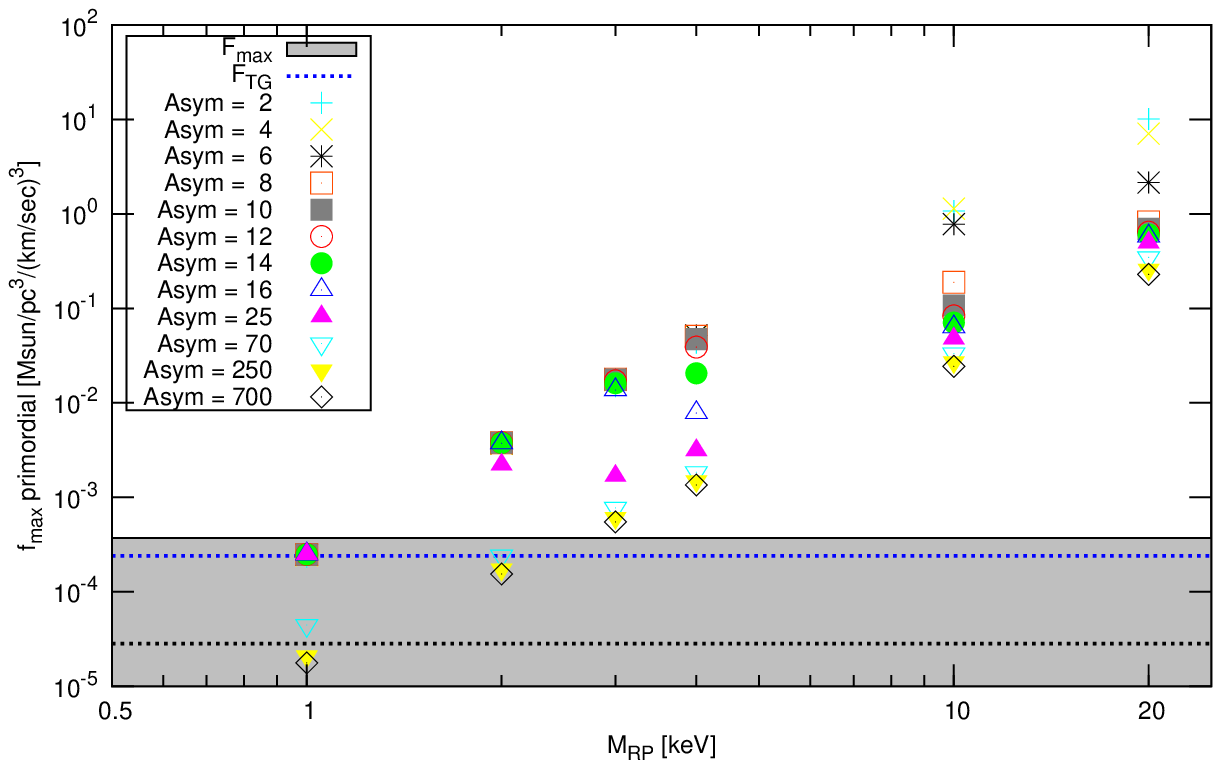} %
  \caption{Restrictions on resonantly produced sterile neutrinos. Primordial
    $f_{max}$ is computed numerically based on the spectra
    from~\cite{Laine:08a,Shaposhnikov:08a}. Different colors parametrize
    different lepton asymmetries for a given mass (see definition of lepton
    asymmetry in~\cite{Laine:08a,Shaposhnikov:08a}). Grey shaded region is
    bounded by the maximal and minimal values of $\bar F$ for Leo~IV (from
    Table~\ref{tab:gilmore}, column~5).  Horizontal dotted lines represent
    central value for $\bar F$ (lower) and $F_\tg$ (upper) for Leo~IV.  The DM
    spectrum is ruled out if the point falls into the shaded region (below the
    dotted line).}
  \label{fig:sf}
\end{figure}

\begin{figure}
  \centering
  \includegraphics[width=\linewidth]{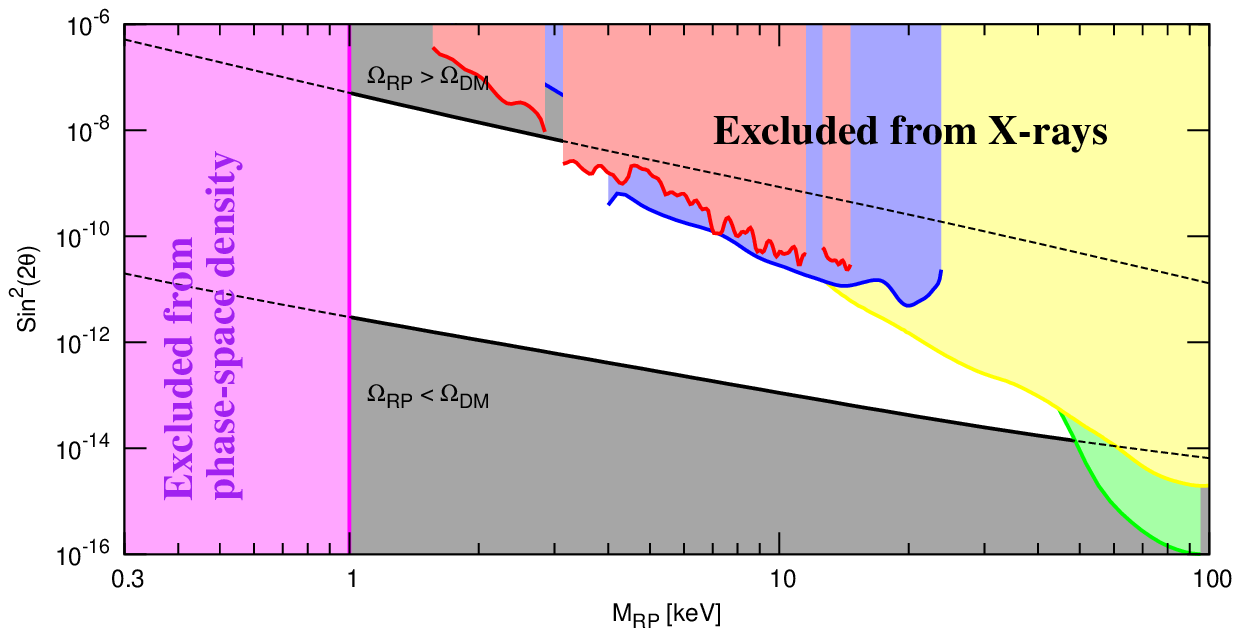} %
  \caption{Allowed window of parameters for sterile neutrinos produced via
    resonant oscillations (white unshaded strip between two black lines). Two
    bounding black lines are obtained for non-resonant (upper line, lepton
    asymmetry $=0$) and resonant production with the maximal lepton asymmetry,
    attainable in the $\nu$MSM~\cite{Laine:08a,Shaposhnikov:08a} (lower line).
    The colored regions in the upper right corner represent X-ray
    bounds~\cite{Boyarsky:06c,Boyarsky:06d,Boyarsky:07a,Boyarsky:07b}. Region
    below $1\kev$ is ruled out from the PSD arguments (this work).}
  \label{fig:sf-window}
\end{figure}

We also performed the analysis for sterile neutrinos, produced in the presence
of lepton asymmetry~(resonant production mechanism,
RP)~\cite{Shi:98,Laine:08a,Shaposhnikov:08a}. This mechanism is more
efficient than the NRP scenario and allows us to achieve the required DM abundance
for weaker mixings (c.f. Fig.~4 in~\cite{Laine:08a}). This lifts the upper
bound on the DM particle mass in RP scenario to $\sim 50\keV$. To estimate the
lower mass bound at this scenario, we have analyzed a number of available
spectra (mass range $1-20\kev$, asymmetries $(2-700)\times 10^{-6}$
(see~\cite{Laine:08a,Shaposhnikov:08a} for the definition of asymmetry). The
result are collected on the Fig.~\ref{fig:sf}.  One can see that based on
$\bar F$, the $M_\sf=1\kev$ is ruled out for lepton asymmetries $L\gtrsim
10^2$ and higher masses $M_\sf \ge 2\kev$ are allowed for all available
asymmetries. Based on the original Tremaine-Gunn bound, $M_\sf = 2\kev$ is
also ruled out for sufficiently high ($L\gtrsim 10^2$) lepton asymmetries.
Thus, resonantly produced sterile neutrinos remain a viable DM candidate (see
Fig.~\ref{fig:sf-window}).

Finally, we would like to notice that our bounds~(\ref{eq:27})--(\ref{eq:30})
are valid under the assumption that the influence of the baryons does not
result in the \emph{increase} of the PSD in the course of structure formation.
If this assumption does not hold, only the bound~(\ref{eq:28}) remains intact.

In this work we chose not to use the bound, based on the ``average PSD''
Q~\cite{Dalcanton:00,Hogan:00} (see discussion in the
Section~\ref{sec:overview}). However, as this bound is widely used in the
literature, we quote analogs of lower limits~(\ref{eq:27}) and~(\ref{eq:29})
based on inequality~(\ref{eq:59}) (which we denote $m_{\fd,\textsc{hd}}$ and
$m_{\dw,\textsc{hd}}$ correspondingly):
\begin{equation}
  \begin{aligned}
    m_{\fd,\textsc{hd}} & = 0.9\kev\;,\\
    m_{\dw,\textsc{hd}} & = 4.0\kev\;.
  \end{aligned}
  \label{eq:13}
\end{equation}
For details see Appendix~\ref{sec:hd}.

\section{Discussion}
\label{sec:discussion}

In this paper we suggested that a conservative way to put the bound on a DM
particle mass may be based on the requirements that the maximum of the
observed coarse-grained phase space density should not exceed the maximum of
the initial distribution function of the DM particles. The maximum of the
coarse-graned distribution function in the final state may be conservatively
estimated from the observed quantities. This bound relies on the assumption
that the maximum of the distribution function was not significantly increased
by the interaction with baryons.

Although DM consists of the non-interacting particles, the remaining part of
the galaxy -- the baryons -- interact with one another and dissipate their
energy, finally concentrating towards the center. The baryons, which are
condensed in the center, influence the shape of DM halo gravitationally,
increasing the central DM density~\cite{Blumenthal:86,Gnedin:04}. The opposite
effect is the energy feedback from SNae, galactic winds and reionization,
which creates the strong outflow, significantly decreasing the mass of the gas
and thereby affecting the DM halo shape. Such a feedback is thought to be
responsible to the formation of dwarf spheroidals from gas-rich dwarf
spiral/irregular
galaxies~\cite{Lin:83,Moore:94,Mastropietro:05,Mayer:06,Mayer:07}. Clearly
both gas condensation and feedback strongly influence the central PSD of
DM~\cite{Read:06}, and in principle can lead to the violation of the
inequality (\ref{ff}).  Numerical studies of galaxy mergers show that baryons
can lead to the increase of the phase-space density during the merger (see
e.g.~\cite{Naab:07a}).  However, the method used in this work --
coarse-graining of the PSD over a large phase-space region -- reduces the
influence of baryons. Indeed, we take the spatial averaging over the radius $R
\sim r_h$, which includes external part of the system, where the amount of
baryons is small.  Additional studies are necessary to estimate effects of
baryons and make our bounds more robust. We plan to address these issues
elsewhere.

We would also like to stress that the initial velocities of DM particles in
our approach are \emph{thermal} velocities and they should not be confused
with the so-called \emph{Zeldovich} velocities~\cite{Zeldovich:70}.  Numerical
simulations of galaxy formation do not start at the time, when the DM
phase-space distribution is spatially uniform (redshifts $z\gtrsim 10^3$).
Instead, the initial (linear) stage of the structure formation is computed
analytically in the framework of the so-called \emph{Zeldovich
  approximation}~\cite{Zeldovich:70}.  This approximation is commonly used to
set up initial conditions for the numerical simulations of non-linear stage of
structure formation~\cite{Bertschinger:95,Klypin:97,Klypin:00}, which start at
redshifts $z\sim 10$. The peculiar (\emph{Zeldovich}) velocities acquired by
DM particles at this stage due to structure formation and included into the
initial conditions are normally $\sigma \sim 10\km/\sec$. Apart from Zeldovich
velocities, DM particles also possess thermal velocities, which are discussed
in this paper.  For cold enough Dark Matter these thermal velocities are much
smaller than Zeldovich ones and, thus, are often neglected and not included
into initial conditions.  Therefore, the numerical studies of PSD
evolution\footnote{Most of these studies use the quantity $Q(r) =
  \rho(r)/\sigma^3(r)$ as a PSD estimator} (see
e.g.~\cite{Taylor:01,Peirani:06,Peirani:07,Romano-Diaz:07,Hoffman:07,Romano-Diaz:06})
essentially investigate the change of PSD from Zeldovich to final stage.  It
was found in some of these works that the PSD changes by $10^2-10^3$ in the
process of collapse~\cite{Peirani:06}.  This change of PSD can be understood
as being simply an evolution from initial Zeldovich velocities $\sigma_i \sim
10\km/\sec$ to the final (virial) ones $\sigma_f \sim 10^2\km/\sec$ (with
$Q_i/Q_f \sim (\sigma_f/\sigma_i)^3 \sim 10^3$).

Because initial thermal velocities may be much smaller than Zeldovich ones, initial
PSD may differ from the final (observed) PSD not by 2--3, but by many orders
of magnitude. This fact does not contradict to the results of simulations,
described in e.g.~\cite{Peirani:06} and, therefore, cannot be used to obtain
an \emph{upper} bound on the mass of DM particles
(c.f~\cite{Boyanovsky:08,Gorbunov:08a}).

This work was mostly concentrated on restrictions on the mass of the sterile
neutrino DM, produced in through the non-resonant oscillations with active
neutrino (NRP scenario). We see that our results (Section~\ref{sec:bounds})
strongly disfavor such sterile neutrinos as the single DM component. This
conclusion is not based on the \lya method and therefore is not subject to its
uncertainties (discussed in the Introduction).  However, several uncertainties
can affect this conclusion, the major being baryonic feedback. To make this
result really robust, apart from further modeling of the baryonic influence,
one needs to strengthen the tension between upper and lower mass bounds
discussed in this paper.  This is plausible and may be done either by
improving the X-ray bounds with new observations or by strengthening the PSD
consideration, which is in the first place related to better measurements of
kinematics of dSphs.

In the presence of lepton asymmetry, the resonant production (RP) of sterile
neutrino DM takes place~\cite{Shi:98}. This mechanism is more
efficient~\cite{Shi:98,Laine:08a,Shaposhnikov:08a} than the NRP scenario and
allows to achieve required DM abundance for weaker mixings (c.f. Fig.~4
in~\cite{Laine:08a}).  This lifts the upper bound on the DM particle mass in
this scenario up to $\sim 50$~keV.  At the same time, for the same mass the
primordial velocity distribution of RP sterile neutrino DM is colder than in
NRP one.  This $f_{max}$ is as much as the order of magnitude bigger
than~(\ref{eq:16}) (c.f.~\cite{Laine:08a}).  This brings down by a factor $\sim
2$ the analog of the mass bound~(\ref{eq:29}).  Analyzing available spectra
for a range of lepton asymmetries, we see that models with $m_\sf \gtrsim 1
\kev$ are allowed. Thus, there is a large open ``window'' of allowed DM masses
(c.f.  Fig.\ref{fig:sf-window}).  However, as the dependence of the velocity
spectrum on the lepton assymetry is not monotonic, to obtain the exact shape
of the lower bound on the mass at given mixing angle more work is needed.
Nevertheless, our results show that the sterile neutrinos, produced in the
presence of lepton asymmetry, are viable DM candidates, allowed by all current
bounds.

{Finally, we would like to comment on the mechanism of production of
  sterile neutrinos from decay of massive scalar field, for example the
  inflaton~\cite{Shaposhnikov:06} (for other models
  see~\cite{Kusenko:06a,Petraki:07,Petraki:08,Gorbunov:08a}).  The primordial
  phase-space distribution function for this case was computed e.g.
  in~\cite{Shaposhnikov:06,Petraki:07,Boyanovsky:08b,Gorbunov:08a}.  Maximal
  value of phase-space density for this distribution is that of degenerate
  Fermi gas. Notice that the distribution functions
  in~\cite{Shaposhnikov:06,Petraki:07,Gorbunov:08a} $f(p)$ is formally
  unbounded for small momenta: $f(p) \sim p^{-1/2}$.  From this one can easily
  find that the fraction of particles, having maximal phase-space density, is
  $\sim 10^{-8}$.  As only this small fraction of all particles has maximal
  phase-space density, we expect the mass bound in this case to be stronger
  than~(\ref{eq:28}). The detailed analysis will be presented elsewhere.  }


\bigskip

After this work has been completed, we received a draft of the
paper~\cite{Gorbunov:08b}, where similar issues have been considered. Our
results are consistent with those of discussed in~\cite{Gorbunov:08b} wherever
they overlap.

\section*{Acknowledgments}

We would like to thank F.~Bezrukov, G.~Gilmore, D.~Gorbunov, C.~Hogan,
A.~Macci\`o, B.~Moore, T.~Naab, V.~Rubakov, M.~Shaposhnikov, T.~Theuns,
I.~Tkachev, M.~Viel for useful comments.  We are grateful to M.~Laine for
providing to us the distribution functions for the RP sterile neutrinos.  D.I.
is grateful to to Scientific and Educational
Centre\footnote{\url{http://sec.bitp.kiev.ua}} of the Bogolyubov Institute for
Theoretical Physics in Kiev, Ukraine, and especially to V.~Shadura, for
creating wonderful atmosphere for young Ukrainian scientists.  This work was
supported by the Swiss National Science Foundation and the Swiss Agency for
Development and Cooperation in the framework of the programme SCOPES -
Scientific co-operation between Eastern Europe and Switzerland. D.I. also
acknowledges support from the ``Cosmomicrophysics'' programme and from the
Program of Fundamental Research of the Physics and Astronomy Division of the
National Academy of Sciences of Ukraine.
O.R. would like to acknowledge support of the Swiss National Science
Foundation.

\appendix

\section{Influence of aspherical shapes of DM halos}
\label{App:aspher}

We analyze the change of the bound eq.~(\ref{eq:23}) due to the deviation of a
DM halo from a spherical shape. Such asphericity affects both the spatial
volume $V$ and the escape velocity $v_\infty$.
We consider the dSph as homogeneous ellipsoid with semi-axes $a$, $b$ and $c$
and assume the ellipticity of its 2D projection\footnote{Throughout this
  paper, we define the \textit{ellipticity} $\epsilon$ in a way similar to
  that in~\cite{Binney-Tremaine} (see also~\cite{Martin:08}), i.e. $\epsilon
  \equiv 1-b/a$, where $a$ and $b$ are the \textit{semi-major} and
  \textit{semi-minor} axis, respectively.  Thus, the case of $\epsilon = 0.5$
  corresponds to axis ratio 1:2.}  $\epsilon \lesssim 0.5$. Because we observe
only 2D projection of such an ellipsoid, there are two possibilities:
\begin{itemize}
\item[\textbf{Prolate dSph:}] $c > b \simeq a$. We see the axes $b$ and $c$,
  related to the ``averaged'' radius $R$ via $b = R (1-\epsilon)^{1/2}, c = R
  (1-\epsilon)^{-1/2}$. The spatial volume $V$ is therefore 
\begin{equation}
  V = \frac43 \pi abc
  \approx \frac43 \pi R^3 (1-\epsilon)^{1/2} \approx \frac43 \pi R^3 (1 - 0.5 \epsilon).
  \end{equation}
  The gravitational potential for $\epsilon \lesssim 0.5$ is dominated by
  monopole and quadrupole components, 
\begin{equation}
  \phi \approx \phi^{(0)} +
  \phi^{(2)}.
  \end{equation}
  The maximal value of the potential occurs near the end of the minor
  semi-axis: 
\begin{equation}
  |\phi_{max}| \equiv \frac{v_{\infty}^2}2 = \frac{G_{N}M}{a} -
  \frac{G_{N}D_{zz}}{4a^3},
  \end{equation}
  where $D_{zz} = \frac{2M (c^2-a^2)}{5}$ -- the quadrupole moment of the
  system~\cite{Landau:v2}.  For $\epsilon \ll 1$ we then obtain
\begin{equation}  
  \frac{Vv_{\infty{}}^3|_{prolate}}{Vv_{\infty{}}^3|_{spherical}} \approx 1+0.05\epsilon,
\end{equation}
which gives us the correction for $m_\deg$ of smaller than $1\%$  (for
$\epsilon = 0.5$).

\item[\textbf{Oblate dSph:}] $c \simeq b > a$. We observe the axes $a$ and
  $c$, therefore the spatial volume $V$ changes by $(1 - \epsilon)^{-1/2}
  \approx 1 + 0.5 \epsilon$. The maximum of the gravitational potential is
  then given by
  \begin{equation}
    |\phi_{max}| \approx \frac{G_{N}M}{a} + \frac{G_N
      D_{xx}}{2a^3} \approx \frac{G_N M}{R} (1+0.1\epsilon).\,
  \end{equation}
  where $D_{xx}$ is given by the same expression, as $D_{zz}$ above.  The
  maximal phase-space volume changes in the oblate case by $ \approx 1 + 0.65
  \epsilon$, so the correction for $m_\deg$ will constitute about 8\% for
  $\epsilon \simeq 0.5$.
\end{itemize}

Thus, the departure from spherical symmetry for DM halos of dSphs changes the
limit on $m_\deg$ by less than $\lesssim$ 10\% for the case of axis ratio 1:2.
This uncertainty is below several others, therefore, we will consider dSphs to
be spherical in what follows.

\section{Entropy for different distributions}

\label{Appendix}

In this Appendix we will calculate the entropy for several phase-space
distributions, including those of~(\ref{eq:3}), (\ref{eq:10}), (\ref{eq:4}),
and explore its relation with the quantity $\qhd$, defined in~(\ref{eq:2}).

The entropy of an ideal Fermi gas is given by the expression~\cite{Landau:v5}
\begin{equation}
  S = -\int d^3 p d^3 r \left[ f(r,p) \log \left(\frac{(2\pi\hbar)^3
        f(r,p)}{g}\right) + \left(\frac{g}{(2\pi\hbar)^3} - f(r,p)\right)
    \log\left(1 - \frac{(2\pi\hbar)^3 f(r,p)}{g}\right) \right]. \label{S_FD}
\end{equation}
If the distribution function $f(r,p) \ll \frac{g}{(2\pi\hbar)^3}$, we obtain
the expression for the entropy of a non-degenerate ideal gas: \eq{S = -\int
  d^3 p d^3 r f(r,p)\left[\log\left(\frac{(2\pi\hbar)^3 f(r,p)}{g}\right) -
    1\right]. \label{eq:S_nondeg}}

\subsection{Ideal Boltzmann gas}
\label{sec:ideal-boltzmann-gas}

We start with the case of ideal Boltzmann gas:
\begin{equation}
  \label{eq:33}
  f(r,p) = f_0 e^{-\frac{p^2}{2m T}}\;.
\end{equation}
Substituting it into Eq.~(\ref{eq:S_nondeg}), we arrive to the well-known
expression (c.f. e.g. \cite[\S 42]{Landau:v5}):
\begin{equation}
  \label{eq:39}
  \frac{S}N = \frac52 + \log\left( \frac{gV}{N} \frac{(m T)^{3/2}}{(2\pi \hbar^2)^{3/2}} \right),
\end{equation}
where $V$ is the volume of the system, $N$ is a number of particles.
Expressing $S/N$ as a function of $\bar\rho$ and $\langle v^2 \rangle$, we
finally obtain relation between the entropy and \qhd in the
form~(\ref{CQ_corr})
%
\begin{equation}
  \label{eq:41}
  \frac SN = \log C_B - \log\frac{\qhd \hbar^3}{m^4},\; C_B = g\frac{e^{5/2}}{(6\pi)^{3/2}}\approx g\times 0.1489\dots
\end{equation}

\subsection{Isothermal phase-space density distrubution}
\label{sec:isothermal}

Next, we consider the case when the PSD distribution can be approximated by
(pseudo)-isothermal sphere (c.f.~(\ref{eq:4})):
\begin{equation}
  f_{iso}(r,p) = \frac{9\sigma^2}{4\pi G_N (2\pi m^2\sigma^2)^{3/2}(r^2
    + r_c^2)}e^{-\frac{p^2}{2m^2\sigma^2}}.
  \label{eq:26}
\end{equation}
The number of particles in such a system,
as well as the total entropy, diverges for large $r$, however the entropy per
particles grows logarithmically at large $r$ and therefore the exact value of
cut-off is not important.



Truncating the expression for the entropy at some $r_{max}$ and taking
$r_{max} \gg r_c$, we obtain
\eq{\frac{S}{N} = - \log \frac{\qhd\hbar^3}{m^4} + \log C_{iso}, \;\;\; C_{iso} = \frac{g\exp(1/2)}{\sqrt{3}(2\pi)^{3/2}} \approx g \times 0.0604\dots }

\subsection{Entropy for Fermi-Dirac and NRP distributions}
\label{sec:dw}


Next, we analyze the case of primordial momentum distribution, which has the
form of (rescaled) relativistic Fermi-Dirac.
\begin{equation}
  \label{eq:11}
  f(p) = \frac{g}{(2\pi\hbar)^3}\frac{F}{e^{\epsilon(p)/T}+1},\;\;\epsilon(p)=p\;.
\end{equation}
For now we keep both $F$ and $T$ to be arbitrary. The distribution in the
form~(\ref{eq:11}) accounts for both~(\ref{eq:3}) and~(\ref{eq:10}) cases. The
entropy of $N$ particles with distribution~(\ref{eq:11}) is given by the
expression~(\ref{S_FD})
, which reduces to
\begin{equation}
  S= \frac{gVT^{3}}{2\pi^{2}\hbar^{3}}I(F), \label{S_DW}
\end{equation}
where function $I(F)$ is given by
\begin{equation}
  \label{eq:43}
  I(F) \equiv -\int_{0}^{\infty}dz z^{2}
  \left[\frac{F}{e^{z}+1}\log\left(\frac{F}{e^{z}+1}\right) +
    \left(1-\frac{F}{e^{z}+1}\right)\log\left(1-\frac{F}{e^{z}+1}\right)\right]\;.
\end{equation}
The integral~(\ref{eq:43}) can be computed numerically. 
At $F\ll 1$ the expression~(\ref{eq:43}) can be approximated by
\begin{equation}
  I(F) \approx \frac32\zeta(3) \left(F-F\log F\right) +
  F \int_{0}^{\infty}\frac{dz z^{2}}{e^{z}+1}\log(e^{z}+1).
\end{equation}

The specific entropy $S/N$ equals to
\begin{equation}
  \label{eq:44}
  \frac SN = \frac{g
    m^4I(F)}{2\pi^2\hbar^3}\left(\frac{\zeta(3)}{15\zeta(5)}\right)^{3/2}
  \frac{\langle v^2\rangle^{3/2}}{\bar\rho}
  = \frac{g\,I(F)}{2\pi^2}\left(\frac{\zeta(3)}{15\zeta(5)}\right)^{3/2}\frac{m^4}{\qhd \hbar^3}\;.
\end{equation}
Therefore, we see that for the distributions of the form~(\ref{eq:11})
\emph{relation between the entropy per particle and $\qhd$ is not given by the
  simple expression}~(\ref{CQ_corr}).

Up until this moment we kept parameters $F$ and $T$ in~(\ref{eq:11})
independent.  However, we are mostly interested in two particular cases: (i)
$F=1$ while $T=T_\fd$ -- arbitrary (distribution~(\ref{eq:3})); and (ii) $F<1$
having arbitrary value, while $T$ being fixed to $T_\nu$ -- the temperature of
neutrino background, related to the temperature of the CMB background today
via $T_{\nu_0} = (4/11)^{1/3} T_\textsc{cmb,0}$ (distribution~(\ref{eq:10})).

We start with the case (i).
Expressing $\rho$ as a function of $T_\fd$, we obtain
\begin{equation}
  \label{eq:50}
  \qhd = \frac{g m^4}{\hbar^3} \mathbf{q}\;,
\end{equation}
where numerical constant $\mathbf{q}$ is given by (c.f.~\cite{Hogan:00}):
\begin{equation}
  \label{eq:51}
  \mathbf{q} = \frac{\zeta^{5/2}(3)}{20\pi^2\sqrt{15}\zeta^{3/2}(5)}\approx 1.96...\times 10^{-3}\;.
\end{equation}
As a result for the distribution~(\ref{eq:3}) and fixed number of particles,
the quantity $\qhd$ is independent on $T_\fd$, volume or $N$.  The entropy per
particle is also independent on both $T_\fd$ and $V$ and is given by
\begin{equation}
  \label{eq:52}
  \frac SN = \mathbf{s} = I(1) \frac{2}{3\zeta(3)} \approx 4.20\dots
\end{equation}
Although both quantities $S/N$ and $\qhd$ are simply constants, we find it
convenient to choose them in the form~(\ref{CQ_corr}):
\begin{equation}
  \label{eq:53}
  \frac SN = -\log \left(\frac{\qhd\hbar^3} {m^4}\right)+\log C_\fd, \quad
  C_\fd = g\cdot \mathbf{q}\cdot e^{\mathbf{s}} \approx g \times 0.1311\dots
\end{equation}

In case (ii)  when $F\ll 1$ we obtain for $S/N$:
\begin{equation}
  \label{eq:54}
  \frac SN = \frac2{3\zeta(3)}\frac{I(F)}{F}\simeq ( 1- \log F) + \frac{2
    l}{3\zeta(3)}\;.
\end{equation}
Similarly to~(\ref{eq:50})--(\ref{eq:51})
\begin{equation}
  \label{eq:55}
  \frac{\qhd \hbar^3}{m^4} = g\,\mathbf{ q}\, F\;.
\end{equation}
Combining~(\ref{eq:54})--(\ref{eq:55}) we can write
\begin{equation}
  \label{eq:56}
  \frac SN = -\log\left( \frac{\qhd \hbar^3}{m^4}\right) + \log C_\dw,\quad
  C_\dw = g \,\mathbf{q} \exp\Bigl(1+\frac{2 l}{3\zeta(3)} \Bigr)\approx
  g\times 0.137\dots
\end{equation}

\section{Mass bounds from the evolution of the average PSD}
\label{sec:hd}

\begin{table}
  \centering
  \begin{tabular}{|l|l|c|c|}
    \hline
    \multicolumn{1}{|c|}{dSph} &  \multicolumn{1}{c|}{$Q_f$} & $m_{\fd,\textsc{hd}}$ & $m_{\dw,\textsc{hd}}$ \\
    \multicolumn{1}{|c|}{}& \multicolumn{1}{c|}{$\left[\frac{M_\odot}{\!\pc^3}\left(\frac{\!\km}{\s}\right)^{-3}\right]$} & [keV] & [keV] \\
    \hline
    Sextans   & 5.68$^{+10.67}_{-3.07}$ $\cdot 10^{-6}$ & 0.324$^{+0.098}_{-0.057}$ & 1.04$^{+0.44}_{-0.24}$ \\
Fornax    & 8.86$^{+12.77}_{-4.40}$ $\cdot 10^{-6}$ & 0.362$^{+0.091}_{-0.051}$ & 1.20$^{+0.42}_{-0.25}$ \\
Leo~I     & 1.55$^{+3.08}_{-0.85}$ $\cdot 10^{-5}$ & 0.416$^{+0.131}_{-0.075}$ & 1.45$^{+0.64}_{-0.34}$ \\
UrsaMinor & 1.78$^{+2.70}_{-0.90}$ $\cdot 10^{-5}$ & 0.431$^{+0.112}_{-0.070}$ & 1.52$^{+0.55}_{-0.32}$ \\
Bootes    & 3.78$^{+2.24}_{-1.48}$ $\cdot 10^{-5}$ & 0.520$^{+0.064}_{-0.061}$ & 1.95$^{+0.33}_{-0.30}$ \\
Draco     & 3.21$^{+1.27}_{-0.82}$ $\cdot 10^{-5}$ & 0.499$^{+0.044}_{-0.036}$ & 1.85$^{+0.22}_{-0.17}$ \\
Carina    & 2.60$^{+3.42}_{-1.25}$ $\cdot 10^{-5}$ & 0.474$^{+0.111}_{-0.072}$ & 1.72$^{+0.56}_{-0.34}$ \\
Sculptor  & 5.93$^{+4.30}_{-2.35}$ $\cdot 10^{-5}$ & 0.582$^{+0.085}_{-0.069}$ & 2.27$^{+0.45}_{-0.35}$ \\
Leo~II    & 6.39$^{+6.60}_{-2.73}$ $\cdot 10^{-5}$ & 0.593$^{+0.115}_{-0.077}$ & 2.32$^{+0.62}_{-0.40}$ \\
Canes~Venatici~I~~ & 6.16$^{+4.11}_{-1.94}$ $\cdot 10^{-6}$ & 0.330$^{+0.045}_{-0.030}$ & 1.07$^{+0.20}_{-0.13}$ \\
Ursa~Major~I~~~~~~ & 1.94$^{+1.74}_{-0.84}$ $\cdot 10^{-5}$ & 0.440$^{+0.077}_{-0.058}$ & 1.56$^{+0.37}_{-0.27}$ \\
Hercules~~~~~~~~~~ & 2.68$^{+4.45}_{-1.53}$ $\cdot 10^{-5}$ & 0.477$^{+0.132}_{-0.091}$ & 1.74$^{+0.67}_{-0.43}$ \\
Leo~T~~~~~~~~~~~~~ & 6.26$^{+9.78}_{-3.16}$ $\cdot 10^{-5}$ & 0.590$^{+0.157}_{-0.095}$ & 2.31$^{+0.85}_{-0.48}$ \\
Ursa~Major~II\footnotemark{}~~~~~ & 1.13$^{+1.61}_{-0.55}$ $\cdot 10^{-4}$ & 0.685$^{+0.169}_{-0.104}$ & 2.81$^{+0.96}_{-0.55}$ \\
Leo~IV~~~~~~~~~~~~ & 3.35$^{+40.91}_{-2.27}$ $\cdot 10^{-4}$ & 0.898$^{+0.814}_{-0.221}$ & 4.04$^{+5.51}_{-1.27}$ \\
Canes~Venatici~II~ & 5.91$^{+10.61}_{-3.16}$ $\cdot 10^{-4}$ & 1.03$^{+0.30}_{-0.18}$ & 4.88$^{+1.99}_{-1.10}$ \\
Coma~Berenices~~~~ & 5.46$^{+8.96}_{-2.61}$ $\cdot 10^{-4}$ & 1.01$^{+0.28}_{-0.15}$ & 4.75$^{+1.82}_{-0.92}$ \\

    \hline
  \end{tabular}
  \caption{The mass bounds, based on the evolution of the average PSD $Q$~\cite{Hogan:00,Dalcanton:00}. The bound is provided for illustration purposes only
    (see Section~\ref{sec:hd} for discussion). }
  \label{tab:dh}
\end{table}

For illustration purposes we provide in Table~\ref{tab:dh} the average PSD
estimator $Q$ for all the dSphs, considered in this work, as well as the lower
mass bounds, based on the inequality~(\ref{eq:59}) for $Q$ during the
evolution~\cite{Hogan:00,Dalcanton:00} (for detailed discussion see
Section~\ref{sec:overview}). The value of $Q_f$, shown in the second column of
the Table~\ref{tab:dh} is calculated from the data in the columns (3--4) of
the Table~\ref{tab:gilmore}, using formula~(\ref{eq:5}) (with $\eta = 1$) and
$Q_i$ is defined via~(\ref{eq:46}) for the momentum distributions~(\ref{eq:3})
and~(\ref{eq:10}) (for the bounds $m_{\fd,\textsc{hd}}$ and
$m_{\dw,\textsc{hd}}$ correspondingly). The results for Leo~IV are quoted
in~(\ref{eq:13}) (Section~\ref{sec:bounds}).


\begin{thebibliography}{159}
\expandafter\ifx\csname natexlab\endcsname\relax\def\natexlab#1{#1}\fi
\expandafter\ifx\csname bibnamefont\endcsname\relax
  \def\bibnamefont#1{#1}\fi
\expandafter\ifx\csname bibfnamefont\endcsname\relax
  \def\bibfnamefont#1{#1}\fi
\expandafter\ifx\csname citenamefont\endcsname\relax
  \def\citenamefont#1{#1}\fi
\expandafter\ifx\csname url\endcsname\relax
  \def\url#1{\texttt{#1}}\fi
\expandafter\ifx\csname urlprefix\endcsname\relax\def\urlprefix{URL }\fi
\providecommand{\bibinfo}[2]{#2}
\providecommand{\eprint}[2][]{\url{#2}}

\bibitem[{\citenamefont{{Calchi Novati}}(2007)}]{Calchi:07}
\bibinfo{author}{\bibfnamefont{S.}~\bibnamefont{{Calchi Novati}}}
  (\bibinfo{year}{2007}), \eprint{0711.4474}.

\bibitem[{\citenamefont{{Zel'dovich}}(1970)}]{Zeldovich:70}
\bibinfo{author}{\bibfnamefont{Y.~B.} \bibnamefont{{Zel'dovich}}},
  \bibinfo{journal}{\aap} \textbf{\bibinfo{volume}{5}}, \bibinfo{pages}{84}
  (\bibinfo{year}{1970}).

\bibitem[{\citenamefont{{Bisnovatyi-Kogan}}(1980)}]{Bisnovatyi:80}
\bibinfo{author}{\bibfnamefont{G.~S.} \bibnamefont{{Bisnovatyi-Kogan}}},
  \bibinfo{journal}{\azh} \textbf{\bibinfo{volume}{57}}, \bibinfo{pages}{899}
  (\bibinfo{year}{1980}).

\bibitem[{\citenamefont{{Bond} et~al.}(1980)\citenamefont{{Bond}, {Efstathiou},
  and {Silk}}}]{Bond:80}
\bibinfo{author}{\bibfnamefont{J.~R.} \bibnamefont{{Bond}}},
  \bibinfo{author}{\bibfnamefont{G.}~\bibnamefont{{Efstathiou}}},
  \bibnamefont{and} \bibinfo{author}{\bibfnamefont{J.}~\bibnamefont{{Silk}}},
  \bibinfo{journal}{Phys. Rev. Lett.} \textbf{\bibinfo{volume}{45}},
  \bibinfo{pages}{1980} (\bibinfo{year}{1980}).

\bibitem[{\citenamefont{{Doroshkevich}
  et~al.}(1981)\citenamefont{{Doroshkevich}, {Khlopov}, {Sunyaev}, {Szalay},
  and {Zeldovich}}}]{Doroshkevich:81}
\bibinfo{author}{\bibfnamefont{A.~G.} \bibnamefont{{Doroshkevich}}},
  \bibinfo{author}{\bibfnamefont{M.~I.} \bibnamefont{{Khlopov}}},
  \bibinfo{author}{\bibfnamefont{R.~A.} \bibnamefont{{Sunyaev}}},
  \bibinfo{author}{\bibfnamefont{A.~S.} \bibnamefont{{Szalay}}},
  \bibnamefont{and} \bibinfo{author}{\bibfnamefont{I.~B.}
  \bibnamefont{{Zeldovich}}}, \bibinfo{journal}{New York Academy Sciences
  Annals} \textbf{\bibinfo{volume}{375}}, \bibinfo{pages}{32}
  (\bibinfo{year}{1981}).

\bibitem[{\citenamefont{{Bond} and {Szalay}}(1983)}]{Bond:83}
\bibinfo{author}{\bibfnamefont{J.~R.} \bibnamefont{{Bond}}} \bibnamefont{and}
  \bibinfo{author}{\bibfnamefont{A.~S.} \bibnamefont{{Szalay}}},
  \bibinfo{journal}{\apj} \textbf{\bibinfo{volume}{274}}, \bibinfo{pages}{443}
  (\bibinfo{year}{1983}).

\bibitem[{\citenamefont{{White} et~al.}(1983)\citenamefont{{White}, {Frenk},
  and {Davis}}}]{White:83}
\bibinfo{author}{\bibfnamefont{S.~D.~M.} \bibnamefont{{White}}},
  \bibinfo{author}{\bibfnamefont{C.~S.} \bibnamefont{{Frenk}}},
  \bibnamefont{and} \bibinfo{author}{\bibfnamefont{M.}~\bibnamefont{{Davis}}},
  \bibinfo{journal}{\apjl} \textbf{\bibinfo{volume}{274}}, \bibinfo{pages}{L1}
  (\bibinfo{year}{1983}).

\bibitem[{\citenamefont{{Peebles}}(1984)}]{Peebles:84a}
\bibinfo{author}{\bibfnamefont{P.~J.~E.} \bibnamefont{{Peebles}}},
  \bibinfo{journal}{Science} \textbf{\bibinfo{volume}{224}},
  \bibinfo{pages}{1385} (\bibinfo{year}{1984}).

\bibitem[{\citenamefont{Bergstrom}(2000)}]{Bergstrom:00}
\bibinfo{author}{\bibfnamefont{L.}~\bibnamefont{Bergstrom}},
  \bibinfo{journal}{Rept.Prog.Phys.} \textbf{\bibinfo{volume}{63}},
  \bibinfo{pages}{793} (\bibinfo{year}{2000}), \eprint{hep-ph/0002126}.

\bibitem[{\citenamefont{{Carr} et~al.}(2006)\citenamefont{{Carr}, {Lamanna},
  and {Lavalle}}}]{Carr:06}
\bibinfo{author}{\bibfnamefont{J.}~\bibnamefont{{Carr}}},
  \bibinfo{author}{\bibfnamefont{G.}~\bibnamefont{{Lamanna}}},
  \bibnamefont{and}
  \bibinfo{author}{\bibfnamefont{J.}~\bibnamefont{{Lavalle}}},
  \bibinfo{journal}{Reports of Progress in Physics}
  \textbf{\bibinfo{volume}{69}}, \bibinfo{pages}{2475} (\bibinfo{year}{2006}).

\bibitem[{\citenamefont{Taoso et~al.}(2008)\citenamefont{Taoso, Bertone, and
  Masiero}}]{Taoso:07}
\bibinfo{author}{\bibfnamefont{M.}~\bibnamefont{Taoso}},
  \bibinfo{author}{\bibfnamefont{G.}~\bibnamefont{Bertone}}, \bibnamefont{and}
  \bibinfo{author}{\bibfnamefont{A.}~\bibnamefont{Masiero}},
  \bibinfo{journal}{JCAP} \textbf{\bibinfo{volume}{0803}}, \bibinfo{pages}{022}
  (\bibinfo{year}{2008}), \eprint{0711.4996}.

\bibitem[{\citenamefont{{Bertone} et~al.}(2005)\citenamefont{{Bertone},
  {Hooper}, and {Silk}}}]{Bertone:05}
\bibinfo{author}{\bibfnamefont{G.}~\bibnamefont{{Bertone}}},
  \bibinfo{author}{\bibfnamefont{D.}~\bibnamefont{{Hooper}}}, \bibnamefont{and}
  \bibinfo{author}{\bibfnamefont{J.}~\bibnamefont{{Silk}}},
  \bibinfo{journal}{\physrep} \textbf{\bibinfo{volume}{405}},
  \bibinfo{pages}{279} (\bibinfo{year}{2005}), \eprint{arXiv:hep-ph/0404175}.

\bibitem[{\citenamefont{{Dubovsky} et~al.}(2005)\citenamefont{{Dubovsky},
  {Tinyakov}, and {Tkachev}}}]{Dubovsky:04}
\bibinfo{author}{\bibfnamefont{S.~L.} \bibnamefont{{Dubovsky}}},
  \bibinfo{author}{\bibfnamefont{P.~G.} \bibnamefont{{Tinyakov}}},
  \bibnamefont{and} \bibinfo{author}{\bibfnamefont{I.~I.}
  \bibnamefont{{Tkachev}}}, \bibinfo{journal}{\prl}
  \textbf{\bibinfo{volume}{94}}, \bibinfo{pages}{181102}
  (\bibinfo{year}{2005}), \eprint{hep-th/0411158}.

\bibitem[{\citenamefont{{Holman} et~al.}(1983)\citenamefont{{Holman},
  {Lazarides}, and {Shafi}}}]{Holman:83}
\bibinfo{author}{\bibfnamefont{R.}~\bibnamefont{{Holman}}},
  \bibinfo{author}{\bibfnamefont{G.}~\bibnamefont{{Lazarides}}},
  \bibnamefont{and} \bibinfo{author}{\bibfnamefont{Q.}~\bibnamefont{{Shafi}}},
  \bibinfo{journal}{\prd} \textbf{\bibinfo{volume}{27}}, \bibinfo{pages}{995}
  (\bibinfo{year}{1983}).

\bibitem[{\citenamefont{Dodelson and Widrow}(1994)}]{Dodelson:93}
\bibinfo{author}{\bibfnamefont{S.}~\bibnamefont{Dodelson}} \bibnamefont{and}
  \bibinfo{author}{\bibfnamefont{L.~M.} \bibnamefont{Widrow}},
  \bibinfo{journal}{Phys. Rev. Lett.} \textbf{\bibinfo{volume}{72}},
  \bibinfo{pages}{17} (\bibinfo{year}{1994}), \eprint{hep-ph/9303287}.

\bibitem[{\citenamefont{{Pagels} and {Primack}}(1982)}]{Pagels:82}
\bibinfo{author}{\bibfnamefont{H.}~\bibnamefont{{Pagels}}} \bibnamefont{and}
  \bibinfo{author}{\bibfnamefont{J.~R.} \bibnamefont{{Primack}}},
  \bibinfo{journal}{\prl} \textbf{\bibinfo{volume}{48}}, \bibinfo{pages}{223}
  (\bibinfo{year}{1982}).

\bibitem[{\citenamefont{{Haber} and {Kane}}(1985)}]{Haber:85}
\bibinfo{author}{\bibfnamefont{H.~E.} \bibnamefont{{Haber}}} \bibnamefont{and}
  \bibinfo{author}{\bibfnamefont{G.~L.} \bibnamefont{{Kane}}},
  \bibinfo{journal}{\physrep} \textbf{\bibinfo{volume}{117}},
  \bibinfo{pages}{75} (\bibinfo{year}{1985}).

\bibitem[{\citenamefont{{Covi} et~al.}(1999)\citenamefont{{Covi}, {Kim}, and
  {Roszkowski}}}]{Covi:99}
\bibinfo{author}{\bibfnamefont{L.}~\bibnamefont{{Covi}}},
  \bibinfo{author}{\bibfnamefont{J.~E.} \bibnamefont{{Kim}}}, \bibnamefont{and}
  \bibinfo{author}{\bibfnamefont{L.}~\bibnamefont{{Roszkowski}}},
  \bibinfo{journal}{\prl} \textbf{\bibinfo{volume}{82}}, \bibinfo{pages}{4180}
  (\bibinfo{year}{1999}), \eprint{arXiv:hep-ph/9905212}.

\bibitem[{\citenamefont{Kusenko and Shaposhnikov}(1998)}]{Kusenko:97b}
\bibinfo{author}{\bibfnamefont{A.}~\bibnamefont{Kusenko}} \bibnamefont{and}
  \bibinfo{author}{\bibfnamefont{M.~E.} \bibnamefont{Shaposhnikov}},
  \bibinfo{journal}{Phys. Lett.} \textbf{\bibinfo{volume}{B418}},
  \bibinfo{pages}{46} (\bibinfo{year}{1998}), \eprint{hep-ph/9709492}.

\bibitem[{\citenamefont{Kuzmin and Tkachev}(1998)}]{Kuzmin:98}
\bibinfo{author}{\bibfnamefont{V.~A.} \bibnamefont{Kuzmin}} \bibnamefont{and}
  \bibinfo{author}{\bibfnamefont{I.~I.} \bibnamefont{Tkachev}},
  \bibinfo{journal}{JETP Lett.} \textbf{\bibinfo{volume}{68}},
  \bibinfo{pages}{271} (\bibinfo{year}{1998}), \eprint{hep-ph/9802304}.

\bibitem[{\citenamefont{{Chung} et~al.}(1999)\citenamefont{{Chung}, {Kolb}, and
  {Riotto}}}]{Chung:99}
\bibinfo{author}{\bibfnamefont{D.~J.~H.} \bibnamefont{{Chung}}},
  \bibinfo{author}{\bibfnamefont{E.~W.} \bibnamefont{{Kolb}}},
  \bibnamefont{and} \bibinfo{author}{\bibfnamefont{A.}~\bibnamefont{{Riotto}}},
  \bibinfo{journal}{\prd} \textbf{\bibinfo{volume}{59}},
  \bibinfo{pages}{023501} (\bibinfo{year}{1999}),
  \eprint{arXiv:hep-ph/9802238}.

\bibitem[{\citenamefont{Tremaine and Gunn}(1979)}]{Tremaine:79}
\bibinfo{author}{\bibfnamefont{S.}~\bibnamefont{Tremaine}} \bibnamefont{and}
  \bibinfo{author}{\bibfnamefont{J.~E.} \bibnamefont{Gunn}},
  \bibinfo{journal}{Phys. Rev. Lett.} \textbf{\bibinfo{volume}{42}},
  \bibinfo{pages}{407} (\bibinfo{year}{1979}).

\bibitem[{\citenamefont{{Madsen} and {Epstein}}(1984)}]{Madsen:84}
\bibinfo{author}{\bibfnamefont{J.}~\bibnamefont{{Madsen}}} \bibnamefont{and}
  \bibinfo{author}{\bibfnamefont{R.~I.} \bibnamefont{{Epstein}}},
  \bibinfo{journal}{\apj} \textbf{\bibinfo{volume}{282}}, \bibinfo{pages}{11}
  (\bibinfo{year}{1984}).

\bibitem[{\citenamefont{{Madsen}}(1991)}]{Madsen:91}
\bibinfo{author}{\bibfnamefont{J.}~\bibnamefont{{Madsen}}},
  \bibinfo{journal}{\prd} \textbf{\bibinfo{volume}{44}}, \bibinfo{pages}{999}
  (\bibinfo{year}{1991}).

\bibitem[{\citenamefont{{Madsen}}(1990)}]{Madsen:90}
\bibinfo{author}{\bibfnamefont{J.}~\bibnamefont{{Madsen}}},
  \bibinfo{journal}{Physical Review Letters} \textbf{\bibinfo{volume}{64}},
  \bibinfo{pages}{2744} (\bibinfo{year}{1990}).

\bibitem[{\citenamefont{Dalcanton and Hogan}(2001)}]{Dalcanton:00}
\bibinfo{author}{\bibfnamefont{J.~J.} \bibnamefont{Dalcanton}}
  \bibnamefont{and} \bibinfo{author}{\bibfnamefont{C.~J.} \bibnamefont{Hogan}},
  \bibinfo{journal}{\apj} \textbf{\bibinfo{volume}{561}}, \bibinfo{pages}{35}
  (\bibinfo{year}{2001}), \eprint{astro-ph/0004381}.

\bibitem[{\citenamefont{Hogan and Dalcanton}(2000)}]{Hogan:00}
\bibinfo{author}{\bibfnamefont{C.~J.} \bibnamefont{Hogan}} \bibnamefont{and}
  \bibinfo{author}{\bibfnamefont{J.~J.} \bibnamefont{Dalcanton}},
  \bibinfo{journal}{Phys. Rev.} \textbf{\bibinfo{volume}{D62}},
  \bibinfo{pages}{063511} (\bibinfo{year}{2000}), \eprint{astro-ph/0002330}.

\bibitem[{\citenamefont{{Madsen}}(2001{\natexlab{a}})}]{Madsen:00}
\bibinfo{author}{\bibfnamefont{J.}~\bibnamefont{{Madsen}}},
  \bibinfo{journal}{\prd} \textbf{\bibinfo{volume}{64}},
  \bibinfo{pages}{027301} (\bibinfo{year}{2001}{\natexlab{a}}),
  \eprint{arXiv:astro-ph/0006074}.

\bibitem[{\citenamefont{{Hui} et~al.}(1997)\citenamefont{{Hui}, {Gnedin}, and
  {Zhang}}}]{Hui:97}
\bibinfo{author}{\bibfnamefont{L.}~\bibnamefont{{Hui}}},
  \bibinfo{author}{\bibfnamefont{N.~Y.} \bibnamefont{{Gnedin}}},
  \bibnamefont{and} \bibinfo{author}{\bibfnamefont{Y.}~\bibnamefont{{Zhang}}},
  \bibinfo{journal}{\apj} \textbf{\bibinfo{volume}{486}}, \bibinfo{pages}{599}
  (\bibinfo{year}{1997}), \eprint{astro-ph/9608157}.

\bibitem[{\citenamefont{Gnedin and Hamilton}(2002)}]{Gnedin:01}
\bibinfo{author}{\bibfnamefont{N.~Y.} \bibnamefont{Gnedin}} \bibnamefont{and}
  \bibinfo{author}{\bibfnamefont{A.~J.~S.} \bibnamefont{Hamilton}},
  \bibinfo{journal}{Mon.Not.Roy.Astron.Soc.} \textbf{\bibinfo{volume}{334}},
  \bibinfo{pages}{107} (\bibinfo{year}{2002}), \eprint{astro-ph/0111194}.

\bibitem[{\citenamefont{{Weinberg} et~al.}(2003)\citenamefont{{Weinberg},
  {Dav{\'e}}, {Katz}, and {Kollmeier}}}]{Weinberg:03}
\bibinfo{author}{\bibfnamefont{D.~H.} \bibnamefont{{Weinberg}}},
  \bibinfo{author}{\bibfnamefont{R.}~\bibnamefont{{Dav{\'e}}}},
  \bibinfo{author}{\bibfnamefont{N.}~\bibnamefont{{Katz}}}, \bibnamefont{and}
  \bibinfo{author}{\bibfnamefont{J.~A.} \bibnamefont{{Kollmeier}}}, in
  \emph{\bibinfo{booktitle}{AIP Conf. Proc. 666: The Emergence of Cosmic
  Structure}}, edited by \bibinfo{editor}{\bibfnamefont{S.~H.}
  \bibnamefont{{Holt}}} \bibnamefont{and} \bibinfo{editor}{\bibfnamefont{C.~S.}
  \bibnamefont{{Reynolds}}} (\bibinfo{year}{2003}), pp.
  \bibinfo{pages}{157--169}.

\bibitem[{\citenamefont{Lewis and Bridle}(2002)}]{Lewis:02}
\bibinfo{author}{\bibfnamefont{A.}~\bibnamefont{Lewis}} \bibnamefont{and}
  \bibinfo{author}{\bibfnamefont{S.}~\bibnamefont{Bridle}},
  \bibinfo{journal}{Phys. Rev.} \textbf{\bibinfo{volume}{D66}},
  \bibinfo{pages}{103511} (\bibinfo{year}{2002}), \eprint{astro-ph/0205436}.

\bibitem[{\citenamefont{Theuns et~al.}(1998)\citenamefont{Theuns, Leonard,
  Efstathiou, Pearce, and Thomas}}]{Theuns:98}
\bibinfo{author}{\bibfnamefont{T.}~\bibnamefont{Theuns}},
  \bibinfo{author}{\bibfnamefont{A.}~\bibnamefont{Leonard}},
  \bibinfo{author}{\bibfnamefont{G.}~\bibnamefont{Efstathiou}},
  \bibinfo{author}{\bibfnamefont{F.~R.} \bibnamefont{Pearce}},
  \bibnamefont{and} \bibinfo{author}{\bibfnamefont{P.~A.}
  \bibnamefont{Thomas}}, \bibinfo{journal}{Mon. Not. Roy. Astron. Soc.}
  \textbf{\bibinfo{volume}{301}}, \bibinfo{pages}{478} (\bibinfo{year}{1998}),
  \eprint{astro-ph/9805119}.

\bibitem[{\citenamefont{{McDonald} et~al.}(2006)\citenamefont{{McDonald},
  {Seljak}, {Burles}, {Schlegel}, {Weinberg}, {Cen}, {Shih}, {Schaye},
  {Schneider}, {Bahcall} et~al.}}]{McDonald:05}
\bibinfo{author}{\bibfnamefont{P.}~\bibnamefont{{McDonald}}},
  \bibinfo{author}{\bibfnamefont{U.}~\bibnamefont{{Seljak}}},
  \bibinfo{author}{\bibfnamefont{S.}~\bibnamefont{{Burles}}},
  \bibinfo{author}{\bibfnamefont{D.~J.} \bibnamefont{{Schlegel}}},
  \bibinfo{author}{\bibfnamefont{D.~H.} \bibnamefont{{Weinberg}}},
  \bibinfo{author}{\bibfnamefont{R.}~\bibnamefont{{Cen}}},
  \bibinfo{author}{\bibfnamefont{D.}~\bibnamefont{{Shih}}},
  \bibinfo{author}{\bibfnamefont{J.}~\bibnamefont{{Schaye}}},
  \bibinfo{author}{\bibfnamefont{D.~P.} \bibnamefont{{Schneider}}},
  \bibinfo{author}{\bibfnamefont{N.~A.} \bibnamefont{{Bahcall}}},
  \bibnamefont{et~al.}, \bibinfo{journal}{\apjs}
  \textbf{\bibinfo{volume}{163}}, \bibinfo{pages}{80} (\bibinfo{year}{2006}),
  \eprint{arXiv:astro-ph/0405013}.

\bibitem[{\citenamefont{Viel et~al.}(2004{\natexlab{a}})\citenamefont{Viel,
  Haehnelt, and Springel}}]{Viel:04}
\bibinfo{author}{\bibfnamefont{M.}~\bibnamefont{Viel}},
  \bibinfo{author}{\bibfnamefont{M.~G.} \bibnamefont{Haehnelt}},
  \bibnamefont{and} \bibinfo{author}{\bibfnamefont{V.}~\bibnamefont{Springel}},
  \bibinfo{journal}{Mon. Not. Roy. Astron. Soc.}
  \textbf{\bibinfo{volume}{354}}, \bibinfo{pages}{684}
  (\bibinfo{year}{2004}{\natexlab{a}}), \eprint{astro-ph/0404600}.

\bibitem[{\citenamefont{Viel et~al.}(2005)\citenamefont{Viel, Lesgourgues,
  Haehnelt, Matarrese, and Riotto}}]{Viel:05}
\bibinfo{author}{\bibfnamefont{M.}~\bibnamefont{Viel}},
  \bibinfo{author}{\bibfnamefont{J.}~\bibnamefont{Lesgourgues}},
  \bibinfo{author}{\bibfnamefont{M.~G.} \bibnamefont{Haehnelt}},
  \bibinfo{author}{\bibfnamefont{S.}~\bibnamefont{Matarrese}},
  \bibnamefont{and} \bibinfo{author}{\bibfnamefont{A.}~\bibnamefont{Riotto}},
  \bibinfo{journal}{Phys. Rev.} \textbf{\bibinfo{volume}{D71}},
  \bibinfo{pages}{063534} (\bibinfo{year}{2005}), \eprint{astro-ph/0501562}.

\bibitem[{\citenamefont{Viel et~al.}(2006{\natexlab{a}})\citenamefont{Viel,
  Haehnelt, and Springel}}]{Viel:05b}
\bibinfo{author}{\bibfnamefont{M.}~\bibnamefont{Viel}},
  \bibinfo{author}{\bibfnamefont{M.~G.} \bibnamefont{Haehnelt}},
  \bibnamefont{and} \bibinfo{author}{\bibfnamefont{V.}~\bibnamefont{Springel}},
  \bibinfo{journal}{Mon.Not.Roy.Astron.Soc.} \textbf{\bibinfo{volume}{367}},
  \bibinfo{pages}{1655} (\bibinfo{year}{2006}{\natexlab{a}}),
  \eprint{astro-ph/0504641}.

\bibitem[{\citenamefont{Viel and Haehnelt}(2006)}]{Viel:05c}
\bibinfo{author}{\bibfnamefont{M.}~\bibnamefont{Viel}} \bibnamefont{and}
  \bibinfo{author}{\bibfnamefont{M.~G.} \bibnamefont{Haehnelt}},
  \bibinfo{journal}{Mon. Not. Roy. Astron. Soc.}
  \textbf{\bibinfo{volume}{365}}, \bibinfo{pages}{231} (\bibinfo{year}{2006}),
  \eprint{astro-ph/0508177}.

\bibitem[{\citenamefont{Regan et~al.}(2007)\citenamefont{Regan, Haehnelt, and
  Viel}}]{Regan:06a}
\bibinfo{author}{\bibfnamefont{J.~A.} \bibnamefont{Regan}},
  \bibinfo{author}{\bibfnamefont{M.~G.} \bibnamefont{Haehnelt}},
  \bibnamefont{and} \bibinfo{author}{\bibfnamefont{M.}~\bibnamefont{Viel}},
  \bibinfo{journal}{Mon. Not. Roy. Astron. Soc.}
  \textbf{\bibinfo{volume}{374}}, \bibinfo{pages}{196} (\bibinfo{year}{2007}),
  \eprint{astro-ph/0606638}.

\bibitem[{\citenamefont{Kim et~al.}(2007)\citenamefont{Kim, Bolton, Viel,
  Haehnelt, and Carswell}}]{Kim:07a}
\bibinfo{author}{\bibfnamefont{T.~S.} \bibnamefont{Kim}},
  \bibinfo{author}{\bibfnamefont{J.~S.} \bibnamefont{Bolton}},
  \bibinfo{author}{\bibfnamefont{M.}~\bibnamefont{Viel}},
  \bibinfo{author}{\bibfnamefont{M.~G.} \bibnamefont{Haehnelt}},
  \bibnamefont{and} \bibinfo{author}{\bibfnamefont{R.~F.}
  \bibnamefont{Carswell}} (\bibinfo{year}{2007}), \eprint{0711.1862}.

\bibitem[{\citenamefont{Bolton et~al.}(2007)\citenamefont{Bolton, Viel, Kim,
  Haehnelt, and Carswell}}]{Bolton:07a}
\bibinfo{author}{\bibfnamefont{J.~S.} \bibnamefont{Bolton}},
  \bibinfo{author}{\bibfnamefont{M.}~\bibnamefont{Viel}},
  \bibinfo{author}{\bibfnamefont{T.~S.} \bibnamefont{Kim}},
  \bibinfo{author}{\bibfnamefont{M.~G.} \bibnamefont{Haehnelt}},
  \bibnamefont{and} \bibinfo{author}{\bibfnamefont{R.~F.}
  \bibnamefont{Carswell}} (\bibinfo{year}{2007}), \eprint{0711.2064}.

\bibitem[{\citenamefont{Viel et~al.}(2002{\natexlab{a}})\citenamefont{Viel,
  Matarrese, Mo, Theuns, and Haehnelt}}]{Viel:2002ui}
\bibinfo{author}{\bibfnamefont{M.}~\bibnamefont{Viel}},
  \bibinfo{author}{\bibfnamefont{S.}~\bibnamefont{Matarrese}},
  \bibinfo{author}{\bibfnamefont{H.~J.} \bibnamefont{Mo}},
  \bibinfo{author}{\bibfnamefont{T.}~\bibnamefont{Theuns}}, \bibnamefont{and}
  \bibinfo{author}{\bibfnamefont{M.~G.} \bibnamefont{Haehnelt}},
  \bibinfo{journal}{Mon. Not. Roy. Astron. Soc.}
  \textbf{\bibinfo{volume}{336}}, \bibinfo{pages}{685}
  (\bibinfo{year}{2002}{\natexlab{a}}), \eprint{astro-ph/0203418}.

\bibitem[{\citenamefont{Viel et~al.}(2002{\natexlab{b}})\citenamefont{Viel,
  Matarrese, Mo, Haehnelt, and Theuns}}]{Viel:2001hd}
\bibinfo{author}{\bibfnamefont{M.}~\bibnamefont{Viel}},
  \bibinfo{author}{\bibfnamefont{S.}~\bibnamefont{Matarrese}},
  \bibinfo{author}{\bibfnamefont{H.~J.} \bibnamefont{Mo}},
  \bibinfo{author}{\bibfnamefont{M.~G.} \bibnamefont{Haehnelt}},
  \bibnamefont{and} \bibinfo{author}{\bibfnamefont{T.}~\bibnamefont{Theuns}},
  \bibinfo{journal}{Mon. Not. Roy. Astron. Soc.}
  \textbf{\bibinfo{volume}{329}}, \bibinfo{pages}{848}
  (\bibinfo{year}{2002}{\natexlab{b}}), \eprint{astro-ph/0105233}.

\bibitem[{\citenamefont{Viel et~al.}(2004{\natexlab{b}})\citenamefont{Viel,
  Haehnelt, Carswell, and Kim}}]{Viel:2003fx}
\bibinfo{author}{\bibfnamefont{M.}~\bibnamefont{Viel}},
  \bibinfo{author}{\bibfnamefont{M.~G.} \bibnamefont{Haehnelt}},
  \bibinfo{author}{\bibfnamefont{R.~F.} \bibnamefont{Carswell}},
  \bibnamefont{and} \bibinfo{author}{\bibfnamefont{T.~S.} \bibnamefont{Kim}},
  \bibinfo{journal}{Mon. Not. Roy. Astron. Soc.}
  \textbf{\bibinfo{volume}{349}}, \bibinfo{pages}{L33}
  (\bibinfo{year}{2004}{\natexlab{b}}), \eprint{astro-ph/0308078}.

\bibitem[{\citenamefont{Biermann and Kusenko}(2006)}]{Biermann:06}
\bibinfo{author}{\bibfnamefont{P.~L.} \bibnamefont{Biermann}} \bibnamefont{and}
  \bibinfo{author}{\bibfnamefont{A.}~\bibnamefont{Kusenko}},
  \bibinfo{journal}{Phys. Rev. Lett.} \textbf{\bibinfo{volume}{96}},
  \bibinfo{pages}{091301} (\bibinfo{year}{2006}), \eprint{astro-ph/0601004}.

\bibitem[{\citenamefont{{Gao} and {Theuns}}(2007)}]{Gao:07}
\bibinfo{author}{\bibfnamefont{L.}~\bibnamefont{{Gao}}} \bibnamefont{and}
  \bibinfo{author}{\bibfnamefont{T.}~\bibnamefont{{Theuns}}},
  \bibinfo{journal}{Science} \textbf{\bibinfo{volume}{317}},
  \bibinfo{pages}{1527} (\bibinfo{year}{2007}), \eprint{arXiv:0709.2165}.

\bibitem[{\citenamefont{{Stasielak} et~al.}(2007)\citenamefont{{Stasielak},
  {Biermann}, and {Kusenko}}}]{Stasielak:07}
\bibinfo{author}{\bibfnamefont{J.}~\bibnamefont{{Stasielak}}},
  \bibinfo{author}{\bibfnamefont{P.~L.} \bibnamefont{{Biermann}}},
  \bibnamefont{and}
  \bibinfo{author}{\bibfnamefont{A.}~\bibnamefont{{Kusenko}}},
  \bibinfo{journal}{\apj} \textbf{\bibinfo{volume}{654}}, \bibinfo{pages}{290}
  (\bibinfo{year}{2007}), \eprint{arXiv:astro-ph/0606435}.

\bibitem[{\citenamefont{Boyarsky
  et~al.}(2008{\natexlab{a}})\citenamefont{Boyarsky, Lesgourgues, Ruchayskiy,
  and Viel}}]{Boyarsky:08c}
\bibinfo{author}{\bibfnamefont{A.}~\bibnamefont{Boyarsky}},
  \bibinfo{author}{\bibfnamefont{J.}~\bibnamefont{Lesgourgues}},
  \bibinfo{author}{\bibfnamefont{O.}~\bibnamefont{Ruchayskiy}},
  \bibnamefont{and} \bibinfo{author}{\bibfnamefont{M.}~\bibnamefont{Viel}}
  (\bibinfo{year}{2008}{\natexlab{a}}), \eprint{0812.0010}.

\bibitem[{\citenamefont{Asaka et~al.}(2005)\citenamefont{Asaka, Blanchet, and
  Shaposhnikov}}]{Asaka:05a}
\bibinfo{author}{\bibfnamefont{T.}~\bibnamefont{Asaka}},
  \bibinfo{author}{\bibfnamefont{S.}~\bibnamefont{Blanchet}}, \bibnamefont{and}
  \bibinfo{author}{\bibfnamefont{M.}~\bibnamefont{Shaposhnikov}},
  \bibinfo{journal}{Phys. Lett.} \textbf{\bibinfo{volume}{B631}},
  \bibinfo{pages}{151} (\bibinfo{year}{2005}), \eprint{hep-ph/0503065}.

\bibitem[{\citenamefont{{Fogli} et~al.}(2006)\citenamefont{{Fogli}, {Lisi},
  {Marrone}, {Palazzo}, and {Rotunno}}}]{Fogli:05}
\bibinfo{author}{\bibfnamefont{G.~L.} \bibnamefont{{Fogli}}},
  \bibinfo{author}{\bibfnamefont{E.}~\bibnamefont{{Lisi}}},
  \bibinfo{author}{\bibfnamefont{A.}~\bibnamefont{{Marrone}}},
  \bibinfo{author}{\bibfnamefont{A.}~\bibnamefont{{Palazzo}}},
  \bibnamefont{and} \bibinfo{author}{\bibfnamefont{A.~M.}
  \bibnamefont{{Rotunno}}}, \bibinfo{journal}{Prog. Part. Nucl. Phys.}
  \textbf{\bibinfo{volume}{57}}, \bibinfo{pages}{71} (\bibinfo{year}{2006}),
  \eprint{arXiv:hep-ph/0506083}.

\bibitem[{\citenamefont{Strumia and Vissani}(2006)}]{Strumia:06}
\bibinfo{author}{\bibfnamefont{A.}~\bibnamefont{Strumia}} \bibnamefont{and}
  \bibinfo{author}{\bibfnamefont{F.}~\bibnamefont{Vissani}}
  (\bibinfo{year}{2006}), \eprint{hep-ph/0606054}.

\bibitem[{\citenamefont{Giunti}(2007)}]{Giunti:06}
\bibinfo{author}{\bibfnamefont{C.}~\bibnamefont{Giunti}},
  \bibinfo{journal}{Nucl. Phys. Proc. Suppl.} \textbf{\bibinfo{volume}{169}},
  \bibinfo{pages}{309} (\bibinfo{year}{2007}), \eprint{hep-ph/0611125}.

\bibitem[{\citenamefont{Asaka and Shaposhnikov}(2005)}]{Asaka:2005pn}
\bibinfo{author}{\bibfnamefont{T.}~\bibnamefont{Asaka}} \bibnamefont{and}
  \bibinfo{author}{\bibfnamefont{M.}~\bibnamefont{Shaposhnikov}},
  \bibinfo{journal}{Phys. Lett.} \textbf{\bibinfo{volume}{B620}},
  \bibinfo{pages}{17} (\bibinfo{year}{2005}), \eprint{hep-ph/0505013}.

\bibitem[{\citenamefont{Dolgov}(1997)}]{Dolgov:97}
\bibinfo{author}{\bibfnamefont{A.~D.} \bibnamefont{Dolgov}}
  (\bibinfo{year}{1997}), \eprint{hep-ph/9707419}.

\bibitem[{\citenamefont{Riotto}(1998)}]{Riotto:98}
\bibinfo{author}{\bibfnamefont{A.}~\bibnamefont{Riotto}}
  (\bibinfo{year}{1998}), \eprint{hep-ph/9807454}.

\bibitem[{\citenamefont{{Sommer-Larsen} and {Dolgov}}(2001)}]{Sommer:99}
\bibinfo{author}{\bibfnamefont{J.}~\bibnamefont{{Sommer-Larsen}}}
  \bibnamefont{and} \bibinfo{author}{\bibfnamefont{A.}~\bibnamefont{{Dolgov}}},
  \bibinfo{journal}{\apj} \textbf{\bibinfo{volume}{551}}, \bibinfo{pages}{608}
  (\bibinfo{year}{2001}), \eprint{arXiv:astro-ph/9912166}.

\bibitem[{\citenamefont{Kusenko}(2006)}]{Kusenko:06a}
\bibinfo{author}{\bibfnamefont{A.}~\bibnamefont{Kusenko}},
  \bibinfo{journal}{Phys. Rev. Lett.} \textbf{\bibinfo{volume}{97}},
  \bibinfo{pages}{241301} (\bibinfo{year}{2006}), \eprint{hep-ph/0609081}.

\bibitem[{\citenamefont{Hidaka and Fuller}(2006)}]{Hidaka:06}
\bibinfo{author}{\bibfnamefont{J.}~\bibnamefont{Hidaka}} \bibnamefont{and}
  \bibinfo{author}{\bibfnamefont{G.~M.} \bibnamefont{Fuller}},
  \bibinfo{journal}{\prd} \textbf{\bibinfo{volume}{74}},
  \bibinfo{pages}{125015} (\bibinfo{year}{2006}), \eprint{astro-ph/0609425}.

\bibitem[{\citenamefont{{Hidaka} and {Fuller}}(2007)}]{Hidaka:07}
\bibinfo{author}{\bibfnamefont{J.}~\bibnamefont{{Hidaka}}} \bibnamefont{and}
  \bibinfo{author}{\bibfnamefont{G.~M.} \bibnamefont{{Fuller}}},
  \bibinfo{journal}{\prd} \textbf{\bibinfo{volume}{76}},
  \bibinfo{pages}{083516} (\bibinfo{year}{2007}), \eprint{arXiv:0706.3886}.

\bibitem[{\citenamefont{Stasielak et~al.}(2007)\citenamefont{Stasielak,
  Biermann, and Kusenko}}]{Stasielak:06}
\bibinfo{author}{\bibfnamefont{J.}~\bibnamefont{Stasielak}},
  \bibinfo{author}{\bibfnamefont{P.~L.} \bibnamefont{Biermann}},
  \bibnamefont{and} \bibinfo{author}{\bibfnamefont{A.}~\bibnamefont{Kusenko}},
  \bibinfo{journal}{\apj} \textbf{\bibinfo{volume}{654}}, \bibinfo{pages}{290}
  (\bibinfo{year}{2007}), \eprint{arXiv:astro-ph/0606435}.

\bibitem[{\citenamefont{Shaposhnikov}(2007)}]{Shaposhnikov:07b}
\bibinfo{author}{\bibfnamefont{M.}~\bibnamefont{Shaposhnikov}}
  (\bibinfo{year}{2007}), \eprint{0708.3550}.

\bibitem[{\citenamefont{Dolgov and Hansen}(2002)}]{Dolgov:00}
\bibinfo{author}{\bibfnamefont{A.~D.} \bibnamefont{Dolgov}} \bibnamefont{and}
  \bibinfo{author}{\bibfnamefont{S.~H.} \bibnamefont{Hansen}},
  \bibinfo{journal}{Astropart. Phys.} \textbf{\bibinfo{volume}{16}},
  \bibinfo{pages}{339} (\bibinfo{year}{2002}), \eprint{hep-ph/0009083}.

\bibitem[{\citenamefont{Abazajian et~al.}(2001)\citenamefont{Abazajian, Fuller,
  and Patel}}]{Abazajian:01a}
\bibinfo{author}{\bibfnamefont{K.}~\bibnamefont{Abazajian}},
  \bibinfo{author}{\bibfnamefont{G.~M.} \bibnamefont{Fuller}},
  \bibnamefont{and} \bibinfo{author}{\bibfnamefont{M.}~\bibnamefont{Patel}},
  \bibinfo{journal}{\prd} \textbf{\bibinfo{volume}{64}},
  \bibinfo{pages}{023501} (\bibinfo{year}{2001}), \eprint{astro-ph/0101524}.

\bibitem[{\citenamefont{Asaka et~al.}(2006)\citenamefont{Asaka, Laine, and
  Shaposhnikov}}]{Asaka:06b}
\bibinfo{author}{\bibfnamefont{T.}~\bibnamefont{Asaka}},
  \bibinfo{author}{\bibfnamefont{M.}~\bibnamefont{Laine}}, \bibnamefont{and}
  \bibinfo{author}{\bibfnamefont{M.}~\bibnamefont{Shaposhnikov}},
  \bibinfo{journal}{JHEP} \textbf{\bibinfo{volume}{06}}, \bibinfo{pages}{053}
  (\bibinfo{year}{2006}), \eprint{hep-ph/0605209}.

\bibitem[{\citenamefont{Asaka et~al.}(2007)\citenamefont{Asaka, Laine, and
  Shaposhnikov}}]{Asaka:06c}
\bibinfo{author}{\bibfnamefont{T.}~\bibnamefont{Asaka}},
  \bibinfo{author}{\bibfnamefont{M.}~\bibnamefont{Laine}}, \bibnamefont{and}
  \bibinfo{author}{\bibfnamefont{M.}~\bibnamefont{Shaposhnikov}},
  \bibinfo{journal}{JHEP} \textbf{\bibinfo{volume}{01}}, \bibinfo{pages}{091}
  (\bibinfo{year}{2007}), \eprint{hep-ph/0612182}.

\bibitem[{\citenamefont{Shi and Fuller}(1999)}]{Shi:98}
\bibinfo{author}{\bibfnamefont{X.-d.} \bibnamefont{Shi}} \bibnamefont{and}
  \bibinfo{author}{\bibfnamefont{G.~M.} \bibnamefont{Fuller}},
  \bibinfo{journal}{Phys. Rev. Lett.} \textbf{\bibinfo{volume}{82}},
  \bibinfo{pages}{2832} (\bibinfo{year}{1999}), \eprint{astro-ph/9810076}.

\bibitem[{\citenamefont{{Shaposhnikov}}(2008)}]{Shaposhnikov:08a}
\bibinfo{author}{\bibfnamefont{M.}~\bibnamefont{{Shaposhnikov}}},
  \bibinfo{journal}{JHEP} \textbf{\bibinfo{volume}{08}}, \bibinfo{pages}{008}
  (\bibinfo{year}{2008}), \eprint{0804.4542}.

\bibitem[{\citenamefont{{Laine} and {Shaposhnikov}}(2008)}]{Laine:08a}
\bibinfo{author}{\bibfnamefont{M.}~\bibnamefont{{Laine}}} \bibnamefont{and}
  \bibinfo{author}{\bibfnamefont{M.}~\bibnamefont{{Shaposhnikov}}},
  \bibinfo{journal}{JCAP} \textbf{\bibinfo{volume}{6}}, \bibinfo{pages}{31}
  (\bibinfo{year}{2008}), \eprint{arXiv:0804.4543}.

\bibitem[{\citenamefont{Shaposhnikov and Tkachev}(2006)}]{Shaposhnikov:06}
\bibinfo{author}{\bibfnamefont{M.}~\bibnamefont{Shaposhnikov}}
  \bibnamefont{and} \bibinfo{author}{\bibfnamefont{I.}~\bibnamefont{Tkachev}},
  \bibinfo{journal}{Phys. Lett.} \textbf{\bibinfo{volume}{B639}},
  \bibinfo{pages}{414} (\bibinfo{year}{2006}), \eprint{hep-ph/0604236}.

\bibitem[{\citenamefont{Petraki and Kusenko}(2008)}]{Petraki:07}
\bibinfo{author}{\bibfnamefont{K.}~\bibnamefont{Petraki}} \bibnamefont{and}
  \bibinfo{author}{\bibfnamefont{A.}~\bibnamefont{Kusenko}},
  \bibinfo{journal}{Phys. Rev.} \textbf{\bibinfo{volume}{D77}},
  \bibinfo{pages}{065014} (\bibinfo{year}{2008}), \eprint{0711.4646}.

\bibitem[{\citenamefont{{Petraki}}(2008)}]{Petraki:08}
\bibinfo{author}{\bibfnamefont{K.}~\bibnamefont{{Petraki}}},
  \bibinfo{journal}{\prd} \textbf{\bibinfo{volume}{77}},
  \bibinfo{pages}{105004} (\bibinfo{year}{2008}), \eprint{arXiv:0801.3470}.

\bibitem[{\citenamefont{Viel et~al.}(2006{\natexlab{b}})\citenamefont{Viel,
  Lesgourgues, Haehnelt, Matarrese, and Riotto}}]{Viel:06}
\bibinfo{author}{\bibfnamefont{M.}~\bibnamefont{Viel}},
  \bibinfo{author}{\bibfnamefont{J.}~\bibnamefont{Lesgourgues}},
  \bibinfo{author}{\bibfnamefont{M.~G.} \bibnamefont{Haehnelt}},
  \bibinfo{author}{\bibfnamefont{S.}~\bibnamefont{Matarrese}},
  \bibnamefont{and} \bibinfo{author}{\bibfnamefont{A.}~\bibnamefont{Riotto}},
  \bibinfo{journal}{Phys. Rev. Lett.} \textbf{\bibinfo{volume}{97}},
  \bibinfo{pages}{071301} (\bibinfo{year}{2006}{\natexlab{b}}),
  \eprint{astro-ph/0605706}.

\bibitem[{\citenamefont{Seljak et~al.}(2006)\citenamefont{Seljak, Makarov,
  McDonald, and Trac}}]{Seljak:06}
\bibinfo{author}{\bibfnamefont{U.}~\bibnamefont{Seljak}},
  \bibinfo{author}{\bibfnamefont{A.}~\bibnamefont{Makarov}},
  \bibinfo{author}{\bibfnamefont{P.}~\bibnamefont{McDonald}}, \bibnamefont{and}
  \bibinfo{author}{\bibfnamefont{H.}~\bibnamefont{Trac}},
  \bibinfo{journal}{Phys. Rev. Lett.} \textbf{\bibinfo{volume}{97}},
  \bibinfo{pages}{191303} (\bibinfo{year}{2006}), \eprint{astro-ph/0602430}.

\bibitem[{\citenamefont{{Viel} et~al.}(2008)\citenamefont{{Viel}, {Becker},
  {Bolton}, {Haehnelt}, {Rauch}, and {Sargent}}}]{Viel:08}
\bibinfo{author}{\bibfnamefont{M.}~\bibnamefont{{Viel}}},
  \bibinfo{author}{\bibfnamefont{G.~D.} \bibnamefont{{Becker}}},
  \bibinfo{author}{\bibfnamefont{J.~S.} \bibnamefont{{Bolton}}},
  \bibinfo{author}{\bibfnamefont{M.~G.} \bibnamefont{{Haehnelt}}},
  \bibinfo{author}{\bibfnamefont{M.}~\bibnamefont{{Rauch}}}, \bibnamefont{and}
  \bibinfo{author}{\bibfnamefont{W.~L.~W.} \bibnamefont{{Sargent}}},
  \bibinfo{journal}{\prl} \textbf{\bibinfo{volume}{100}},
  \bibinfo{pages}{041304} (\bibinfo{year}{2008}), \eprint{arXiv:0709.0131}.

\bibitem[{\citenamefont{Dunkley et~al.}(2008)}]{Dunkley:2008ie}
\bibinfo{author}{\bibfnamefont{J.}~\bibnamefont{Dunkley}} \bibnamefont{et~al.}
  (\bibinfo{collaboration}{WMAP}) (\bibinfo{year}{2008}), \eprint{0803.0586}.

\bibitem[{\citenamefont{Palazzo et~al.}(2007)\citenamefont{Palazzo,
  Cumberbatch, Slosar, and Silk}}]{Palazzo:07}
\bibinfo{author}{\bibfnamefont{A.}~\bibnamefont{Palazzo}},
  \bibinfo{author}{\bibfnamefont{D.}~\bibnamefont{Cumberbatch}},
  \bibinfo{author}{\bibfnamefont{A.}~\bibnamefont{Slosar}}, \bibnamefont{and}
  \bibinfo{author}{\bibfnamefont{J.}~\bibnamefont{Silk}}
  (\bibinfo{year}{2007}), \eprint{arXiv:0707.1495 [astro-ph]}.

\bibitem[{\citenamefont{Boyarsky
  et~al.}(2008{\natexlab{b}})\citenamefont{Boyarsky, Lesgourgues, Ruchayskiy,
  and Viel}}]{Boyarsky:08d}
\bibinfo{author}{\bibfnamefont{A.}~\bibnamefont{Boyarsky}},
  \bibinfo{author}{\bibfnamefont{J.}~\bibnamefont{Lesgourgues}},
  \bibinfo{author}{\bibfnamefont{O.}~\bibnamefont{Ruchayskiy}},
  \bibnamefont{and} \bibinfo{author}{\bibfnamefont{M.}~\bibnamefont{Viel}}
  (\bibinfo{year}{2008}{\natexlab{b}}), \eprint{0812.3256}.

\bibitem[{\citenamefont{Hansen et~al.}(2002)\citenamefont{Hansen, Lesgourgues,
  Pastor, and Silk}}]{Hansen:01}
\bibinfo{author}{\bibfnamefont{S.~H.} \bibnamefont{Hansen}},
  \bibinfo{author}{\bibfnamefont{J.}~\bibnamefont{Lesgourgues}},
  \bibinfo{author}{\bibfnamefont{S.}~\bibnamefont{Pastor}}, \bibnamefont{and}
  \bibinfo{author}{\bibfnamefont{J.}~\bibnamefont{Silk}},
  \bibinfo{journal}{\mnras} \textbf{\bibinfo{volume}{333}},
  \bibinfo{pages}{544} (\bibinfo{year}{2002}), \eprint{astro-ph/0106108}.

\bibitem[{\citenamefont{{Simon} and {Geha}}(2007)}]{Simon:07}
\bibinfo{author}{\bibfnamefont{J.~D.} \bibnamefont{{Simon}}} \bibnamefont{and}
  \bibinfo{author}{\bibfnamefont{M.}~\bibnamefont{{Geha}}},
  \bibinfo{journal}{\apj} \textbf{\bibinfo{volume}{670}}, \bibinfo{pages}{313}
  (\bibinfo{year}{2007}), \eprint{arXiv:0706.0516}.

\bibitem[{\citenamefont{{Gilmore} et~al.}(2007)\citenamefont{{Gilmore},
  {Wilkinson}, {Wyse}, {Kleyna}, {Koch}, {Evans}, and {Grebel}}}]{Gilmore:07}
\bibinfo{author}{\bibfnamefont{G.}~\bibnamefont{{Gilmore}}},
  \bibinfo{author}{\bibfnamefont{M.~I.} \bibnamefont{{Wilkinson}}},
  \bibinfo{author}{\bibfnamefont{R.~F.~G.} \bibnamefont{{Wyse}}},
  \bibinfo{author}{\bibfnamefont{J.~T.} \bibnamefont{{Kleyna}}},
  \bibinfo{author}{\bibfnamefont{A.}~\bibnamefont{{Koch}}},
  \bibinfo{author}{\bibfnamefont{N.~W.} \bibnamefont{{Evans}}},
  \bibnamefont{and} \bibinfo{author}{\bibfnamefont{E.~K.}
  \bibnamefont{{Grebel}}}, \bibinfo{journal}{\apj}
  \textbf{\bibinfo{volume}{663}}, \bibinfo{pages}{948} (\bibinfo{year}{2007}),
  \eprint{arXiv:astro-ph/0703308}.

\bibitem[{\citenamefont{{Bilic} and {Viollier}}(1997)}]{Bilic:97}
\bibinfo{author}{\bibfnamefont{N.}~\bibnamefont{{Bilic}}} \bibnamefont{and}
  \bibinfo{author}{\bibfnamefont{R.~D.} \bibnamefont{{Viollier}}},
  \bibinfo{journal}{Phys. Lett. B} \textbf{\bibinfo{volume}{408}},
  \bibinfo{pages}{75} (\bibinfo{year}{1997}), \eprint{arXiv:astro-ph/9607077}.

\bibitem[{\citenamefont{{Viollier}}(1994)}]{Viollier:94}
\bibinfo{author}{\bibfnamefont{R.~D.} \bibnamefont{{Viollier}}},
  \bibinfo{journal}{Progress in Particle and Nuclear Physics}
  \textbf{\bibinfo{volume}{32}}, \bibinfo{pages}{51} (\bibinfo{year}{1994}).

\bibitem[{\citenamefont{{Martin} et~al.}(2008)\citenamefont{{Martin}, {de
  Jong}, and {Rix}}}]{Martin:08}
\bibinfo{author}{\bibfnamefont{N.~F.} \bibnamefont{{Martin}}},
  \bibinfo{author}{\bibfnamefont{J.~T.~A.} \bibnamefont{{de Jong}}},
  \bibnamefont{and} \bibinfo{author}{\bibfnamefont{H.-W.} \bibnamefont{{Rix}}},
  \bibinfo{journal}{ArXiv e-prints} \textbf{\bibinfo{volume}{805}}
  (\bibinfo{year}{2008}), \eprint{0805.2945}.

\bibitem[{\citenamefont{{Kuhlen} et~al.}(2007)\citenamefont{{Kuhlen},
  {Diemand}, and {Madau}}}]{Kuhlen:07}
\bibinfo{author}{\bibfnamefont{M.}~\bibnamefont{{Kuhlen}}},
  \bibinfo{author}{\bibfnamefont{J.}~\bibnamefont{{Diemand}}},
  \bibnamefont{and} \bibinfo{author}{\bibfnamefont{P.}~\bibnamefont{{Madau}}},
  \bibinfo{journal}{\apj} \textbf{\bibinfo{volume}{671}}, \bibinfo{pages}{1135}
  (\bibinfo{year}{2007}), \eprint{0705.2037}.

\bibitem[{\citenamefont{{Cowsik} and {Ghosh}}(1987)}]{Cowsik:87}
\bibinfo{author}{\bibfnamefont{R.}~\bibnamefont{{Cowsik}}} \bibnamefont{and}
  \bibinfo{author}{\bibfnamefont{P.}~\bibnamefont{{Ghosh}}},
  \bibinfo{journal}{\apj} \textbf{\bibinfo{volume}{317}}, \bibinfo{pages}{26}
  (\bibinfo{year}{1987}).

\bibitem[{\citenamefont{{Tremaine} et~al.}(1986)\citenamefont{{Tremaine},
  {Henon}, and {Lynden-Bell}}}]{Tremaine:86}
\bibinfo{author}{\bibfnamefont{S.}~\bibnamefont{{Tremaine}}},
  \bibinfo{author}{\bibfnamefont{M.}~\bibnamefont{{Henon}}}, \bibnamefont{and}
  \bibinfo{author}{\bibfnamefont{D.}~\bibnamefont{{Lynden-Bell}}},
  \bibinfo{journal}{\mnras} \textbf{\bibinfo{volume}{219}},
  \bibinfo{pages}{285} (\bibinfo{year}{1986}).

\bibitem[{\citenamefont{{Madsen}}(2001{\natexlab{b}})}]{Madsen:01}
\bibinfo{author}{\bibfnamefont{J.}~\bibnamefont{{Madsen}}},
  \bibinfo{journal}{\prd} \textbf{\bibinfo{volume}{64}},
  \bibinfo{pages}{027301} (\bibinfo{year}{2001}{\natexlab{b}}),
  \eprint{arXiv:astro-ph/0006074}.

\bibitem[{\citenamefont{{Boyanovsky} et~al.}(2008)\citenamefont{{Boyanovsky},
  {de Vega}, and {Sanchez}}}]{Boyanovsky:08}
\bibinfo{author}{\bibfnamefont{D.}~\bibnamefont{{Boyanovsky}}},
  \bibinfo{author}{\bibfnamefont{H.~J.} \bibnamefont{{de Vega}}},
  \bibnamefont{and} \bibinfo{author}{\bibfnamefont{N.~G.}
  \bibnamefont{{Sanchez}}}, \bibinfo{journal}{\prd}
  \textbf{\bibinfo{volume}{77}}, \bibinfo{pages}{043518}
  (\bibinfo{year}{2008}), \eprint{arXiv:0710.5180}.

\bibitem[{\citenamefont{{Binney} and {Tremaine}}(2008)}]{Binney-Tremaine}
\bibinfo{author}{\bibfnamefont{J.}~\bibnamefont{{Binney}}} \bibnamefont{and}
  \bibinfo{author}{\bibfnamefont{S.}~\bibnamefont{{Tremaine}}},
  \emph{\bibinfo{title}{{Galactic Dynamics: Second Edition}}}
  (\bibinfo{publisher}{Galactic Dynamics: Second Edition, by James Binney and
  Scott Tremaine.~ISBN 978-0-691-13026-2 (HB).~Published by Princeton
  University Press, Princeton, NJ USA, 2008.}, \bibinfo{year}{2008}).

\bibitem[{\citenamefont{{Hansen} et~al.}(2005)\citenamefont{{Hansen}, {Egli},
  {Hollenstein}, and {Salzmann}}}]{Hansen:04}
\bibinfo{author}{\bibfnamefont{S.~H.} \bibnamefont{{Hansen}}},
  \bibinfo{author}{\bibfnamefont{D.}~\bibnamefont{{Egli}}},
  \bibinfo{author}{\bibfnamefont{L.}~\bibnamefont{{Hollenstein}}},
  \bibnamefont{and}
  \bibinfo{author}{\bibfnamefont{C.}~\bibnamefont{{Salzmann}}},
  \bibinfo{journal}{New Astronomy} \textbf{\bibinfo{volume}{10}},
  \bibinfo{pages}{379} (\bibinfo{year}{2005}), \eprint{arXiv:astro-ph/0407111}.

\bibitem[{\citenamefont{Bolz et~al.}(2001)\citenamefont{Bolz, Brandenburg, and
  Buchmuller}}]{Bolz:00}
\bibinfo{author}{\bibfnamefont{M.}~\bibnamefont{Bolz}},
  \bibinfo{author}{\bibfnamefont{A.}~\bibnamefont{Brandenburg}},
  \bibnamefont{and}
  \bibinfo{author}{\bibfnamefont{W.}~\bibnamefont{Buchmuller}},
  \bibinfo{journal}{Nucl.Phys. B} \textbf{\bibinfo{volume}{606}},
  \bibinfo{pages}{518} (\bibinfo{year}{2001}), \eprint{hep-ph/0012052}.

\bibitem[{\citenamefont{Rychkov and Strumia}(2007)}]{Rychkov:07}
\bibinfo{author}{\bibfnamefont{V.~S.} \bibnamefont{Rychkov}} \bibnamefont{and}
  \bibinfo{author}{\bibfnamefont{A.}~\bibnamefont{Strumia}},
  \bibinfo{journal}{Phys. Rev.} \textbf{\bibinfo{volume}{D75}},
  \bibinfo{pages}{075011} (\bibinfo{year}{2007}), \eprint{hep-ph/0701104}.

\bibitem[{\citenamefont{Borgani et~al.}(1996)\citenamefont{Borgani, Masiero,
  and Yamaguchi}}]{Borgani:96}
\bibinfo{author}{\bibfnamefont{S.}~\bibnamefont{Borgani}},
  \bibinfo{author}{\bibfnamefont{A.}~\bibnamefont{Masiero}}, \bibnamefont{and}
  \bibinfo{author}{\bibfnamefont{M.}~\bibnamefont{Yamaguchi}},
  \bibinfo{journal}{Phys. Lett.} \textbf{\bibinfo{volume}{B386}},
  \bibinfo{pages}{189} (\bibinfo{year}{1996}), \eprint{hep-ph/9605222}.

\bibitem[{\citenamefont{{Lynden-Bell} and {Wood}}(1968)}]{Lynden-Bell:68}
\bibinfo{author}{\bibfnamefont{D.}~\bibnamefont{{Lynden-Bell}}}
  \bibnamefont{and} \bibinfo{author}{\bibfnamefont{R.}~\bibnamefont{{Wood}}},
  \bibinfo{journal}{\mnras} \textbf{\bibinfo{volume}{138}},
  \bibinfo{pages}{495} (\bibinfo{year}{1968}).

\bibitem[{\citenamefont{{Antonov}}(1962)}]{Antonov:62}
\bibinfo{author}{\bibfnamefont{V.~A.} \bibnamefont{{Antonov}}},
  \emph{\bibinfo{title}{{Solution of the problem of stability of stellar system
  Emden's density law and the spherical distribution of velocities}}}
  (\bibinfo{publisher}{Vestnik Leningradskogo Universiteta, Leningrad:
  University, 1962}, \bibinfo{year}{1962}).

\bibitem[{\citenamefont{{Pryor} and {Kormendy}}(1990)}]{Pryor:90}
\bibinfo{author}{\bibfnamefont{C.}~\bibnamefont{{Pryor}}} \bibnamefont{and}
  \bibinfo{author}{\bibfnamefont{J.}~\bibnamefont{{Kormendy}}},
  \bibinfo{journal}{\aj} \textbf{\bibinfo{volume}{100}}, \bibinfo{pages}{127}
  (\bibinfo{year}{1990}).

\bibitem[{\citenamefont{{Cole} and {Lacey}}(1996)}]{Cole:95}
\bibinfo{author}{\bibfnamefont{S.}~\bibnamefont{{Cole}}} \bibnamefont{and}
  \bibinfo{author}{\bibfnamefont{C.}~\bibnamefont{{Lacey}}},
  \bibinfo{journal}{\mnras} \textbf{\bibinfo{volume}{281}},
  \bibinfo{pages}{716} (\bibinfo{year}{1996}), \eprint{arXiv:astro-ph/9510147}.

\bibitem[{\citenamefont{{Carlberg} et~al.}(1997)\citenamefont{{Carlberg},
  {Yee}, {Ellingson}, {Morris}, {Abraham}, {Gravel}, {Pritchet},
  {Smecker-Hane}, {Hartwick}, {Hesser} et~al.}}]{Carlberg:97}
\bibinfo{author}{\bibfnamefont{R.~G.} \bibnamefont{{Carlberg}}},
  \bibinfo{author}{\bibfnamefont{H.~K.~C.} \bibnamefont{{Yee}}},
  \bibinfo{author}{\bibfnamefont{E.}~\bibnamefont{{Ellingson}}},
  \bibinfo{author}{\bibfnamefont{S.~L.} \bibnamefont{{Morris}}},
  \bibinfo{author}{\bibfnamefont{R.}~\bibnamefont{{Abraham}}},
  \bibinfo{author}{\bibfnamefont{P.}~\bibnamefont{{Gravel}}},
  \bibinfo{author}{\bibfnamefont{C.~J.} \bibnamefont{{Pritchet}}},
  \bibinfo{author}{\bibfnamefont{T.}~\bibnamefont{{Smecker-Hane}}},
  \bibinfo{author}{\bibfnamefont{F.~D.~A.} \bibnamefont{{Hartwick}}},
  \bibinfo{author}{\bibfnamefont{J.~E.} \bibnamefont{{Hesser}}},
  \bibnamefont{et~al.}, \bibinfo{journal}{\apjl}
  \textbf{\bibinfo{volume}{485}}, \bibinfo{pages}{L13+} (\bibinfo{year}{1997}),
  \eprint{arXiv:astro-ph/9703107}.

\bibitem[{\citenamefont{{Hansen} and {Moore}}(2006)}]{Hansen:06}
\bibinfo{author}{\bibfnamefont{S.~H.} \bibnamefont{{Hansen}}} \bibnamefont{and}
  \bibinfo{author}{\bibfnamefont{B.}~\bibnamefont{{Moore}}},
  \bibinfo{journal}{New Astronomy} \textbf{\bibinfo{volume}{11}},
  \bibinfo{pages}{333} (\bibinfo{year}{2006}).

\bibitem[{\citenamefont{{Zait} et~al.}(2008)\citenamefont{{Zait}, {Hoffman},
  and {Shlosman}}}]{Zait:07}
\bibinfo{author}{\bibfnamefont{A.}~\bibnamefont{{Zait}}},
  \bibinfo{author}{\bibfnamefont{Y.}~\bibnamefont{{Hoffman}}},
  \bibnamefont{and}
  \bibinfo{author}{\bibfnamefont{I.}~\bibnamefont{{Shlosman}}},
  \bibinfo{journal}{\apj} \textbf{\bibinfo{volume}{682}}, \bibinfo{pages}{835}
  (\bibinfo{year}{2008}), \eprint{0711.3791}.

\bibitem[{\citenamefont{{Van Hese} et~al.}(2008)\citenamefont{{Van Hese},
  {Baes}, and {Dejonghe}}}]{Van_Hese:08}
\bibinfo{author}{\bibfnamefont{E.}~\bibnamefont{{Van Hese}}},
  \bibinfo{author}{\bibfnamefont{M.}~\bibnamefont{{Baes}}}, \bibnamefont{and}
  \bibinfo{author}{\bibfnamefont{H.}~\bibnamefont{{Dejonghe}}},
  \bibinfo{journal}{ArXiv e-prints}  (\bibinfo{year}{2008}),
  \eprint{0809.0901}.

\bibitem[{\citenamefont{{An} and {Evans}}(2006)}]{An:05}
\bibinfo{author}{\bibfnamefont{J.~H.} \bibnamefont{{An}}} \bibnamefont{and}
  \bibinfo{author}{\bibfnamefont{N.~W.} \bibnamefont{{Evans}}},
  \bibinfo{journal}{\apj} \textbf{\bibinfo{volume}{642}}, \bibinfo{pages}{752}
  (\bibinfo{year}{2006}), \eprint{arXiv:astro-ph/0511686}.

\bibitem[{\citenamefont{{Evans} et~al.}(2008)\citenamefont{{Evans}, {An}, and
  {Walker}}}]{Evans:08}
\bibinfo{author}{\bibfnamefont{N.~W.} \bibnamefont{{Evans}}},
  \bibinfo{author}{\bibfnamefont{J.}~\bibnamefont{{An}}}, \bibnamefont{and}
  \bibinfo{author}{\bibfnamefont{M.~G.} \bibnamefont{{Walker}}},
  \bibinfo{journal}{ArXiv e-prints}  (\bibinfo{year}{2008}),
  \eprint{0811.1488}.

\bibitem[{\citenamefont{{Strigari} et~al.}(2007)\citenamefont{{Strigari},
  {Bullock}, {Kaplinghat}, {Diemand}, {Kuhlen}, and {Madau}}}]{Strigari:07}
\bibinfo{author}{\bibfnamefont{L.~E.} \bibnamefont{{Strigari}}},
  \bibinfo{author}{\bibfnamefont{J.~S.} \bibnamefont{{Bullock}}},
  \bibinfo{author}{\bibfnamefont{M.}~\bibnamefont{{Kaplinghat}}},
  \bibinfo{author}{\bibfnamefont{J.}~\bibnamefont{{Diemand}}},
  \bibinfo{author}{\bibfnamefont{M.}~\bibnamefont{{Kuhlen}}}, \bibnamefont{and}
  \bibinfo{author}{\bibfnamefont{P.}~\bibnamefont{{Madau}}},
  \textbf{\bibinfo{volume}{704}} (\bibinfo{year}{2007}), \eprint{0704.1817}.

\bibitem[{\citenamefont{{Wu}}(2007)}]{Wu:07}
\bibinfo{author}{\bibfnamefont{X.}~\bibnamefont{{Wu}}} (\bibinfo{year}{2007}),
  \eprint{astro-ph/0702233}.

\bibitem[{\citenamefont{{Strigari} et~al.}(2008)\citenamefont{{Strigari},
  {Koushiappas}, {Bullock}, {Kaplinghat}, {Simon}, {Geha}, and
  {Willman}}}]{Strigari:07a}
\bibinfo{author}{\bibfnamefont{L.~E.} \bibnamefont{{Strigari}}},
  \bibinfo{author}{\bibfnamefont{S.~M.} \bibnamefont{{Koushiappas}}},
  \bibinfo{author}{\bibfnamefont{J.~S.} \bibnamefont{{Bullock}}},
  \bibinfo{author}{\bibfnamefont{M.}~\bibnamefont{{Kaplinghat}}},
  \bibinfo{author}{\bibfnamefont{J.~D.} \bibnamefont{{Simon}}},
  \bibinfo{author}{\bibfnamefont{M.}~\bibnamefont{{Geha}}}, \bibnamefont{and}
  \bibinfo{author}{\bibfnamefont{B.}~\bibnamefont{{Willman}}},
  \bibinfo{journal}{\apj} \textbf{\bibinfo{volume}{678}}, \bibinfo{pages}{614}
  (\bibinfo{year}{2008}), \eprint{arXiv:0709.1510}.

\bibitem[{\citenamefont{{Belokurov} et~al.}(2007)\citenamefont{{Belokurov},
  {Zucker}, {Evans}, {Kleyna}, {Koposov}, {Hodgkin}, {Irwin}, {Gilmore},
  {Wilkinson}, {Fellhauer} et~al.}}]{Belokurov:07}
\bibinfo{author}{\bibfnamefont{V.}~\bibnamefont{{Belokurov}}},
  \bibinfo{author}{\bibfnamefont{D.~B.} \bibnamefont{{Zucker}}},
  \bibinfo{author}{\bibfnamefont{N.~W.} \bibnamefont{{Evans}}},
  \bibinfo{author}{\bibfnamefont{J.~T.} \bibnamefont{{Kleyna}}},
  \bibinfo{author}{\bibfnamefont{S.}~\bibnamefont{{Koposov}}},
  \bibinfo{author}{\bibfnamefont{S.~T.} \bibnamefont{{Hodgkin}}},
  \bibinfo{author}{\bibfnamefont{M.~J.} \bibnamefont{{Irwin}}},
  \bibinfo{author}{\bibfnamefont{G.}~\bibnamefont{{Gilmore}}},
  \bibinfo{author}{\bibfnamefont{M.~I.} \bibnamefont{{Wilkinson}}},
  \bibinfo{author}{\bibfnamefont{M.}~\bibnamefont{{Fellhauer}}},
  \bibnamefont{et~al.}, \bibinfo{journal}{\apj} \textbf{\bibinfo{volume}{654}},
  \bibinfo{pages}{897} (\bibinfo{year}{2007}), \eprint{arXiv:astro-ph/0608448}.

\bibitem[{\citenamefont{{Koposov} et~al.}(2007)\citenamefont{{Koposov},
  {Belokurov}, {Evans}, {Hewett}, {Irwin}, {Gilmore}, {Zucker}, {Rix},
  {Fellhauer}, {Bell} et~al.}}]{Koposov:07}
\bibinfo{author}{\bibfnamefont{S.}~\bibnamefont{{Koposov}}},
  \bibinfo{author}{\bibfnamefont{V.}~\bibnamefont{{Belokurov}}},
  \bibinfo{author}{\bibfnamefont{N.~W.} \bibnamefont{{Evans}}},
  \bibinfo{author}{\bibfnamefont{P.~C.} \bibnamefont{{Hewett}}},
  \bibinfo{author}{\bibfnamefont{M.~J.} \bibnamefont{{Irwin}}},
  \bibinfo{author}{\bibfnamefont{G.}~\bibnamefont{{Gilmore}}},
  \bibinfo{author}{\bibfnamefont{D.~B.} \bibnamefont{{Zucker}}},
  \bibinfo{author}{\bibfnamefont{H.~.} \bibnamefont{{Rix}}},
  \bibinfo{author}{\bibfnamefont{M.}~\bibnamefont{{Fellhauer}}},
  \bibinfo{author}{\bibfnamefont{E.~F.} \bibnamefont{{Bell}}},
  \bibnamefont{et~al.}, \bibinfo{journal}{\apj} \textbf{\bibinfo{volume}{663}},
  \bibinfo{pages}{948} (\bibinfo{year}{2007}), \eprint{0706.2687}.

\bibitem[{\citenamefont{{Irwin} et~al.}(2007)\citenamefont{{Irwin},
  {Belokurov}, {Evans}, {Ryan-Weber}, {de Jong}, {Koposov}, {Zucker},
  {Hodgkin}, {Gilmore}, {Prema} et~al.}}]{Irwin:07}
\bibinfo{author}{\bibfnamefont{M.~J.} \bibnamefont{{Irwin}}},
  \bibinfo{author}{\bibfnamefont{V.}~\bibnamefont{{Belokurov}}},
  \bibinfo{author}{\bibfnamefont{N.~W.} \bibnamefont{{Evans}}},
  \bibinfo{author}{\bibfnamefont{E.~V.} \bibnamefont{{Ryan-Weber}}},
  \bibinfo{author}{\bibfnamefont{J.~T.~A.} \bibnamefont{{de Jong}}},
  \bibinfo{author}{\bibfnamefont{S.}~\bibnamefont{{Koposov}}},
  \bibinfo{author}{\bibfnamefont{D.~B.} \bibnamefont{{Zucker}}},
  \bibinfo{author}{\bibfnamefont{S.~T.} \bibnamefont{{Hodgkin}}},
  \bibinfo{author}{\bibfnamefont{G.}~\bibnamefont{{Gilmore}}},
  \bibinfo{author}{\bibfnamefont{P.}~\bibnamefont{{Prema}}},
  \bibnamefont{et~al.}, \bibinfo{journal}{\apjl}
  \textbf{\bibinfo{volume}{656}}, \bibinfo{pages}{L13} (\bibinfo{year}{2007}),
  \eprint{arXiv:astro-ph/0701154}.

\bibitem[{\citenamefont{{Zucker} et~al.}(2006)\citenamefont{{Zucker},
  {Belokurov}, {Evans}, {Kleyna}, {Irwin}, {Wilkinson}, {Fellhauer}, {Bramich},
  {Gilmore}, {Newberg} et~al.}}]{Zucker:06}
\bibinfo{author}{\bibfnamefont{D.~B.} \bibnamefont{{Zucker}}},
  \bibinfo{author}{\bibfnamefont{V.}~\bibnamefont{{Belokurov}}},
  \bibinfo{author}{\bibfnamefont{N.~W.} \bibnamefont{{Evans}}},
  \bibinfo{author}{\bibfnamefont{J.~T.} \bibnamefont{{Kleyna}}},
  \bibinfo{author}{\bibfnamefont{M.~J.} \bibnamefont{{Irwin}}},
  \bibinfo{author}{\bibfnamefont{M.~I.} \bibnamefont{{Wilkinson}}},
  \bibinfo{author}{\bibfnamefont{M.}~\bibnamefont{{Fellhauer}}},
  \bibinfo{author}{\bibfnamefont{D.~M.} \bibnamefont{{Bramich}}},
  \bibinfo{author}{\bibfnamefont{G.}~\bibnamefont{{Gilmore}}},
  \bibinfo{author}{\bibfnamefont{H.~J.} \bibnamefont{{Newberg}}},
  \bibnamefont{et~al.}, \bibinfo{journal}{\apjl}
  \textbf{\bibinfo{volume}{650}}, \bibinfo{pages}{L41} (\bibinfo{year}{2006}),
  \eprint{arXiv:astro-ph/0606633}.

\bibitem[{\citenamefont{{Belokurov} et~al.}(2006)\citenamefont{{Belokurov},
  {Zucker}, {Evans}, {Wilkinson}, {Irwin}, {Hodgkin}, {Bramich}, {Irwin},
  {Gilmore}, {Willman} et~al.}}]{Belokurov:06}
\bibinfo{author}{\bibfnamefont{V.}~\bibnamefont{{Belokurov}}},
  \bibinfo{author}{\bibfnamefont{D.~B.} \bibnamefont{{Zucker}}},
  \bibinfo{author}{\bibfnamefont{N.~W.} \bibnamefont{{Evans}}},
  \bibinfo{author}{\bibfnamefont{M.~I.} \bibnamefont{{Wilkinson}}},
  \bibinfo{author}{\bibfnamefont{M.~J.} \bibnamefont{{Irwin}}},
  \bibinfo{author}{\bibfnamefont{S.}~\bibnamefont{{Hodgkin}}},
  \bibinfo{author}{\bibfnamefont{D.~M.} \bibnamefont{{Bramich}}},
  \bibinfo{author}{\bibfnamefont{J.~M.} \bibnamefont{{Irwin}}},
  \bibinfo{author}{\bibfnamefont{G.}~\bibnamefont{{Gilmore}}},
  \bibinfo{author}{\bibfnamefont{B.}~\bibnamefont{{Willman}}},
  \bibnamefont{et~al.}, \bibinfo{journal}{\apjl}
  \textbf{\bibinfo{volume}{647}}, \bibinfo{pages}{L111} (\bibinfo{year}{2006}),
  \eprint{arXiv:astro-ph/0604355}.

\bibitem[{\citenamefont{{Mateo} et~al.}(1991)\citenamefont{{Mateo},
  {Olszewski}, {Welch}, {Fischer}, and {Kunkel}}}]{Mateo:91}
\bibinfo{author}{\bibfnamefont{M.}~\bibnamefont{{Mateo}}},
  \bibinfo{author}{\bibfnamefont{E.}~\bibnamefont{{Olszewski}}},
  \bibinfo{author}{\bibfnamefont{D.~L.} \bibnamefont{{Welch}}},
  \bibinfo{author}{\bibfnamefont{P.}~\bibnamefont{{Fischer}}},
  \bibnamefont{and} \bibinfo{author}{\bibfnamefont{W.}~\bibnamefont{{Kunkel}}},
  \bibinfo{journal}{\aj} \textbf{\bibinfo{volume}{102}}, \bibinfo{pages}{914}
  (\bibinfo{year}{1991}).

\bibitem[{\citenamefont{{Mateo}}(1998)}]{Mateo:98}
\bibinfo{author}{\bibfnamefont{M.~L.} \bibnamefont{{Mateo}}},
  \bibinfo{journal}{\araa} \textbf{\bibinfo{volume}{36}}, \bibinfo{pages}{435}
  (\bibinfo{year}{1998}), \eprint{astro-ph/9810070}.

\bibitem[{\citenamefont{{Bonanos} et~al.}(2004)\citenamefont{{Bonanos},
  {Stanek}, {Szentgyorgyi}, {Sasselov}, and {Bakos}}}]{Bonanos:03}
\bibinfo{author}{\bibfnamefont{A.~Z.} \bibnamefont{{Bonanos}}},
  \bibinfo{author}{\bibfnamefont{K.~Z.} \bibnamefont{{Stanek}}},
  \bibinfo{author}{\bibfnamefont{A.~H.} \bibnamefont{{Szentgyorgyi}}},
  \bibinfo{author}{\bibfnamefont{D.~D.} \bibnamefont{{Sasselov}}},
  \bibnamefont{and} \bibinfo{author}{\bibfnamefont{G.~{\'A}.}
  \bibnamefont{{Bakos}}}, \bibinfo{journal}{\aj}
  \textbf{\bibinfo{volume}{127}}, \bibinfo{pages}{861} (\bibinfo{year}{2004}),
  \eprint{arXiv:astro-ph/0310477}.

\bibitem[{\citenamefont{{Dall'Ora} et~al.}(2006)\citenamefont{{Dall'Ora},
  {Clementini}, {Kinemuchi}, {Ripepi}, {Marconi}, {Di Fabrizio}, {Greco},
  {Rodgers}, {Kuehn}, and {Smith}}}]{Dall'Ora:06}
\bibinfo{author}{\bibfnamefont{M.}~\bibnamefont{{Dall'Ora}}},
  \bibinfo{author}{\bibfnamefont{G.}~\bibnamefont{{Clementini}}},
  \bibinfo{author}{\bibfnamefont{K.}~\bibnamefont{{Kinemuchi}}},
  \bibinfo{author}{\bibfnamefont{V.}~\bibnamefont{{Ripepi}}},
  \bibinfo{author}{\bibfnamefont{M.}~\bibnamefont{{Marconi}}},
  \bibinfo{author}{\bibfnamefont{L.}~\bibnamefont{{Di Fabrizio}}},
  \bibinfo{author}{\bibfnamefont{C.}~\bibnamefont{{Greco}}},
  \bibinfo{author}{\bibfnamefont{C.~T.} \bibnamefont{{Rodgers}}},
  \bibinfo{author}{\bibfnamefont{C.}~\bibnamefont{{Kuehn}}}, \bibnamefont{and}
  \bibinfo{author}{\bibfnamefont{H.~A.} \bibnamefont{{Smith}}},
  \bibinfo{journal}{\apjl} \textbf{\bibinfo{volume}{653}},
  \bibinfo{pages}{L109} (\bibinfo{year}{2006}),
  \eprint{arXiv:astro-ph/0611285}.

\bibitem[{\citenamefont{{Coleman} et~al.}(2007)\citenamefont{{Coleman}, {de
  Jong}, {Martin}, {Rix}, {Sand}, {Bell}, {Pogge}, {Thompson}, {Hippelein},
  {Giallongo} et~al.}}]{Coleman:07a}
\bibinfo{author}{\bibfnamefont{M.~G.} \bibnamefont{{Coleman}}},
  \bibinfo{author}{\bibfnamefont{J.~T.~A.} \bibnamefont{{de Jong}}},
  \bibinfo{author}{\bibfnamefont{N.~F.} \bibnamefont{{Martin}}},
  \bibinfo{author}{\bibfnamefont{H.-W.} \bibnamefont{{Rix}}},
  \bibinfo{author}{\bibfnamefont{D.~J.} \bibnamefont{{Sand}}},
  \bibinfo{author}{\bibfnamefont{E.~F.} \bibnamefont{{Bell}}},
  \bibinfo{author}{\bibfnamefont{R.~W.} \bibnamefont{{Pogge}}},
  \bibinfo{author}{\bibfnamefont{D.~J.} \bibnamefont{{Thompson}}},
  \bibinfo{author}{\bibfnamefont{H.}~\bibnamefont{{Hippelein}}},
  \bibinfo{author}{\bibfnamefont{E.}~\bibnamefont{{Giallongo}}},
  \bibnamefont{et~al.}, \bibinfo{journal}{\apjl}
  \textbf{\bibinfo{volume}{668}}, \bibinfo{pages}{L43} (\bibinfo{year}{2007}).

\bibitem[{\citenamefont{{de Jong} et~al.}(2008)\citenamefont{{de Jong},
  {Harris}, {Coleman}, {Martin}, {Bell}, {Rix}, {Hill}, {Skillman}, {Sand},
  {Olszewski} et~al.}}]{Jong:08}
\bibinfo{author}{\bibfnamefont{J.~T.~A.} \bibnamefont{{de Jong}}},
  \bibinfo{author}{\bibfnamefont{J.}~\bibnamefont{{Harris}}},
  \bibinfo{author}{\bibfnamefont{M.~G.} \bibnamefont{{Coleman}}},
  \bibinfo{author}{\bibfnamefont{N.~F.} \bibnamefont{{Martin}}},
  \bibinfo{author}{\bibfnamefont{E.~F.} \bibnamefont{{Bell}}},
  \bibinfo{author}{\bibfnamefont{H.-W.} \bibnamefont{{Rix}}},
  \bibinfo{author}{\bibfnamefont{J.~M.} \bibnamefont{{Hill}}},
  \bibinfo{author}{\bibfnamefont{E.~D.} \bibnamefont{{Skillman}}},
  \bibinfo{author}{\bibfnamefont{D.~J.} \bibnamefont{{Sand}}},
  \bibinfo{author}{\bibfnamefont{E.~W.} \bibnamefont{{Olszewski}}},
  \bibnamefont{et~al.}, \bibinfo{journal}{\apj} \textbf{\bibinfo{volume}{680}},
  \bibinfo{pages}{1112} (\bibinfo{year}{2008}), \eprint{arXiv:0801.4027}.

\bibitem[{\citenamefont{{Okamoto} et~al.}(2008)\citenamefont{{Okamoto},
  {Arimoto}, {Yamada}, and {Onodera}}}]{Okamoto:08}
\bibinfo{author}{\bibfnamefont{S.}~\bibnamefont{{Okamoto}}},
  \bibinfo{author}{\bibfnamefont{N.}~\bibnamefont{{Arimoto}}},
  \bibinfo{author}{\bibfnamefont{Y.}~\bibnamefont{{Yamada}}}, \bibnamefont{and}
  \bibinfo{author}{\bibfnamefont{M.}~\bibnamefont{{Onodera}}},
  \bibinfo{journal}{ArXiv e-prints} \textbf{\bibinfo{volume}{804}}
  (\bibinfo{year}{2008}), \eprint{0804.2976}.

\bibitem[{\citenamefont{{Lee} et~al.}(2003)\citenamefont{{Lee}, {Park}, {Park},
  {Sohn}, {Oh}, {Yuk}, {Rey}, {Lee}, {Lee}, {Kim} et~al.}}]{Lee:03}
\bibinfo{author}{\bibfnamefont{M.~G.} \bibnamefont{{Lee}}},
  \bibinfo{author}{\bibfnamefont{H.~S.} \bibnamefont{{Park}}},
  \bibinfo{author}{\bibfnamefont{J.-H.} \bibnamefont{{Park}}},
  \bibinfo{author}{\bibfnamefont{Y.-J.} \bibnamefont{{Sohn}}},
  \bibinfo{author}{\bibfnamefont{S.~J.} \bibnamefont{{Oh}}},
  \bibinfo{author}{\bibfnamefont{I.-S.} \bibnamefont{{Yuk}}},
  \bibinfo{author}{\bibfnamefont{S.-C.} \bibnamefont{{Rey}}},
  \bibinfo{author}{\bibfnamefont{S.-G.} \bibnamefont{{Lee}}},
  \bibinfo{author}{\bibfnamefont{Y.-W.} \bibnamefont{{Lee}}},
  \bibinfo{author}{\bibfnamefont{H.-I.} \bibnamefont{{Kim}}},
  \bibnamefont{et~al.}, \bibinfo{journal}{\aj} \textbf{\bibinfo{volume}{126}},
  \bibinfo{pages}{2840} (\bibinfo{year}{2003}).

\bibitem[{\citenamefont{{Rizzi} et~al.}(2007)\citenamefont{{Rizzi}, {Held},
  {Saviane}, {Tully}, and {Gullieuszik}}}]{Rizzi:07}
\bibinfo{author}{\bibfnamefont{L.}~\bibnamefont{{Rizzi}}},
  \bibinfo{author}{\bibfnamefont{E.~V.} \bibnamefont{{Held}}},
  \bibinfo{author}{\bibfnamefont{I.}~\bibnamefont{{Saviane}}},
  \bibinfo{author}{\bibfnamefont{R.~B.} \bibnamefont{{Tully}}},
  \bibnamefont{and}
  \bibinfo{author}{\bibfnamefont{M.}~\bibnamefont{{Gullieuszik}}},
  \bibinfo{journal}{\mnras} \textbf{\bibinfo{volume}{380}},
  \bibinfo{pages}{1255} (\bibinfo{year}{2007}), \eprint{arXiv:0707.0521}.

\bibitem[{\citenamefont{{Bellazzini} et~al.}(2004)\citenamefont{{Bellazzini},
  {Gennari}, {Ferraro}, and {Sollima}}}]{Bellazzini:04}
\bibinfo{author}{\bibfnamefont{M.}~\bibnamefont{{Bellazzini}}},
  \bibinfo{author}{\bibfnamefont{N.}~\bibnamefont{{Gennari}}},
  \bibinfo{author}{\bibfnamefont{F.~R.} \bibnamefont{{Ferraro}}},
  \bibnamefont{and}
  \bibinfo{author}{\bibfnamefont{A.}~\bibnamefont{{Sollima}}},
  \bibinfo{journal}{\mnras} \textbf{\bibinfo{volume}{354}},
  \bibinfo{pages}{708} (\bibinfo{year}{2004}), \eprint{arXiv:astro-ph/0407444}.

\bibitem[{\citenamefont{{Bellazzini} et~al.}(2005)\citenamefont{{Bellazzini},
  {Gennari}, and {Ferraro}}}]{Bellazzini:05}
\bibinfo{author}{\bibfnamefont{M.}~\bibnamefont{{Bellazzini}}},
  \bibinfo{author}{\bibfnamefont{N.}~\bibnamefont{{Gennari}}},
  \bibnamefont{and} \bibinfo{author}{\bibfnamefont{F.~R.}
  \bibnamefont{{Ferraro}}}, \bibinfo{journal}{\mnras}
  \textbf{\bibinfo{volume}{360}}, \bibinfo{pages}{185} (\bibinfo{year}{2005}),
  \eprint{arXiv:astro-ph/0503418}.

\bibitem[{\citenamefont{{Pietrzy{\'n}ski}
  et~al.}(2008)\citenamefont{{Pietrzy{\'n}ski}, {Gieren}, {Szewczyk}, {Walker},
  {Rizzi}, {Bresolin}, {Kudritzki}, {Nalewajko}, {Storm}, {Dall'Ora}
  et~al.}}]{Pietrzynski:08}
\bibinfo{author}{\bibfnamefont{G.}~\bibnamefont{{Pietrzy{\'n}ski}}},
  \bibinfo{author}{\bibfnamefont{W.}~\bibnamefont{{Gieren}}},
  \bibinfo{author}{\bibfnamefont{O.}~\bibnamefont{{Szewczyk}}},
  \bibinfo{author}{\bibfnamefont{A.}~\bibnamefont{{Walker}}},
  \bibinfo{author}{\bibfnamefont{L.}~\bibnamefont{{Rizzi}}},
  \bibinfo{author}{\bibfnamefont{F.}~\bibnamefont{{Bresolin}}},
  \bibinfo{author}{\bibfnamefont{R.-P.} \bibnamefont{{Kudritzki}}},
  \bibinfo{author}{\bibfnamefont{K.}~\bibnamefont{{Nalewajko}}},
  \bibinfo{author}{\bibfnamefont{J.}~\bibnamefont{{Storm}}},
  \bibinfo{author}{\bibfnamefont{M.}~\bibnamefont{{Dall'Ora}}},
  \bibnamefont{et~al.}, \bibinfo{journal}{\aj} \textbf{\bibinfo{volume}{135}},
  \bibinfo{pages}{1993} (\bibinfo{year}{2008}), \eprint{arXiv:0804.0347}.

\bibitem[{\citenamefont{{Mu{\~n}oz} et~al.}(2006)\citenamefont{{Mu{\~n}oz},
  {Carlin}, {Frinchaboy}, {Nidever}, {Majewski}, and {Patterson}}}]{Munoz:06}
\bibinfo{author}{\bibfnamefont{R.~R.} \bibnamefont{{Mu{\~n}oz}}},
  \bibinfo{author}{\bibfnamefont{J.~L.} \bibnamefont{{Carlin}}},
  \bibinfo{author}{\bibfnamefont{P.~M.} \bibnamefont{{Frinchaboy}}},
  \bibinfo{author}{\bibfnamefont{D.~L.} \bibnamefont{{Nidever}}},
  \bibinfo{author}{\bibfnamefont{S.~R.} \bibnamefont{{Majewski}}},
  \bibnamefont{and} \bibinfo{author}{\bibfnamefont{R.~J.}
  \bibnamefont{{Patterson}}}, \bibinfo{journal}{\apjl}
  \textbf{\bibinfo{volume}{650}}, \bibinfo{pages}{L51} (\bibinfo{year}{2006}),
  \eprint{arXiv:astro-ph/0606271}.

\bibitem[{\citenamefont{{Spergel} et~al.}(2007)\citenamefont{{Spergel}, {Bean},
  {Dor{\'e}}, {Nolta}, {Bennett}, {Dunkley}, {Hinshaw}, {Jarosik}, {Komatsu},
  {Page} et~al.}}]{Spergel:07}
\bibinfo{author}{\bibfnamefont{D.~N.} \bibnamefont{{Spergel}}},
  \bibinfo{author}{\bibfnamefont{R.}~\bibnamefont{{Bean}}},
  \bibinfo{author}{\bibfnamefont{O.}~\bibnamefont{{Dor{\'e}}}},
  \bibinfo{author}{\bibfnamefont{M.~R.} \bibnamefont{{Nolta}}},
  \bibinfo{author}{\bibfnamefont{C.~L.} \bibnamefont{{Bennett}}},
  \bibinfo{author}{\bibfnamefont{J.}~\bibnamefont{{Dunkley}}},
  \bibinfo{author}{\bibfnamefont{G.}~\bibnamefont{{Hinshaw}}},
  \bibinfo{author}{\bibfnamefont{N.}~\bibnamefont{{Jarosik}}},
  \bibinfo{author}{\bibfnamefont{E.}~\bibnamefont{{Komatsu}}},
  \bibinfo{author}{\bibfnamefont{L.}~\bibnamefont{{Page}}},
  \bibnamefont{et~al.}, \bibinfo{journal}{\apjs}
  \textbf{\bibinfo{volume}{170}}, \bibinfo{pages}{377} (\bibinfo{year}{2007}),
  \eprint{arXiv:astro-ph/0603449}.

\bibitem[{\citenamefont{Boyarsky
  et~al.}(2006{\natexlab{a}})\citenamefont{Boyarsky, Neronov, Ruchayskiy, and
  Shaposhnikov}}]{Boyarsky:05}
\bibinfo{author}{\bibfnamefont{A.}~\bibnamefont{Boyarsky}},
  \bibinfo{author}{\bibfnamefont{A.}~\bibnamefont{Neronov}},
  \bibinfo{author}{\bibfnamefont{O.}~\bibnamefont{Ruchayskiy}},
  \bibnamefont{and}
  \bibinfo{author}{\bibfnamefont{M.}~\bibnamefont{Shaposhnikov}},
  \bibinfo{journal}{\mnras} \textbf{\bibinfo{volume}{370}},
  \bibinfo{pages}{213} (\bibinfo{year}{2006}{\natexlab{a}}),
  \eprint{astro-ph/0512509}.

\bibitem[{\citenamefont{Boyarsky
  et~al.}(2006{\natexlab{b}})\citenamefont{Boyarsky, Neronov, Ruchayskiy, and
  Shaposhnikov}}]{Boyarsky:06b}
\bibinfo{author}{\bibfnamefont{A.}~\bibnamefont{Boyarsky}},
  \bibinfo{author}{\bibfnamefont{A.}~\bibnamefont{Neronov}},
  \bibinfo{author}{\bibfnamefont{O.}~\bibnamefont{Ruchayskiy}},
  \bibnamefont{and}
  \bibinfo{author}{\bibfnamefont{M.}~\bibnamefont{Shaposhnikov}},
  \bibinfo{journal}{\prd} \textbf{\bibinfo{volume}{74}},
  \bibinfo{pages}{103506} (\bibinfo{year}{2006}{\natexlab{b}}),
  \eprint{astro-ph/0603368}.

\bibitem[{\citenamefont{Boyarsky
  et~al.}(2006{\natexlab{c}})\citenamefont{Boyarsky, Neronov, Ruchayskiy,
  Shaposhnikov, and Tkachev}}]{Boyarsky:06c}
\bibinfo{author}{\bibfnamefont{A.}~\bibnamefont{Boyarsky}},
  \bibinfo{author}{\bibfnamefont{A.}~\bibnamefont{Neronov}},
  \bibinfo{author}{\bibfnamefont{O.}~\bibnamefont{Ruchayskiy}},
  \bibinfo{author}{\bibfnamefont{M.}~\bibnamefont{Shaposhnikov}},
  \bibnamefont{and} \bibinfo{author}{\bibfnamefont{I.}~\bibnamefont{Tkachev}},
  \bibinfo{journal}{\prl} \textbf{\bibinfo{volume}{97}},
  \bibinfo{pages}{261302} (\bibinfo{year}{2006}{\natexlab{c}}),
  \eprint{astro-ph/0603660}.

\bibitem[{\citenamefont{{Riemer-S{\o}rensen}
  et~al.}(2006)\citenamefont{{Riemer-S{\o}rensen}, {Hansen}, and
  {Pedersen}}}]{Riemer:06}
\bibinfo{author}{\bibfnamefont{S.}~\bibnamefont{{Riemer-S{\o}rensen}}},
  \bibinfo{author}{\bibfnamefont{S.~H.} \bibnamefont{{Hansen}}},
  \bibnamefont{and}
  \bibinfo{author}{\bibfnamefont{K.}~\bibnamefont{{Pedersen}}},
  \bibinfo{journal}{\apjl} \textbf{\bibinfo{volume}{644}}, \bibinfo{pages}{L33}
  (\bibinfo{year}{2006}), \eprint{astro-ph/0603661}.

\bibitem[{\citenamefont{Watson et~al.}(2006)\citenamefont{Watson, Beacom,
  Yuksel, and Walker}}]{Watson:06}
\bibinfo{author}{\bibfnamefont{C.~R.} \bibnamefont{Watson}},
  \bibinfo{author}{\bibfnamefont{J.~F.} \bibnamefont{Beacom}},
  \bibinfo{author}{\bibfnamefont{H.}~\bibnamefont{Yuksel}}, \bibnamefont{and}
  \bibinfo{author}{\bibfnamefont{T.~P.} \bibnamefont{Walker}},
  \bibinfo{journal}{Phys. Rev.} \textbf{\bibinfo{volume}{D74}},
  \bibinfo{pages}{033009} (\bibinfo{year}{2006}), \eprint{astro-ph/0605424}.

\bibitem[{\citenamefont{Boyarsky
  et~al.}(2008{\natexlab{c}})\citenamefont{Boyarsky, Ruchayskiy, and
  Markevitch}}]{Boyarsky:06e}
\bibinfo{author}{\bibfnamefont{A.}~\bibnamefont{Boyarsky}},
  \bibinfo{author}{\bibfnamefont{O.}~\bibnamefont{Ruchayskiy}},
  \bibnamefont{and}
  \bibinfo{author}{\bibfnamefont{M.}~\bibnamefont{Markevitch}},
  \bibinfo{journal}{\apj} \textbf{\bibinfo{volume}{673}}, \bibinfo{pages}{752}
  (\bibinfo{year}{2008}{\natexlab{c}}), \eprint{astro-ph/0611168}.

\bibitem[{\citenamefont{{Abazajian} et~al.}(2007)\citenamefont{{Abazajian},
  {Markevitch}, {Koushiappas}, and {Hickox}}}]{Abazajian:06b}
\bibinfo{author}{\bibfnamefont{K.~N.} \bibnamefont{{Abazajian}}},
  \bibinfo{author}{\bibfnamefont{M.}~\bibnamefont{{Markevitch}}},
  \bibinfo{author}{\bibfnamefont{S.~M.} \bibnamefont{{Koushiappas}}},
  \bibnamefont{and} \bibinfo{author}{\bibfnamefont{R.~C.}
  \bibnamefont{{Hickox}}}, \bibinfo{journal}{\prd}
  \textbf{\bibinfo{volume}{75}}, \bibinfo{pages}{063511}
  (\bibinfo{year}{2007}), \eprint{arXiv:astro-ph/0611144}.

\bibitem[{\citenamefont{Boyarsky
  et~al.}(2007{\natexlab{a}})\citenamefont{Boyarsky, Nevalainen, and
  Ruchayskiy}}]{Boyarsky:06d}
\bibinfo{author}{\bibfnamefont{A.}~\bibnamefont{Boyarsky}},
  \bibinfo{author}{\bibfnamefont{J.}~\bibnamefont{Nevalainen}},
  \bibnamefont{and}
  \bibinfo{author}{\bibfnamefont{O.}~\bibnamefont{Ruchayskiy}},
  \bibinfo{journal}{\aap} \textbf{\bibinfo{volume}{471}}, \bibinfo{pages}{51}
  (\bibinfo{year}{2007}{\natexlab{a}}), \eprint{astro-ph/0610961}.

\bibitem[{\citenamefont{Boyarsky
  et~al.}(2007{\natexlab{b}})\citenamefont{Boyarsky, den Herder, Neronov, and
  Ruchayskiy}}]{Boyarsky:06f}
\bibinfo{author}{\bibfnamefont{A.}~\bibnamefont{Boyarsky}},
  \bibinfo{author}{\bibfnamefont{J.~W.} \bibnamefont{den Herder}},
  \bibinfo{author}{\bibfnamefont{A.}~\bibnamefont{Neronov}}, \bibnamefont{and}
  \bibinfo{author}{\bibfnamefont{O.}~\bibnamefont{Ruchayskiy}},
  \bibinfo{journal}{Astropart.\ Phys.} \textbf{\bibinfo{volume}{28}},
  \bibinfo{pages}{303} (\bibinfo{year}{2007}{\natexlab{b}}),
  \eprint{astro-ph/0612219}.

\bibitem[{\citenamefont{{Boyarsky} et~al.}(2008)\citenamefont{{Boyarsky},
  {Iakubovskyi}, {Ruchayskiy}, and {Savchenko}}}]{Boyarsky:07a}
\bibinfo{author}{\bibfnamefont{A.}~\bibnamefont{{Boyarsky}}},
  \bibinfo{author}{\bibfnamefont{D.}~\bibnamefont{{Iakubovskyi}}},
  \bibinfo{author}{\bibfnamefont{O.}~\bibnamefont{{Ruchayskiy}}},
  \bibnamefont{and}
  \bibinfo{author}{\bibfnamefont{V.}~\bibnamefont{{Savchenko}}},
  \bibinfo{journal}{\mnras} \textbf{\bibinfo{volume}{387}},
  \bibinfo{pages}{1361} (\bibinfo{year}{2008}), \eprint{arXiv:0709.2301}.

\bibitem[{\citenamefont{Boyarsky et~al.}(2008)\citenamefont{Boyarsky, Malyshev,
  Neronov, and Ruchayskiy}}]{Boyarsky:07b}
\bibinfo{author}{\bibfnamefont{A.}~\bibnamefont{Boyarsky}},
  \bibinfo{author}{\bibfnamefont{D.}~\bibnamefont{Malyshev}},
  \bibinfo{author}{\bibfnamefont{A.}~\bibnamefont{Neronov}}, \bibnamefont{and}
  \bibinfo{author}{\bibfnamefont{O.}~\bibnamefont{Ruchayskiy}},
  \bibinfo{journal}{\mnras} \textbf{\bibinfo{volume}{387}},
  \bibinfo{pages}{1345} (\bibinfo{year}{2008}), \eprint{0710.4922}.

\bibitem[{\citenamefont{{Blumenthal} et~al.}(1986)\citenamefont{{Blumenthal},
  {Faber}, {Flores}, and {Primack}}}]{Blumenthal:86}
\bibinfo{author}{\bibfnamefont{G.~R.} \bibnamefont{{Blumenthal}}},
  \bibinfo{author}{\bibfnamefont{S.~M.} \bibnamefont{{Faber}}},
  \bibinfo{author}{\bibfnamefont{R.}~\bibnamefont{{Flores}}}, \bibnamefont{and}
  \bibinfo{author}{\bibfnamefont{J.~R.} \bibnamefont{{Primack}}},
  \bibinfo{journal}{\apj} \textbf{\bibinfo{volume}{301}}, \bibinfo{pages}{27}
  (\bibinfo{year}{1986}).

\bibitem[{\citenamefont{{Gnedin} et~al.}(2004)\citenamefont{{Gnedin},
  {Kravtsov}, {Klypin}, and {Nagai}}}]{Gnedin:04}
\bibinfo{author}{\bibfnamefont{O.~Y.} \bibnamefont{{Gnedin}}},
  \bibinfo{author}{\bibfnamefont{A.~V.} \bibnamefont{{Kravtsov}}},
  \bibinfo{author}{\bibfnamefont{A.~A.} \bibnamefont{{Klypin}}},
  \bibnamefont{and} \bibinfo{author}{\bibfnamefont{D.}~\bibnamefont{{Nagai}}},
  \bibinfo{journal}{\apj} \textbf{\bibinfo{volume}{616}}, \bibinfo{pages}{16}
  (\bibinfo{year}{2004}), \eprint{arXiv:astro-ph/0406247}.

\bibitem[{\citenamefont{Lin and Faber}(1983)}]{Lin:83}
\bibinfo{author}{\bibfnamefont{D.~N.~C.} \bibnamefont{Lin}} \bibnamefont{and}
  \bibinfo{author}{\bibfnamefont{S.~M.} \bibnamefont{Faber}},
  \bibinfo{journal}{\apj} \textbf{\bibinfo{volume}{266}}, \bibinfo{pages}{L21}
  (\bibinfo{year}{1983}).

\bibitem[{\citenamefont{{Moore}}(1994)}]{Moore:94}
\bibinfo{author}{\bibfnamefont{B.}~\bibnamefont{{Moore}}},
  \bibinfo{journal}{\nat} \textbf{\bibinfo{volume}{370}}, \bibinfo{pages}{629}
  (\bibinfo{year}{1994}).

\bibitem[{\citenamefont{{Mastropietro}
  et~al.}(2005)\citenamefont{{Mastropietro}, {Moore}, {Mayer}, {Debattista},
  {Piffaretti}, and {Stadel}}}]{Mastropietro:05}
\bibinfo{author}{\bibfnamefont{C.}~\bibnamefont{{Mastropietro}}},
  \bibinfo{author}{\bibfnamefont{B.}~\bibnamefont{{Moore}}},
  \bibinfo{author}{\bibfnamefont{L.}~\bibnamefont{{Mayer}}},
  \bibinfo{author}{\bibfnamefont{V.~P.} \bibnamefont{{Debattista}}},
  \bibinfo{author}{\bibfnamefont{R.}~\bibnamefont{{Piffaretti}}},
  \bibnamefont{and} \bibinfo{author}{\bibfnamefont{J.}~\bibnamefont{{Stadel}}},
  \bibinfo{journal}{\mnras} \textbf{\bibinfo{volume}{364}},
  \bibinfo{pages}{607} (\bibinfo{year}{2005}), \eprint{arXiv:astro-ph/0411648}.

\bibitem[{\citenamefont{{Mayer} et~al.}(2006)\citenamefont{{Mayer},
  {Mastropietro}, {Wadsley}, {Stadel}, and {Moore}}}]{Mayer:06}
\bibinfo{author}{\bibfnamefont{L.}~\bibnamefont{{Mayer}}},
  \bibinfo{author}{\bibfnamefont{C.}~\bibnamefont{{Mastropietro}}},
  \bibinfo{author}{\bibfnamefont{J.}~\bibnamefont{{Wadsley}}},
  \bibinfo{author}{\bibfnamefont{J.}~\bibnamefont{{Stadel}}}, \bibnamefont{and}
  \bibinfo{author}{\bibfnamefont{B.}~\bibnamefont{{Moore}}},
  \bibinfo{journal}{\mnras} \textbf{\bibinfo{volume}{369}},
  \bibinfo{pages}{1021} (\bibinfo{year}{2006}),
  \eprint{arXiv:astro-ph/0504277}.

\bibitem[{\citenamefont{{Mayer} et~al.}(2007)\citenamefont{{Mayer},
  {Kazantzidis}, {Mastropietro}, and {Wadsley}}}]{Mayer:07}
\bibinfo{author}{\bibfnamefont{L.}~\bibnamefont{{Mayer}}},
  \bibinfo{author}{\bibfnamefont{S.}~\bibnamefont{{Kazantzidis}}},
  \bibinfo{author}{\bibfnamefont{C.}~\bibnamefont{{Mastropietro}}},
  \bibnamefont{and}
  \bibinfo{author}{\bibfnamefont{J.}~\bibnamefont{{Wadsley}}},
  \bibinfo{journal}{\nat} \textbf{\bibinfo{volume}{445}}, \bibinfo{pages}{738}
  (\bibinfo{year}{2007}), \eprint{arXiv:astro-ph/0702495}.

\bibitem[{\citenamefont{{Read} et~al.}(2006)\citenamefont{{Read}, {Pontzen},
  and {Viel}}}]{Read:06}
\bibinfo{author}{\bibfnamefont{J.~I.} \bibnamefont{{Read}}},
  \bibinfo{author}{\bibfnamefont{A.~P.} \bibnamefont{{Pontzen}}},
  \bibnamefont{and} \bibinfo{author}{\bibfnamefont{M.}~\bibnamefont{{Viel}}},
  \bibinfo{journal}{\mnras} \textbf{\bibinfo{volume}{371}},
  \bibinfo{pages}{885} (\bibinfo{year}{2006}), \eprint{arXiv:astro-ph/0606391}.

\bibitem[{\citenamefont{{Naab} et~al.}(2007)\citenamefont{{Naab}, {Johansson},
  {Ostriker}, and {Efstathiou}}}]{Naab:07a}
\bibinfo{author}{\bibfnamefont{T.}~\bibnamefont{{Naab}}},
  \bibinfo{author}{\bibfnamefont{P.~H.} \bibnamefont{{Johansson}}},
  \bibinfo{author}{\bibfnamefont{J.~P.} \bibnamefont{{Ostriker}}},
  \bibnamefont{and}
  \bibinfo{author}{\bibfnamefont{G.}~\bibnamefont{{Efstathiou}}},
  \bibinfo{journal}{\apj} \textbf{\bibinfo{volume}{658}}, \bibinfo{pages}{710}
  (\bibinfo{year}{2007}), \eprint{arXiv:astro-ph/0512235}.

\bibitem[{\citenamefont{{Bertschinger}}(1995)}]{Bertschinger:95}
\bibinfo{author}{\bibfnamefont{E.}~\bibnamefont{{Bertschinger}}}
  (\bibinfo{year}{1995}), \eprint{astro-ph/9506070}.

\bibitem[{\citenamefont{Klypin and Holtzman}(1997)}]{Klypin:97}
\bibinfo{author}{\bibfnamefont{A.}~\bibnamefont{Klypin}} \bibnamefont{and}
  \bibinfo{author}{\bibfnamefont{J.}~\bibnamefont{Holtzman}}
  (\bibinfo{year}{1997}), \eprint{astro-ph/9712217}.

\bibitem[{\citenamefont{Klypin}(2000)}]{Klypin:00}
\bibinfo{author}{\bibfnamefont{A.}~\bibnamefont{Klypin}}
  (\bibinfo{year}{2000}), \eprint{astro-ph/0005502}.

\bibitem[{\citenamefont{Taylor and Navarro}(2001)}]{Taylor:01}
\bibinfo{author}{\bibfnamefont{J.~E.} \bibnamefont{Taylor}} \bibnamefont{and}
  \bibinfo{author}{\bibfnamefont{J.~F.} \bibnamefont{Navarro}},
  \bibinfo{journal}{Astrophys. J.} \textbf{\bibinfo{volume}{563}},
  \bibinfo{pages}{483} (\bibinfo{year}{2001}), \eprint{astro-ph/0104002}.

\bibitem[{\citenamefont{{Peirani} et~al.}(2006)\citenamefont{{Peirani},
  {Durier}, and {de Freitas Pacheco}}}]{Peirani:06}
\bibinfo{author}{\bibfnamefont{S.}~\bibnamefont{{Peirani}}},
  \bibinfo{author}{\bibfnamefont{F.}~\bibnamefont{{Durier}}}, \bibnamefont{and}
  \bibinfo{author}{\bibfnamefont{J.~A.} \bibnamefont{{de Freitas Pacheco}}},
  \bibinfo{journal}{\mnras} \textbf{\bibinfo{volume}{367}},
  \bibinfo{pages}{1011} (\bibinfo{year}{2006}),
  \eprint{arXiv:astro-ph/0512482}.

\bibitem[{\citenamefont{{Peirani} and {de Freitas Pacheco}}(2007)}]{Peirani:07}
\bibinfo{author}{\bibfnamefont{S.}~\bibnamefont{{Peirani}}} \bibnamefont{and}
  \bibinfo{author}{\bibfnamefont{J.~A.} \bibnamefont{{de Freitas Pacheco}}},
  \bibinfo{journal}{ArXiv Astrophysics e-prints}  (\bibinfo{year}{2007}),
  \eprint{astro-ph/0701292}.

\bibitem[{\citenamefont{{Romano-D{\'{\i}}az}
  et~al.}(2007)\citenamefont{{Romano-D{\'{\i}}az}, {Hoffman}, {Heller},
  {Faltenbacher}, {Jones}, and {Shlosman}}}]{Romano-Diaz:07}
\bibinfo{author}{\bibfnamefont{E.}~\bibnamefont{{Romano-D{\'{\i}}az}}},
  \bibinfo{author}{\bibfnamefont{Y.}~\bibnamefont{{Hoffman}}},
  \bibinfo{author}{\bibfnamefont{C.}~\bibnamefont{{Heller}}},
  \bibinfo{author}{\bibfnamefont{A.}~\bibnamefont{{Faltenbacher}}},
  \bibinfo{author}{\bibfnamefont{D.}~\bibnamefont{{Jones}}}, \bibnamefont{and}
  \bibinfo{author}{\bibfnamefont{I.}~\bibnamefont{{Shlosman}}},
  \bibinfo{journal}{\apj} \textbf{\bibinfo{volume}{657}}, \bibinfo{pages}{56}
  (\bibinfo{year}{2007}).

\bibitem[{\citenamefont{{Hoffman} et~al.}(2007)\citenamefont{{Hoffman},
  {Romano-D{\'{\i}}az}, {Shlosman}, and {Heller}}}]{Hoffman:07}
\bibinfo{author}{\bibfnamefont{Y.}~\bibnamefont{{Hoffman}}},
  \bibinfo{author}{\bibfnamefont{E.}~\bibnamefont{{Romano-D{\'{\i}}az}}},
  \bibinfo{author}{\bibfnamefont{I.}~\bibnamefont{{Shlosman}}},
  \bibnamefont{and} \bibinfo{author}{\bibfnamefont{C.}~\bibnamefont{{Heller}}},
  \bibinfo{journal}{\apj} \textbf{\bibinfo{volume}{671}}, \bibinfo{pages}{1108}
  (\bibinfo{year}{2007}), \eprint{arXiv:0706.0006}.

\bibitem[{\citenamefont{{Romano-Diaz} et~al.}(2006)\citenamefont{{Romano-Diaz},
  {Faltenbacher}, {Jones}, {Heller}, {Hoffman}, and
  {Shlosman}}}]{Romano-Diaz:06}
\bibinfo{author}{\bibfnamefont{E.}~\bibnamefont{{Romano-Diaz}}},
  \bibinfo{author}{\bibfnamefont{A.}~\bibnamefont{{Faltenbacher}}},
  \bibinfo{author}{\bibfnamefont{D.}~\bibnamefont{{Jones}}},
  \bibinfo{author}{\bibfnamefont{C.}~\bibnamefont{{Heller}}},
  \bibinfo{author}{\bibfnamefont{Y.}~\bibnamefont{{Hoffman}}},
  \bibnamefont{and}
  \bibinfo{author}{\bibfnamefont{I.}~\bibnamefont{{Shlosman}}},
  \bibinfo{journal}{\apjl} \textbf{\bibinfo{volume}{637}}, \bibinfo{pages}{L93}
  (\bibinfo{year}{2006}), \eprint{arXiv:astro-ph/0508272}.

\bibitem[{\citenamefont{{Gorbunov} et~al.}(2008)\citenamefont{{Gorbunov},
  {Khmelnitsky}, and {Rubakov}}}]{Gorbunov:08a}
\bibinfo{author}{\bibfnamefont{D.}~\bibnamefont{{Gorbunov}}},
  \bibinfo{author}{\bibfnamefont{A.}~\bibnamefont{{Khmelnitsky}}},
  \bibnamefont{and} \bibinfo{author}{\bibfnamefont{V.}~\bibnamefont{{Rubakov}}}
  (\bibinfo{year}{2008}), \eprint{0805.2836}.

\bibitem[{\citenamefont{Boyanovsky}(2008)}]{Boyanovsky:08b}
\bibinfo{author}{\bibfnamefont{D.}~\bibnamefont{Boyanovsky}},
  \bibinfo{journal}{Phys. Rev.} \textbf{\bibinfo{volume}{D78}},
  \bibinfo{pages}{103505} (\bibinfo{year}{2008}), \eprint{0807.0646}.

\bibitem[{\citenamefont{Gorbunov et~al.}(2008)\citenamefont{Gorbunov,
  Khmelnitsky, and Rubakov}}]{Gorbunov:08b}
\bibinfo{author}{\bibfnamefont{D.}~\bibnamefont{Gorbunov}},
  \bibinfo{author}{\bibfnamefont{A.}~\bibnamefont{Khmelnitsky}},
  \bibnamefont{and} \bibinfo{author}{\bibfnamefont{V.}~\bibnamefont{Rubakov}},
  \bibinfo{journal}{JCAP} \textbf{\bibinfo{volume}{0810}}, \bibinfo{pages}{041}
  (\bibinfo{year}{2008}), \eprint{0808.3910}.

\bibitem[{\citenamefont{{Landau} and {Lifshitz}}(1975)}]{Landau:v2}
\bibinfo{author}{\bibfnamefont{L.~D.} \bibnamefont{{Landau}}} \bibnamefont{and}
  \bibinfo{author}{\bibfnamefont{E.~M.} \bibnamefont{{Lifshitz}}},
  \emph{\bibinfo{title}{{The classical theory of fields}}}
  (\bibinfo{publisher}{Course of theoretical physics - Pergamon International
  Library of Science, Technology, Engineering and Social Studies, Oxford:
  Pergamon Press, 1975, 4th rev.engl.ed.}, \bibinfo{year}{1975}).

\bibitem[{\citenamefont{{Landau} and {Lifshitz}}(1980)}]{Landau:v5}
\bibinfo{author}{\bibfnamefont{L.~D.} \bibnamefont{{Landau}}} \bibnamefont{and}
  \bibinfo{author}{\bibfnamefont{E.~M.} \bibnamefont{{Lifshitz}}},
  \emph{\bibinfo{title}{{Statistical physics. Pt.1, Pt.2}}}
  (\bibinfo{publisher}{Course of theoretical physics, Pergamon International
  Library of Science, Technology, Engineering and Social Studies, Oxford:
  Pergamon Press, 1980|c1980, 3rd rev.and enlarg.~ed.}, \bibinfo{year}{1980}).

\end{thebibliography}

\let\jnlstyle=\rm\def\jref#1{{\jnlstyle#1}}\def\aj{\jref{AJ}}
  \def\araa{\jref{ARA\&A}} \def\apj{\jref{ApJ}} \def\apjl{\jref{ApJ}}
  \def\apjs{\jref{ApJS}} \def\ao{\jref{Appl.~Opt.}} \def\apss{\jref{Ap\&SS}}
  \def\aap{\jref{A\&A}} \def\aapr{\jref{A\&A~Rev.}} \def\aaps{\jref{A\&AS}}
  \def\azh{\jref{AZh}} \def\baas{\jref{BAAS}} \def\jrasc{\jref{JRASC}}
  \def\memras{\jref{MmRAS}} \def\mnras{\jref{MNRAS}}
  \def\pra{\jref{Phys.~Rev.~A}} \def\prb{\jref{Phys.~Rev.~B}}
  \def\prc{\jref{Phys.~Rev.~C}} \def\prd{\jref{Phys.~Rev.~D}}
  \def\pre{\jref{Phys.~Rev.~E}} \def\prl{\jref{Phys.~Rev.~Lett.}}
  \def\pasp{\jref{PASP}} \def\pasj{\jref{PASJ}} \def\qjras{\jref{QJRAS}}
  \def\skytel{\jref{S\&T}} \def\solphys{\jref{Sol.~Phys.}}
  \def\sovast{\jref{Soviet~Ast.}} \def\ssr{\jref{Space~Sci.~Rev.}}
  \def\zap{\jref{ZAp}} \def\nat{\jref{Nature}} \def\iaucirc{\jref{IAU~Circ.}}
  \def\aplett{\jref{Astrophys.~Lett.}}
  \def\apspr{\jref{Astrophys.~Space~Phys.~Res.}}
  \def\bain{\jref{Bull.~Astron.~Inst.~Netherlands}}
  \def\fcp{\jref{Fund.~Cosmic~Phys.}} \def\gca{\jref{Geochim.~Cosmochim.~Acta}}
  \def\grl{\jref{Geophys.~Res.~Lett.}} \def\jcp{\jref{J.~Chem.~Phys.}}
  \def\jgr{\jref{J.~Geophys.~Res.}}
  \def\jqsrt{\jref{J.~Quant.~Spec.~Radiat.~Transf.}}
  \def\memsai{\jref{Mem.~Soc.~Astron.~Italiana}}
  \def\nphysa{\jref{Nucl.~Phys.~A}} \def\physrep{\jref{Phys.~Rep.}}
  \def\physscr{\jref{Phys.~Scr}} \def\planss{\jref{Planet.~Space~Sci.}}
  \def\procspie{\jref{Proc.~SPIE}} \let\astap=\aap \let\apjlett=\apjl
  \let\apjsupp=\apjs \let\applopt=\ao

\end{document}